\documentclass[a4paper,11pt]{article}
\usepackage{jheppub} 
\usepackage{lineno}
\usepackage{tikz}
\usepackage{amsmath,amsfonts,amssymb,bm}
\usetikzlibrary{shapes.geometric}
\usetikzlibrary{arrows.meta}
\usetikzlibrary{decorations.pathmorphing}

\title{$20'$ Five-Point Function of $\mathcal{N}=4$ SYM and Stringy Corrections}

\author{Joao Vilas Boas}
\affiliation{School of Mathematical Sciences, Queen Mary University of London,\\
Mile End Road, London, E1 4NS, United Kingdom}

\emailAdd{joaomiguelvb@gmail.com}

\abstract{We set up a bootstrap approach to compute the first stringy correction to the supergravity regime of the correlation function of five 20' operators in $\mathcal{N}=4$ super Yang-Mills. We use factorization of Mellin amplitudes, supersymmetric constraints and protected observables to refine our ansatz, leaving only a single undetermined coefficient. We identify non-protected CFT data that is sensitive to the residual ambiguity of our ansatz, and we verify the compatibility of our result with the anticipated behaviour of the Mellin amplitude in the flat-space limit. As a byproduct of our analysis, we also compute the first stringy correction to the four-point correlators of three 20' operators and either one R-symmetry current or one stress tensor.  }

\begin{document}
\maketitle
\flushbottom

\section{Introduction}
Holographic correlators are a cornerstone of the AdS/CFT correspondence, providing a concrete computational framework for understanding the duality. 
Perhaps the most well-studied example of the correspondence states the duality between type IIB superstring theory on AdS$_5\times S^5$ and $\mathcal{N}=4$ super Yang-Mills (SYM) theory with $SU(N)$ gauge group. In the weak form of the duality, at large $N$ and large 't Hooft coupling $\lambda=g_{\text{YM}}^2 N$, the CFT observables can be computed using classical 10d type IIB supergravity. Higher-derivative corrections to the supergravity effective action correspond to stringy corrections in $1/\lambda$ and bulk loop diagrams correspond to corrections to the planar limit.

The traditional methods of computing holographic correlators rely on diagrammatic expansions in AdS that make use of explicit vertex interactions. However, this makes these methods impractical for most purposes. While all vertices can in principle be obtained from expanding the effective action on AdS$_5\times S^5$, this computation is extremely cumbersome. Furthermore, in this work, we will be interested in studying a five-point correlator that requires this expansion to be performed to quintic order. This computation has never been attempted in literature, and rightly so.~\footnote{See, however, recent progress on five-point couplings in type IIB around 10d flat space~\cite{Liu:2025uqu}.} On the other hand, more modern techniques avoid the complicated details of traditional methods and resort to supersymmetry and consistency conditions to fix the form of the holographic correlators. The bootstrap approach to compute these objects has been applied with tremendous success at the level of four-point correlators. For instance, at tree level, the correlations functions of four half-BPS operators of arbitrary Kaluza-Klein modes have been determined for all maximal~\cite{Rastelli:2017udc,Rastelli:2016nze,Alday:2020lbp,Alday:2020dtb} and half-maximal~\cite{Rastelli:2019gtj,Giusto:2020neo,Alday:2021odx} superconformal field theories. These objects have also been studied in the presence of $1/\lambda$~\cite{Goncalves:2014ffa,Binder:2019jwn,Chester:2018aca,Drummond:2020dwr,Alday:2022uxp, Alday:2022xwz, Alday:2023mvu,Alday:2024yax,Alday:2024rjs,Alday:2024ksp,Fardelli:2023fyq,Wang:2025pjo,Wang:2025owf} and $1/N$~\cite{Aharony:2016dwx,Alday:2017xua,Aprile:2017bgs,Alday:2018kkw,Alday:2018pdi,Alday:2019nin,Chester:2019pvm,Drummond:2019hel,Aprile:2019rep,Chester:2019jas,Drummond:2020uni,Alday:2020tgi,Chester:2020dja,Bissi:2020wtv,Bissi:2020woe,Komatsu:2020sag,Alday:2021vfb,Alday:2021ymb,Alday:2021ajh,Alday:2022rly,Chester:2025ssu,Huang:2021xws} corrections. 

Much less has been done at level of higher-point functions, but recent years have come with the advent of the study of these objects in CFTs. Despite the short timeframe, there is already a list of remarkable results in holographic theories that deserve mention. At the level of five-point functions of scalar half-BPS operators, both type IIB supergravity on AdS$_5\times S^5$~\cite{Goncalves:2019znr,Goncalves:2023oyx,Fernandes:2025eqe} and SYM on AdS$_5\times S^3$~\cite{Alday:2022lkk,Huang:2024dxr} were studied, including correlators with generic Kaluza-Klein (KK) modes. On the other hand, for six-point correlators only the lowest KK modes of supergravitons~\cite{Goncalves:2025jcg} and supergluons~\cite{Alday:2023kfm} were computed. For supergluon amplitudes however a recursive and constructive method for computing $n$-point correlators was proposed in~\cite{Cao:2023cwa,Cao:2024bky}.~\footnote{See also~\cite{Green:2020eyj} for maximal $U(1)$-violating $n$-point correlators.}

The increasing number of results for higher-point correlation functions should not come as a surprise. While it is easy to understand that these objects transcend a psychological barrier for many, the challenge they present does not come without reward. In fact, studying higher-point functions in CFTs and in the holographic context is of great importance. The first reason is that these observables provide access to CFT data that is not available at the level of four-point correlators that are within our reach. Indeed, the use of the operator product expansion (OPE) on these functions shows that they contain, in generic cases, an infinite amount of four-point functions and, necessarily, the data within. In the holographic context, in particular, the OPE coefficient between two double-trace and one single-trace operators is not available in a correlation function of four single-traces, but can be obtained by considering a five-point function of those same operators. Furthermore, selective use of OPE at higher-point functions allows one to extract less studied four-point correlators such as those with double-traces as external operators~\cite{Aprile:2024lwy,Aprile:2025hlt}.

A second reason to consider higher-point functions is related with the use of constructive AdS unitarity methods~\cite{Meltzer:2019nbs,Meltzer:2020qbr} to find higher-loop correlators. At two loops, for instance, the use of these methods already requires the knowledge of tree-level five-point functions. Furthermore, in the AdS/CFT correspondence, holographic correlators correspond to on-shell scattering amplitudes in AdS. Naturally, in recent years, there has been an effort to find AdS generalizations of properties of flat-space scattering amplitudes~\cite{Alday:2021odx,Zhou:2021gnu,Farrow:2018yni,Armstrong:2020woi,Albayrak:2020fyp,Jain:2021qcl,Diwakar:2021juk,Cheung:2022pdk,Herderschee:2022ntr,Drummond:2022dxd,Bissi:2022wuh,Armstrong:2022mfr,Lee:2022fgr,Li:2022tby}. Studying higher-point correlators provides then new tests for these generalizations and offers a new realm for searching for new identities and relations that might only become apparent at higher multiplicities. 

Finally, in~\cite{Caron-Huot:2018kta}, it was shown that all tree-level four-point correlators of chiral primaries of arbitrary KK modes in type IIB supergravity on AdS$_5\times S^5$ can be organized by a 10-dimensional hidden conformal symmetry.~\footnote{A higher-dimensional hidden symmetry was also observed in four-point correlators with two giant gravitons and two half-BPS light operators in~\cite{Chen:2025yxg}.} Very recently, the results of~\cite{Huang:2024dxr,Fernandes:2025eqe}  in AdS$_5\times S^3$ and AdS$_5\times S^5$ show that five-point correlation functions of arbitrary KK modes of supergluons and supergravitons admit a similar organization by 8- and 10-dimensional hidden symmetries. This is a strong indication that the existence of these hidden symmetries is not fortuitous and exclusive of four-point correlators and calls for a deeper understanding of the origin of these symmetries.

For all these reasons, we continue the exploration of higher-point correlators. In particular, we consider a five-point function of supergravitons in the lowest KK mode on AdS$_5\times S^5$ and compute the first stringy correction to the result of~\cite{Goncalves:2019znr}. To do so, we write an ansatz for this correction in Mellin space, which enjoys a simple analytic structure in holographic theories~\cite{Penedones:2010ue}, and we fix its coefficients by demanding that it satisfies consistency conditions. In our bootstrap, similarly to what was observed in~\cite{Goncalves:2023oyx}, factorization of Mellin amplitudes plays a pivotal role in determining the singular behaviour of our ansatz. The remaining regular terms are further   restricted by supersymmetric constraints known as Drukker-Plefka~\cite{Drukker:2009sf} and chiral algebra~\cite{Beem:2013sza} twists. While the former can be applied entirely in Mellin space following the prescription of~\cite{Goncalves:2023oyx}, the latter restricts the kinematics of the correlation function to a 2d plane and therefore needs to be imposed in position space. We find that cumulatively these constraints fix almost entirely our ansatz up to two undetermined coefficients. Via the OPE, we then extract contributions from our ansatz to some four-point correlators. These observables, being protected, impose further constraints on our ansatz, fixing one more coefficient.

For convenience, we summarize here the structure of the paper.  In section~\ref{sec:setup}, we define more precisely the five-point correlator we want to study and succinctly present R-symmetry and conformal constraints that it obeys. In section~\ref{sec:Mellinspace}, we review some of the basics of Mellin space formalism, highlighting its key simplifications for holographic theories, and we present the factorization property of Mellin amplitudes in~\ref{sec:Factorization}. Section~\ref{sec:bootstrap} is devoted to writing down an ansatz for the first stringy correction of our correlator and explaining the steps we take to fix it. We end up with a discussion in section~\ref{sec:discussion} including an outlook for future directions. For improved readability, some technical details and ancillary results were not included in the main text but are shared in the appendices. In Appendix~\ref{sec:fourpoints}, we derive the form of spinning correlators with three chiral primaries and either one R-symmetry current or the stress tensor up to order $1/\lambda^{3/2}$. Appendix~\ref{sec:Dfuncs} is dedicated to reviewing some basics of D-functions, and Appendix~\ref{sec:Rsympoly} uses the Casimir operator to determine the R-symmetry structures corresponding to the various representations exchanged in the OPE. For convenience, the form of the fixed ansatz and of the spinning correlators is shared in a Mathematica notebook.

\section{Setup}
\label{sec:setup}

We consider $\tfrac{1}{2}$-BPS operators in the $20'$ representation of the $SU(4)\simeq SO(6)$ R-symmetry group,
\begin{equation}
    O_2^{I_1 I_2}=\text{tr}(\phi^{\{I_1}\phi^{I_2\}})
\end{equation}
where the $\phi^{I}$, with $I=1,\dots,6$, are the adjoint scalar fields of the vector multiplet of $\mathcal{N}=4$ SYM. The curly brackets $\{\}$ symbolize the projection to the the rank-2 symmetric and traceless representation of the R-symmetry group.
As customary, we introduce null polarization vectors $t_I$ that contract the R-symmetry index dependence leading to
\begin{equation}
    O_2(x,t)=O_2^{I_1 I_2}\, t_{I_1} t_{I_2}\,,\quad t\cdot t=0\,.
\end{equation}
These scalar operators have protected scaling dimensions $\Delta=2$ due to supersymmetry. They are the bottom-components of the stress tensor multiplet, that also includes the R-symmetry current. Via the AdS/CFT correspondence, these are dual to scalar fields in $\text{AdS}_5$ and, in particular, correspond to the lowest Kaluza-Klein (KK) mode obtained from type IIB reduction on $S_5$.

In this work, our main object of interest is the five-point correlator
\begin{equation}
\label{eq:correlatorgoal}
    G(x_i,t_i)=\langle O_2(x_1,t_1)O_2(x_2,t_2)O_2(x_3,t_3)O_2(x_4,t_4)O_2(x_5,t_5)\rangle\,.
\end{equation}
For generic value of the 't Hooft coupling $\lambda=g_{YM}^2 N$, where $N$ is the number of colours of the $SU(N)$ gauge group of $\mathcal{N}=4$, these correlators are extremely complicated and encode a large amount of dynamical data. 
We will always consider the large $N$ planar limit of $\mathcal{N}=4$ and the connected part of our correlator of order $1/N^3$. The disconnected part is protected as it factorizes into a three-point and a two-point function, which are protected themselves. In the supergravity regime, $N\to \infty$ followed by  $\lambda\to \infty$, the form of the connected part of the correlator~\eqref{eq:correlatorgoal} was studied and fixed via bootstrap methods in~\cite{Goncalves:2019znr}. Later, other KK five-point correlators were bootstrapped in~\cite{Goncalves:2023oyx,Fernandes:2025eqe}. 
In this paper, we extend the bootstrap methods of these papers to compute subleading contributions in $1/\lambda$ in this correlator. These correspond to higher derivative corrections of the type IIB supergravity action. On the other hand, $1/N$ corrections, that we do not explore here, are associated with bulk loop diagrams.

Conformal symmetry of the correlator allows us to write the five-point function as a function of five conformal cross ratios after extracting a kinematic factor that takes care of the scaling of the correlator,
\begin{equation}
    G(x_i,t_i)= \frac{x_{13}^2}{x_{12}^4 x_{35}^4 x_{14}^2 x_{34}^2} G(u_i, t_i)\,,
\end{equation}
where we use the choice of cyclic cross-ratios
\begin{equation}
\label{eq:crossratios}
    u_1 = \frac{x_{12}^2 x_{35}^2}{x_{13}^2 x_{25}^2}\,,\quad u_{i+1}=u_i\big\vert_{x_i\to x_{i+1}}\,,
    \end{equation}
with $x_{ij}=x_i-x_j$.

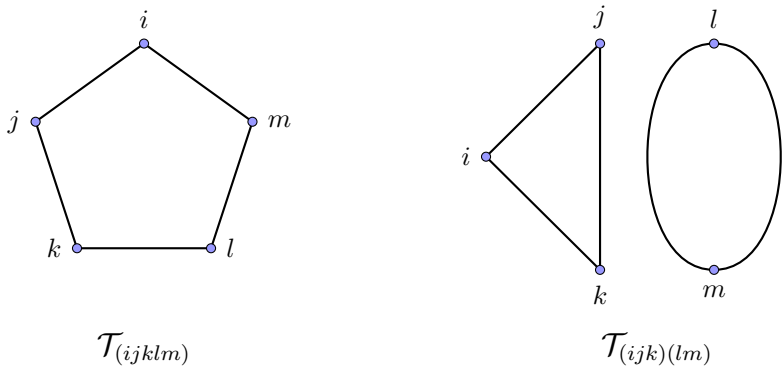
\begin{figure}
    \centering
\begin{tikzpicture}[scale=1.5, every node/.style={font=\small}]

\tikzset{
    vtx/.style={circle, fill=blue!40, draw=black, inner sep=1.2pt}
}

\foreach \i/\name in {1/i, 2/j, 3/k, 4/l, 5/m} {
    \coordinate (\name) at ({cos(90 + (\i-1)*72)}, {sin(90 + (\i-1)*72)});
}

\draw[thick]
    (i) -- (j) -- (k) -- (l) -- (m) -- cycle;

\node[vtx, label=above:$i$] at (i) {};
\node[vtx, label= left:$j$] at (j) {};
\node[vtx, label= left:$k$] at (k) {};
\node[vtx, label=right:$l$] at (l) {};
\node[vtx, label=right:$m$] at (m) {};

\node at (0,-1.7) {\large $\mathcal{T}_{(ijklm)}$};

\begin{scope}[xshift=3.cm]

\coordinate (i2) at (0,0);
\coordinate (j2) at (1,1.);
\coordinate (k2) at (1,-1.);

\draw[ thick] (i2) -- (j2) -- (k2) -- cycle;

\node[vtx, label=left:$i$] at (i2) {};
\node[vtx, label=above:$j$] at (j2) {};
\node[vtx, label=below:$k$] at (k2) {};

\coordinate (l2) at (2,1.);
\coordinate (m2) at (2,-1.);

\draw[thick] (l2) to[out=180,in=-180] (m2);
\draw[thick] (l2) to[out=0,in=0] (m2);

\node[vtx, label=above:$l$] at (l2) {};
\node[vtx, label=below:$m$] at (m2) {};

\node at (1.5, -1.7) {\large $\mathcal{T}_{(ijk)(lm)}$};

\end{scope}

\end{tikzpicture}

    \caption{R-symmetry structures with one and two cycles. Each line between points $i$ and $j$ corresponds to a factor of $t_{ij}$. The number of independent structures is determined by the number of ways to distribute the points into the cycles modulo invariance under cyclic permutations and reflections.}
    \label{fig:Rsymstructures}
\end{figure}

Similarly, we can also use symmetry to impose constraints on the R-symmetry dependence. Since the polarization vectors $t_i$ are introduced in order to suppress the R-symmetry indices, it is clear that $G(x_i,t_i)$ depends on them only through monomials of the form $\prod_{i<j} t_{ij}^{a_{ij}}$, $t_{ij}=t_i\cdot t_j$, satisfying the conditions
\begin{equation}
    a_{ij}=a_{j i}\,,\quad \sum_{j\ne i} a_{i j}=2\,.
\end{equation}
These follow from the homogeneity condition of $G(x_i,t_i)$ under the rescaling $t_i \to \lambda_i t_i$
\begin{equation}
    G(x_i,\lambda_i t_i)=\lambda_1^2\lambda_2^2\lambda_3^2\lambda_4^2\lambda_5^2G(x_i, t_i)\,,
\end{equation}
for independent $\lambda_i$. There are 22 of such structures obeying these conditions. Using permutation symmetry, we can divide them into two categories
\begin{align}
\label{eq:Rsymstructures}
    &\mathcal{T}_{(ijklm)} = t_{ij} t_{jk} t_{kl} t_{lm} t_{mi}\,,\nonumber\\
    &\mathcal{T}_{(ijk)(lm)} = t_{ij} t_{jk} t_{ki} t_{lm}^2\,
\end{align}
which are depicted in figure~\ref{fig:Rsymstructures}.

R-symmetry imposes further constraints on the form of the correlator that follow from the fermionic generators of the superconformal group $PSU(2,2|4)$. While the full fermionic implications are not known for higher-point correlators, there are two constraints that can be applied generically. One such constraint occurs for a choice of polarization known as chiral algebra twist~\cite{Beem:2013sza}. When all operators are restricted to a 2d plane with coordinates $(z_i,\bar{z}_i)$ and the polarization vectors are chosen such that
\begin{equation}
    t_{ij}=(z_i-z_j)(v_i-v_j)\,,
\end{equation}
for some arbitrary $v_i$ vectors, the correlator becomes independent of $z_i$
\begin{equation}
    G(z_i,\bar{z}_i,t_{i})|_{t_{ij}=(z_i-z_j)(v_i-v_j)}=g(\bar{z}_i,v_i)\,.
\end{equation}
Moreover, the correlator with this choice of twist does not depend on the marginal coupling and can therefore be computed solely from free theory,
\begin{equation}
    g(\bar{z}_i,v_i)=g_{\text{free}}(\bar{z}_i,v_i)\,.
\end{equation}
A second constraint is of topological nature and we will refer to it as the Drukker-Plefka twist~\cite{Drukker:2009sf}. When one chooses the R-symmetry polarization vectors such that
\begin{equation}
    t_{ij}=x_{ij}^2\,,
\end{equation}
the correlator becomes topological and therefore is given by
\begin{equation}
    G(x_i,t_i)|_{t_{ij}=x_{ij}^2}=\text{constant}\,,
\end{equation}
where the constant can be computed from free theory. Note that unlike the chiral algebra twist that restricts the kinematics to a 2d plane, no such constraint is required by Drukker-Plefka twist.

\section{Mellin space}
\label{sec:Mellinspace}
It will be useful to use Mellin space to express the connected part of the holographic correlator. In Mellin formalism, the connected part of a scalar $n$-point function can be written as~\cite{Mack:2009mi,Penedones:2010ue}
\begin{equation}
\label{eq:Mellindef}
    \langle \mathcal{O}_1(x_1,t_1)\dots\mathcal{O}_n(x_n,t_n)\rangle_{\text{conn}}=\int [d\delta] M(\delta_{ij},t_{ij})\prod_{1\le i<j\le n} (x_{ij}^2)^{-\delta_{ij}}\, \Gamma(\delta_{ij})\,,
\end{equation}
where $M(\delta_{ij},t_{ij})$ is called Mellin amplitude and it is function of Mellin-Mandelstam variables satisfying
\begin{equation}
\label{eq:deltaconstraints}
    \delta_{ij}=\delta_{ji}\,,\quad \delta_{ii}=-\Delta_i\,,\quad \sum_{j}\delta_{ij}=0\,.
\end{equation}
 These conditions lead to $n(n-3)/2$ independent Mellin variables, a number that matches the number of cross ratios of a $n$-point correlator for large enough spacetime dimensions. In our case of interest, the dependence of the Mellin amplitude in $t_{ij}$ is completely determined by the R-symmetry structures of~\eqref{eq:Rsymstructures} that multiply a function of $\delta_{ij}$ alone.

In Mellin space, the integration contours are placed parallel to the imaginary axis in such a way that semi-infinite sequences of poles lie separately to the left and right of it. The integration is performed over the independent Mellin variables with an extra $1/2\pi i$ per variable.

We will also have to deal with Mellin amplitudes corresponding to correlators where one of the external operators is an operator of spin $J$, say $\mathcal{O}^{\mu_1\dots \mu_J}(x)$. As usual, we conveniently suppress the Lorentz indices by contracting them against some null polarization vector $z$
\begin{equation}
    \mathcal{O}(x,z) \equiv \mathcal{O}^{\mu_1\dots \mu_J}(x) z_{\mu_1}\dots z_{\mu_J}\,,\qquad z\cdot z=0\,.
\end{equation}
As explained in~\cite{Costa:2011mg}, this packaging ensures that the operators is indeed in a spin $J$ symmetric and traceless representation. 

We consider a correlator with a spin $J$ operator at position $x_0$ and $n$ scalar operators. In Mellin space, the connected part of this correlator can be written as~\cite{Goncalves:2014rfa}
\begin{equation}
\label{eq:spinningMellin}
    \langle \mathcal{O}(x_0, z_0) \cdots \mathcal{O}_n \rangle = \sum_{a_1, \ldots, a_J = 1}^n \prod_{i=1}^J (z_0\cdot x_{a_i 0}) \int [d\delta] M^{\{a\}}(\delta_{ij},t_{ij}) \prod_{i=1}^n \frac{\Gamma(\delta_i + \{a\}_i)}{(x_{i0}^2)^{\delta_i + \{a\}_i}} \prod_{1 \leq i < j \leq n} \frac{\Gamma(\delta_{ij})}{(x_{ij}^2)^{\delta_{ij}}},
\end{equation}
where $\{a\}$ represents the indices $a_1,\cdots,a_J$ and
\begin{equation}
\quad \{a\}_i = {\bm \delta}_i^{a_1} + \cdots + {\bm\delta}_i^{a_J}\,,\quad \delta_i = -\sum_{j=1}^n \delta_{ij}, \quad \sum_{i,j=1}^n \delta_{ij} = J - \Delta_0.
\end{equation}
Here we use the bold symbol $\bm\delta$ to denote Kronecker delta, such that $\{a\}_i$ counts the number of times index $i$ appears in the set $a_1,\cdots, a_J$. It also remains true $\delta_{ij}=\delta_{ji}\,, \delta_{ii}=-\Delta_i$.

The Mellin amplitudes $M^{\{a\}}$ satisfy tranversality conditions that relate different components~\cite{Goncalves:2014rfa}
\begin{equation}
\label{eq:Mellintransverse}
     \sum_{a_1=1}^{n} \left( \delta_{a_1} + {\bm \delta}_{a_1}^{a_2} + {\bm \delta}_{a_1}^{a_3} + \cdots + {\bm \delta}_{a_1}^{a_J} \right) M^{a_1 a_2 \ldots a_J} = 0\,.
\end{equation}

If the spinning operator is moreover conserved ($\Delta_0=d-2+J$), there is an extra condition that needs to be satisfied
\begin{equation}
    (2J + d - 4) \sum_{\substack{a, b = 1 \\ a \neq b}}^n \delta_{ab} [M^{ac_2 \ldots c_J}]^{ab} = (J - 1) \sum_{\substack{a, b = 1 \\ a \neq b}}^n \delta_{ab} [M^{abc_3 \ldots c_J}]^{ab},
\end{equation}
where we should set $d=4$ for $\mathcal{N}=4$ SYM. Here we use
\begin{equation}
    [M(\delta_{ij})]^{ab} \equiv M(\delta_{ij} + {\bm\delta}_i^a {\bm\delta}_j^b + {\bm\delta}_j^a {\bm\delta}_i^b)\,.
    \label{eq:placeholder}
\end{equation}

Mellin formalism is known to be a natural language to study holographic correlators. There are several reasons for that. The first and main reason is the simplicity of the analytic structure of Mellin amplitudes compared to position space analogues. Indeed, Mellin amplitudes resemble the form of tree-level scattering amplitudes in flat space, enjoying similar factorization properties at the location of poles~\cite{Goncalves:2014rfa}. These properties will be crucial for our bootstrap method and will be reviewed in section~\ref{sec:Factorization}. Importantly, in the large $N$ limit, the Mellin amplitudes of correlation functions have a polar behaviour completely determined by the exchange of single-trace operators and multi-trace contributions are captured by the gamma factors in the definition~\eqref{eq:Mellindef}. 

Another remarkable property of Mellin amplitudes is the mapping between their high-energy behaviour and flat-space amplitudes~\cite{Penedones:2010ue,Fitzpatrick:2011hu}. 
This relation has been a powerful tool to compute correlation functions both in the supergravity regime as well as including stringy corrections, see for instance~\cite{Goncalves:2014ffa,Binder:2019jwn}. However, as it will be discussed in section~\ref{sec:Flatspacelimit}, this relation is less useful in the case five-point correlators and, particularly, in our case of interest. This happens because the mapping occurs for a specific choice of polarizations of the flat-space amplitude in which it vanishes.

Having introduced Mellin space formalism, we can initiate a bootstrap program to compute the Mellin amplitude of the holographic correlator~\eqref{eq:correlatorgoal} at large $N$ but including stringy corrections in an expansion around large $\lambda$. The correlator takes the form
\begin{equation}
    M(\delta_{ij},t_{ij})=\frac{1}{N^3}\left(M^{\text{SUGRA}}(\delta_{ij},t_{ij})+\frac{1}{\lambda^{3/2}}M^{R^4}(\delta_{ij},t_{ij})+\mathcal{O}\left(\frac{1}{\lambda^{5/2}}\right)\right)+\mathcal{O}\left(\frac{1}{N^4}\right)\,,
\end{equation}
where the first term was fixed in~\cite{Goncalves:2019znr,Goncalves:2023oyx} and it corresponds to the two derivative effective interaction of classical 10d type IIB supergravity on $\text{AdS}_5\times S^5$. The superscript in $M^{R^4}$ refers to the $R^4$ effective bulk interaction vertex with 8 derivatives. 

In this work, we use the available constraints explored in~\cite{Goncalves:2019znr,Goncalves:2023oyx} to bootstrap the correlator to first subleading order $\mathcal{O}(1/\lambda^{3/2})$ up to two undetermined coefficients. Using the OPE, we also consider the contribution of those coefficients to protected four-point correlators, allowing us to further determine one of the remaining coefficients. The methods presented here can be generalized to put constraints at higher orders. It would be interesting to see how much of the corresponding ansätze could be fixed in this way. 

In what follows, factorization of Mellin amplitudes and known stringy corrections of the four-point correlators of 20' operators are fundamental ingredients to fix the singular behaviour of the stringy correction.

\subsection{Factorization of Mellin amplitudes}
\label{sec:Factorization}
The operator product expansion is one of the most important tools to study conformal field theories. It states that the product of $k$ local primary operators at different positions can be written as an infinite sum over local operators
\begin{equation}
    \mathcal{O}_1(x_1)\dots\mathcal{O}_k(x_k)=\sum_{p}C_{\mu_1\dots\mu_J}^{(1\dots k, p)}(x_1,\dots,x_k,y,\partial_y)\mathcal{O}_p^{\mu_1\dots\mu_J}(y)\,,
\end{equation}
where the operator $\mathcal{O}_p^{\mu_1\dots\mu_J}$ has spin $J$ and it is located at a generic point $y$  contained within a sphere that encircles also all the other $k$ operators. Convergence of the OPE at finite radius requires that no other insertion of operators is contained in the sphere. Here the sum is over primary operators of the CFT and the contribution of descendants is recovered by the action of the operator $C_{\mu_1\dots\mu_J}^{(1\dots k, p)}(x_1,\dots,x_k,y,\partial_y)$.

The OPE can be applied within $n$-point correlation functions, effectively reducing them to sums of products of $k+1$ and $n-k+1$-point functions. In Mellin space, the existence of OPE constrains the analytic form of Mellin amplitudes. For the remaining of this section, we suppress the dependence of the Mellin amplitude on R-symmetry structures to draw the attention to the analytic structure related with spacetime.

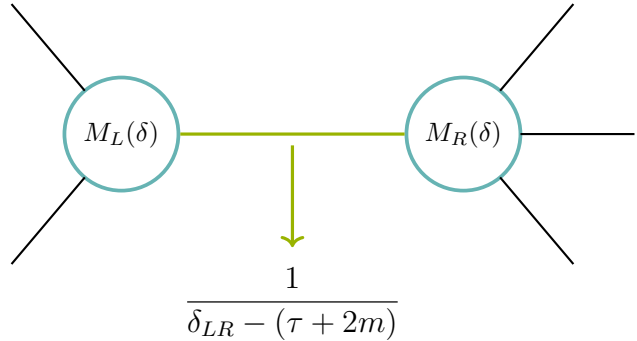
\begin{figure}
    \centering
    \begin{tikzpicture}[thick, scale=1.5, every node/.style={font=\small}]

\tikzset{
    blob/.style={circle, draw=teal!60, line width=1.4pt, minimum size=1.5cm},
    ext/.style={thick, black},
    intline/.style={very thick, orange!60!green},
    arrowstyle/.style={->, very thick, draw=orange!60!green},
    arrowlabel/.style={font=\large}
}

\node[blob, label=center:$M_L(\delta)$] (L) at (0,0) {};
\node[blob, label=center:$M_R(\delta)$] (R) at (3,0) {};

\draw[intline] (L) -- (R);

\foreach \angle in {130, 230} {
    \draw[ext] (L) ++(\angle:0.5cm) -- ++(\angle:1.cm);
}

\foreach \angle in {50, 0, -50} {
    \draw[ext] (R) ++(\angle:0.5cm) -- ++(\angle:1.cm);
}

\draw[arrowstyle] (1.5,-0.1) to[out=-90,in=90] (1.5,-1.);

\node[arrowlabel] at (1.5,-1.5) {%
    $\displaystyle\frac{1}{\delta_{LR} - (\tau + 2m)}$
};

\end{tikzpicture}
    \caption{Factorization of a Mellin amplitude with $n=5$ and $k=2$. The poles of the Mellin amplitude are associated with the OPE exchange of single-trace operators of twist $\tau$. The residues are related to lower-point functions $M_L$ and $M_R$.}
    \label{fig:Mellinfactorization}
\end{figure}

Mellin amplitudes are analytic functions of $\delta_{ij}$ with simple poles at
\begin{align}
\label{eq:Mellinfactorization}
    M(\delta_{ij}) \approx \frac{Q_m(\delta_{ij})}{\delta_{LR} - (\tau + 2m)}, \quad \delta_{LR} = \sum_{a=1}^{k} \sum_{b=k+1}^{n} \delta_{ab}\,,
\end{align}
where $\tau=\Delta-J$ is the twist of an exchanged primary operator of scaling dimension $\Delta$ and spin $J$ and $m=0,1,2\dots$ labels the conformal descendants. Here we divide the external operators into two sets: from 1 to $k$ and from $k+1$ to $n$, labelled by $L$ and $R$, respectively. The residues $Q_m(\delta_{ij})$ are nontrivial functions of the Mellin variables and are related to lower-point functions $M_L$ and $M_R$ each with $k+1$ and $n-k+1$ operators. In particular, they involve the external operators of each of the sets and an extra one which is the exchanged primary operator. An example of the factorization of Mellin amplitudes is given in figure~\ref{fig:Mellinfactorization}.

Here we always consider the OPE between two external operators and fix $k=2$. Due to the permutation symmetry of the correlator~\eqref{eq:correlatorgoal}, there is in fact only one possible factorization to consider, schematically
\begin{equation}
    \langle O_2 O_2 X\rangle\langle X O_2 O_2 O_2\rangle\,,
\end{equation}
where $X$ is some primary operator being exchanged. To apply factorization, we need to sum over the possible contributions.

As we will momentarily discuss in the next section, it will be enough for us to consider $m=0$ contributions and operators with spin 0, 1 and 2. In this case, we can compute explicitly the residue function in terms of lower-point functions~\cite{Goncalves:2014rfa}.

For the exchange of a scalar, we have
\begin{equation}
\label{eq:residuescalar}
    Q_0(\delta_{ij})\propto M_L M_R\,,\qquad \text{(scalar)}
\end{equation}
while for spin 1 and 2 operators one reads
\begin{align}
    & Q_{0} \propto \sum_{a=1}^{k} \sum_{i=k+1}^{n} \delta_{ai} M^{a}_{L} M^{i}_{R}\,,\qquad &\text{(spin 1)}\\
    &Q_0 \propto
\sum_{\substack{a,b=1 \\}}^{k}
\sum_{\substack{i,j=k+1 \\}}^{n}
\delta_{ai} \left( \delta_{bj} + {\bm\delta}^a_b {\bm\delta}^i_j \right)
M_L^{ab} M_R^{ij}\,. &\text{(spin 2)}
\end{align}
The precise normalization constants were omitted here but were determined in~\cite{Goncalves:2014rfa}. In what follows, we absorb these constants in the OPE coefficients appearing in the lower-point functions and we fix them by comparing the resulting contributions with the supergravity result of~\cite{Goncalves:2023oyx}.

\subsection{OPE exchanges and four-point correlators}
\label{sec:OPEand4pts}
The use of factorization requires the knowledge of the possible single-trace operators that can appear in the OPE of two 20' operators. These determine the location of poles as well as the lower-point functions that contribute to the residue in~\eqref{eq:Mellinfactorization}. Recall that in the large $N$ limit, double-trace contributions are entirely captured by the Gamma factors in the definition of Mellin amplitude~\eqref{eq:Mellindef}. 

At large $\lambda$, the discussion of what single-trace operators can be exchanged replicates the one for the supergravity regime in~\cite{Rastelli:2017udc,Goncalves:2019znr}. The analysis amounts to list all the non-vanishing cubic vertices of the form $s_2^is_2^jX$ where $s_2^i$ is the scalar field dual to the superprimary operator $O_2(x_i)$ and $X$ is the exchanged field to be determined. R-symmetry selection rules and the requirement of non-extremality of cubic vertices\footnote{A cubic vertex of the form $s_{p_1} s_{p_2}X$, with $s_{p}$ the scalar field dual to a KK mode operator $O_p$, can only be non-vanishing if the parent vertex $s_{p_1} s_{p_2} s_{p_3}$ does not vanish.  One says the vertex $s_{p_1} s_{p_2} s_{p_3}$ is extremal if $p_1+p_2=p_3$, where we assume $p_3$ to be the largest scaling dimension.} impose that, at large $\lambda$, $X$ can only be:
the scalar $s_2$ dual to the chiral primary $O_2$ of scaling dimension $\Delta=2$; the graviphoton $A_\mu$ dual to the R-symmetry current $J_\mu$ with $\Delta=3$; or the graviton $\varphi_{\mu\nu}$ dual to the stress tensor $T_{\mu\nu}$ with $\Delta=4$. 
Note that all of these operators have twist $\tau=2$. 
Similarly to what happens in the supergravity regime, the contribution of descendants truncates at $m=0$. For concreteness, let us consider the factorization in the (12) channel (the remaining channels are equivalent under cyclicity) and a twist 2 exchange for which we have
\begin{equation}
    \delta_{LR}-(\tau+2m)\to 2(1-m-\delta_{12})\,.
\end{equation}
It becomes clear that for $m=1,2,\dots$, the location of these poles overlaps with the poles of double-traces arising from the Gamma factor $\Gamma(\delta_{12})$. These double-poles translate in position space to logarithmic terms associated with anomalous dimensions of the double-trace operators. These anomalous dimensions appear at order $\mathcal{O}(1/N^2)$~\cite{Aprile:2018efk} and therefore, at first nontrivial order at large $N$, we can ignore those contributions.

In order to use factorization to compute the first stringy corrections of the five-point correlation function, we then need to know the following correlators up to order $\mathcal{O}(1/\lambda^{3/2})$~\footnote{Note that the three-point functions involved in the factorization are all protected and therefore only the four-point correlators can have stringy corrections.}
\begin{align}
    &\langle O_2 O_2 O_2 O_2\rangle\,,\nonumber\\
    &\langle J_\mu O_2 O_2 O_2\rangle\,,\\
    &\langle T_{\mu \nu} O_2 O_2 O_2\rangle\nonumber\,.
\end{align}
The first of these correlators has been extensively studied both at weak and strong coupling $\lambda$. Superconformal ward identities allow us to write it as~\cite{Eden:2000bk,Nirschl:2004pa,Dolan:2001tt}
\begin{align}
\label{eq:fourpointO2}
& \langle O_2 O_2 O_2 O_2\rangle= \frac{t_{12}^2 t_{34}^2}{x_{12}^4 x_{34}^4}\mathcal{G}_{2222}\left(u,v;\sigma,\tau\right)\,,\\
&\mathcal{G}_{2222}\left(u,v;\sigma,\tau\right)= \mathcal{G}_{2222}^{\text{free}}\left(u,v;\sigma,\tau\right)+ R(u,v;\sigma,\tau) H(u,v)\,,
\end{align}
where
\begin{equation}
\label{eq:RSWard}
 R(u,v;\sigma,\tau)=\tau +\sigma  \tau  u^2+\tau  u (\tau-\sigma-1 )+\sigma  u v (\sigma -\tau -1)+\sigma  v^2+v (1-\sigma -\tau)\,,
\end{equation}
which is written in terms of the cross ratios
\begin{equation}
\label{eq:4ptcrossratios}
u=\frac{x_{12}^2 x_{34}^2}{x_{13}^2 x_{24}^2}\,,\quad v=\frac{x_{14}^2 x_{23}^2}{x_{13}^2 x_{24}^2}\,,\quad \sigma=\frac{t_{13} t_{24}}{t_{12} t_{34}}\,,\quad \tau=\frac{t_{14} t_{23}}{t_{12} t_{34}}\,.
\end{equation}
Above $\mathcal{G}_{2222}^{\text{free}}$ corresponds to the free theory correlator
\begin{equation}
	\mathcal{G}_{2222}^{\text{free}}=1+\sigma ^2 u^2+\frac{\tau ^2 u^2}{v^2}+\frac{1}{c}\left(\frac{\sigma  \tau  u^2}{v}+\sigma  u+\frac{\tau  u}{v}\right)\,,
\end{equation}
with central charge $c=\frac{1}{4}\left(N^2-1\right)$. On the other hand, $H(u,v)$ is the reduced correlator and contains the dynamical data of the theory. Using Mellin space formalism of section~\ref{sec:Mellinspace}, one can conveniently consider the Mellin transform of the reduced correlator
\begin{equation}
\label{eq:reducedmellin}
\mathcal{M}(s,t) \Gamma \left(\frac{4-s}{2}\right)^2 \Gamma \left(\frac{4-t}{2}\right)^2\Gamma \left(\frac{4-\tilde{u}}{2}\right)^2   = \int_0^\infty \mathrm{d}u \int_0^\infty \mathrm{d}v \, u^{-\frac{s}{2}-1} v^{1-\frac{ t}{2}} H(u,v)\,,
\end{equation}
where we use the parametrization
\begin{align}
\label{eq:mellinvariables4pt}
&\delta_{ij} = \frac{\Delta_i + \Delta_j - s_{ij}}{2}\,,\quad
s_{12} = s_{34} = s\,, \quad
s_{14} = s_{23} = t\,, \nonumber\\
&s_{13} = s_{24} = 8 - s - t=4-\tilde{u}\,.
\end{align}
 Up to order $\mathcal{O}(1/\lambda^{3/2})$, the reduced Mellin amplitude takes the simple form
\begin{equation}
\label{eq:reducedMellinscalar}
\mathcal{M}(s,t)=\frac{1}{N^2}\frac{32}{(s-2) (t-2) (2-s-t)}+\frac{1}{N^2 \lambda^{3/2}}480\zeta_3+\dots\,.
\end{equation}
where the first term is the supergravity result of~\cite{Rastelli:2017udc,Rastelli:2016nze} and the stringy correction is determined by requiring the consistency of the flat-space limit of the Mellin amplitude and the flat-space scattering amplitude of four gravitons~\cite{Goncalves:2014ffa}.~\footnote{Here we use the same normalization as in~\cite{Goncalves:2023oyx}.}  The $\dots$ represent subleading contributions both in $1/\lambda$ and $1/N$ that will not be important for us here but that have been studied using localisation~\cite{Binder:2019jwn,Chester:2020dja} and dispersive sum rules~\cite{Alday:2022uxp,Alday:2022xwz}. Recently, it has been proposed a constructive method to compute the AdS Virasoro-Shapiro amplitude, order by order in AdS curvature corrections, in terms of world-sheet integrals involving single-valued multiple polylogarithms~\cite{Alday:2023jdk,Alday:2023mvu,Wang:2025pjo}. This proposal passes several tests by matching known data, while making a plethora of predictions for new CFT data of $\mathcal{N}=4$ SYM at large $N$.

In contrast to the extensively studied four-point correlator $\langle O_2 O_2 O_2 O_2\rangle$, the large $\lambda$ expansions of the correlators with one R-symmetry current or one stress tensor have not yet been computed. However, the three operators belong to the same supermultiplet. One can use this information to derive the two last correlators from the first by applying differential operators. The explicit logic for this derivation has already been used in~\cite{Goncalves:2019znr,Goncalves:2023oyx} in the supergravity regime. In Appendix~\ref{sec:fourpoints}, we briefly review the logic and we use it to obtain these correlators up to order $\mathcal{O}(1/\lambda^{3/2})$.  The explicit expressions are made available in this Appendix and in an ancillary file. 

In this chapter we have mostly ignored the role of the global R-symmetry in factorization. Indeed, this contribution is stored in monomials of the form $\prod_{i<j} t_{ij}^{a_{ij}}$ that appear in both the three and four-point functions contributing to the residue function in~\eqref{eq:Mellinfactorization}. The two lower-point functions share an internal common point associated with the exchanged primary operator. Its label dependence should not appear in the five-point correlator. We need then a R-symmetry gluing procedure that gets rid of this extra point. This is a group theoretic exercise that was developed in~\cite{Goncalves:2023oyx}. For our interests here, it is enough to use the following rules
\begin{align}
\label{eq:rulegluing}
&t_{0i}t_{0j}t_{0k}t_{0\ell}\longrightarrow t_{i\ell } t_{jk}+t_{ik} t_{j\ell }-\frac{1}{3} t_{ij} t_{k\ell }\,,\\
&Y_{0, ij}Y_{0, k\ell}\longrightarrow t_{ik} t_{j\ell}-t_{i\ell} t_{jk}\,,
\end{align}
where we label the internal point by $0$. Here we also have $Y_{0, ij}$, which appears in the four-point correlation function with one R-symmetry current, and it is defined in \eqref{eq:Ysdef} in Appendix~\ref{sec:fourpoints}. Later, in equation~\eqref{eq:factlowerpts} and below, we provide an explicit example of the use of factorization formulas and of the R-symmetry gluing procedure.

\subsection*{A comment on stringy states}
There is a word to be said about single-trace operators whose scaling dimensions are not protected. We refer to the so-called stringy states whose twist $\tau(\lambda)\sim \tau_0\lambda^{1/4}+\dots$ for some function $\tau_0$ of the quantum numbers of the operator itself but independent of $\lambda$ (see for instance~\cite{Alday:2022uxp}). The $\dots$ represent subleading contributions at large $\lambda$. The OPE coefficients involving two chiral primaries and a stringy state are also exponentially suppressed in $\lambda$.\footnote{Away from very special configurations where the twist of the stringy state is $\tau(\lambda)=2n$ for integer $n$. At these locations, the OPE coefficients develop a pole related to the mixing of stringy states and double-trace operators with $n\gg1$. See~\cite{Alday:2022uxp} .} Ultimately both facts mean that, for an expansion around large $\lambda$, these operators decouple and we do not have to include them in the simple poles of~\eqref{eq:Mellinfactorization}.\footnote{In~\cite{Alday:2022uxp,Alday:2022xwz}, the bound on chaos~\cite{Maldacena:2015waa} was used to derive dispersive sum rules for the reduced Mellin amplitude of a four-point correlator of 20' operators that are valid at large $N$ but finite $\lambda$. The sum rules relate low twist data from protected single-trace operators and double-trace operators to the CFT data of stringy states. This latter data was indispensable to find a solution to the sum rules and to fix the Wilson coefficients of a large $\lambda$ expansion. There is no contradiction here. This only means that low twist data must be appropriately UV completed in order to satisfy the dispersive rules. }

\section{The bootstrap}
\label{sec:bootstrap}
After presenting the required ingredients for our bootstrap, we are now prepared to formulate an ansatz for the first stringy correction to the Mellin amplitude of five 20’ operators. The singular behaviour of the correlator is, in fact, completely determined by the factorization of Mellin amplitudes. On the other hand, the regular contributions are determined by the use of two supersymmetric constraints that follow from two choices of superconformal twist and by considering protected quantities contained in the five-point correlator.
In the next subsections, we explain how to impose these constraints in practice.
\subsection{The ansatz}
\label{sec:theansatz}
Following the discussion we had so far, it is clear that the first stringy correction should correspond to a generic ansatz of the form
\begin{align}
&M^{R^4}(\delta_{ij},t_{ij})=\sum_{\substack{k=1\\ \ell=k+1}}^{5}\frac{A^{k\ell}(\delta_{ij},t_{ij})}{\delta_{k\ell}-1}+R(\delta_{ij},t_{ij})\,,
\end{align}
where
\begin{align}
&A^{12}(\delta_{ij},t_{ij})=\sum_{\text{R-sym}}\sum_{n_i=0}^{n_1+\cdots+n_4\leq 4}p^{12}_{\{n_i\}}(t_{ij})\delta_{15}^{n_1}\delta_{23}^{n_2}\delta_{34}^{n_3}\delta_{45}^{n_4}\,,\\
&R(\delta_{ij},t_{ij})=\sum_{I}\sum_{n_i=0}^{n_1+\cdots+n_5\leq 4}r_{\{n_i\}, I} \mathcal{T}^I \delta_{12}^{n_1}\delta_{15}^{n_2}\delta_{23}^{n_3}\delta_{34}^{n_4}\delta_{45}^{n_5}\,,
\end{align}
where the coefficients $p^{k\ell}_{\{n_i\}}$ and $r_{\{n_i\}, I}$ are to be determined.~\footnote{Above we only write explicitly the contribution $A^{12}(\delta_{ij},t_{ij})$. When considering the remaining channels, we should be careful to not introduce redundancies in the ansatz due to relations between Mellin variables. } Here we choose $\delta_{12}, \delta_{15}, \delta_{23}, \delta_{34}$ and $\delta_{45}$ as the independent Mellin variables. The sum over R-symmetry structures for the regular terms contains the 22 independent structures presented in~\eqref{eq:Rsymstructures}. On the other hand, the R-symmetry dependence of the singular terms can be made more refined because we know what the possible single-trace contributions to the poles are. This is the subject of the next subsection. 

In the ansatz above, the maximum power of the polynomials in the Mellin variables is simple to justify. For the singular terms, it follows from the application of factorization formulas of section~\ref{sec:Factorization} and the necessary four-point correlators discussed in appendix~\ref{sec:fourpoints}. On the other hand, the regular terms are associated with contact Witten diagrams. Here, the maximum power of the polynomials is fixed to be 4 in accordance with an 8-derivative effective action due to the $R^4$ term. Note that the counting works just as in flat space, as Mellin and Mandelstam variables are related by the flat-space formula of Penedones~\cite{Penedones:2010ue}.

\subsection*{R-symmetry Casimir}
In section~\ref{sec:OPEand4pts}, we discussed the single-trace OPE exchanges contributing to the poles of the Mellin amplitude of the correlator we are studying at both large $N$ and large $\lambda$. These exchanges transform in specific R-symmetry representations, and this information is necessarily encoded in the R-symmetry dependence of the correlator.

Similarly to what is done to study conformal blocks of the conformal group $SO(d+1,1)$~\cite{Dolan:2003hv}, one can use Casimir operators to fix the R-symmetry dependence associated with each operator exchange. We consider the two-particle $SO(6)$ R-symmetry Casimir
\begin{equation}
\label{eq:CasRsym}
\mathcal{C}_{ij}= \frac{1}{2}\left(L_{IJ}^{(i)} + L_{IJ}^{(j)}\right)\left(L^{(i),IJ} + L^{(j),IJ}\right)\,,
\end{equation}
where we use the R-symmetry generators written in terms of the null vectors $t_i^I$ as 
\begin{equation}
L_{IJ}^{(i)} = t_{i,I} \frac{\partial}{\partial t_i^J} - t_{i,J} \frac{\partial}{\partial t_i^I}\,.
\end{equation}
When acting with the Casimir operator in states in the same R-symmetry representation, one gets the same eigenvalue. Here we are focusing in the allowed single-trace exchanges: the 20' operator, the R-symmetry current and the stress tensor. For each of these exchanges, one has the respective Casimir eigenvalue $\lambda_c$ of $-12,-8$ and $0$.  More generally, a R-symmetry representation $[q,p,q]$ has $\lambda_c=-p^2-2 p (q+2)-2 q (q+3)$.

To find the correct R-symmetry dependence associated with each exchange becomes then a straightforward task of solving a Casimir equation with the appropriate eigenvalue. For each equation, we propose a generic solution for the eigenstates of the form 
\begin{equation}
R_{ij}^{\lambda_c}=\sum_{I} a_{ij,I}^{\lambda_c}\mathcal{T}^I
\end{equation}
where the sum is over the 22 independent R-symmetry structures and the $a_{ij,I}^{\lambda_c}$ are unfixed coefficients.

For concreteness, let us consider the constraints in the (12) channel, for instance. The other channels are treated similarly. The two-particle Casimir equations read
\begin{align}
\mathcal{C}_{12} R_{12}^{\lambda_c}=\lambda_c R_{12}^{\lambda_c}\,,
\end{align}
for $\lambda_c= -12,-8,0$. One finds the solutions
\begin{align}
R_{12}^{0}&=a_{12,1}^{0} t_{12}^2 t_{34} t_{35} t_{45}\,,\\
R_{12}^{-8}&=a_{12,2}^{-8} \left(t_{12} t_{15} t_{24} t_{34} t_{35}-t_{12} t_{14} t_{25} t_{34}
   t_{35}\right)+a_{12,3}^{-8} \left(t_{12} t_{15} t_{23} t_{34} t_{45}-t_{12} t_{13} t_{25} t_{34}
   t_{45}\right)\nonumber\\
&+a_{12,4}^{-8} \left(t_{12} t_{14} t_{23} t_{35} t_{45}-t_{12} t_{13} t_{24} t_{35}
   t_{45}\right)\,,\\
R_{12}^{-12}&=a_{12,5}^{-12} \left( t_{15} t_{24}
   t_{34} t_{35} t_{12}+ t_{14} t_{25} t_{34} t_{35} t_{12}-\tfrac{1}{3} t_{34} t_{35} t_{45} t_{12}^2\right)+a_{12,6}^{-12} t_{12} t_{15} t_{25} t_{34}^2\nonumber\\
   &+ a_{12,7}^{-12}
   \left( t_{15} t_{23} t_{34} t_{45} t_{12}+ t_{13} t_{25}
   t_{34} t_{45} t_{12}-\tfrac{1}{3} t_{34} t_{35} t_{45} t_{12}^2\right)+a_{12,8}^{-12} t_{12} t_{14} t_{24} t_{35}^2\nonumber\\
   &+a_{12,9}^{-12} \left( t_{14}
   t_{23} t_{35} t_{45} t_{12}+ t_{13} t_{24} t_{35} t_{45} t_{12}-\tfrac{1}{3}t_{34} t_{35} t_{45} t_{12}^2\right)+a_{12,10}^{-12} t_{12}
   t_{13} t_{23} t_{45}^2\,,
\end{align}
where the coefficients $a_{12,i}^{\lambda_c}$ are left undetermined.~\footnote{Here we relabel the $a_{12,I}^{\lambda_c}$ coefficients for I from 1 to 10.} It is easy to understand why the Casimir equation does not completely determine the R-symmetry dependence since different operator exchanges can carry the same representation.

We can now explicitly write down the R-symmetry dependence of the singular terms in our ansatz. In the (12) channel, we have
\begin{align}
&\sum_{\text{R-sym}}p^{12}_{\{n_i\}}(t_{ij})=p^{12,0}_{\{n_i\}} R_{12}^{0}+p^{12,-8}_{\{n_i\}} R_{12}^{-8}+p^{12,-12}_{\{n_i\}} R_{12}^{-12}\\
&=p^{12,0}_{\{n_i\},1} t_{12}^2 t_{34} t_{35} t_{45}+p^{12,-8}_{\{n_i\},2} \left(t_{12} t_{15} t_{24} t_{34} t_{35}-t_{12} t_{14} t_{25} t_{34}
   t_{35}\right)\nonumber\\
&+p^{12,-8}_{\{n_i\},3}\left(t_{12} t_{15} t_{23} t_{34} t_{45}-t_{12} t_{13} t_{25} t_{34}
   t_{45}\right)+p^{12,-8}_{\{n_i\},4} \left(t_{12} t_{14} t_{23} t_{35} t_{45}-t_{12} t_{13} t_{24} t_{35}
   t_{45}\right)+\cdots\nonumber
\end{align}
where $\cdots$ represent the remaining terms of the sum in the first line. Here the unfixed coefficients $a_{12,i}^{\lambda_c}$ are absorbed in the coefficients $p^{12,\lambda_c}_{\{n_i\},j}$ that we aim to fix with the bootstrap.

The use of the two-particle Casimir equations to determine the R-symmetry dependence of the singular terms of the ansatz substantially decreases the number of undetermined coefficients in it. Applying this analysis for all the channels, we are left with an ansatz with 2772 undetermined coefficients for the regular terms and 7000 coefficients for the singular terms.

\subsection*{Permutation symmetry}
To this moment, we have not used the permutation symmetry of exchanging identical operators in the correlator $\langle O_2 O_2 O_2 O_2 O_2\rangle$.  The equivalence of the correlator under the permutation of pairs of operators highly constrains the number of undetermined coefficients of the ansatz. It is enough to impose that the correlation function remains invariant under permutations with the generators of the $S_5$ permutation symmetry, i.e. the adjacent transpositions $\{(12),(23),(34),(45)\}$. By doing so, our ansatz is then a function of 39 undetermined regular coefficients and 69 undetermined coefficients for singular terms.

\subsection{Factorization}
\label{sec:Factorizationboots}

The explicit knowledge of the necessary four-point correlators at large $N$ and large $\lambda$ up to order $\mathcal{O}(\lambda^{-3/2})$ in appendix~\ref{sec:fourpoints} and of the factorization formulas in Mellin space presented in section~\ref{sec:Factorization} completely determines the singular dependence of the ansatz and thus all the so far undetermined $p^{k\ell ,\lambda_c}_{\{n_i\},j}$ coefficients.

For completeness, we provide here an explicit example of factorization in the (45) OPE channel.  We consider for simplicity the case of an exchange of a scalar 20' operator. The necessary ingredients are thus the form of the four-point Mellin amplitude for $\langle O_2 O_2 O_2 O_2 \rangle$ up to order $\mathcal{O}(\lambda^{-3/2})$ and the three-point Mellin amplitude for the protected $\langle O_2 O_2 O_2 \rangle$. As we are focusing on fixing the singular part of the ansatz for the stringy corrections, it is enough here to focus on those contributions to the four-point Mellin amplitude. We have~\footnote{We strip off the $N^{-2}\lambda^{-3/2}$ suppression of the four-point amplitude.}
\begin{align}
\label{eq:factlowerpts}
&M_{222}=f_{OOO}t_{45} t_{04} t_{05}\,\\
&M_{2222}^{\text{str}}={480\zeta_3}\left(\delta _{23}^4 t_{03}^2 t_{12}^2-2 \delta _{12} \delta _{23}^3 t_{02} t_{03} t_{12} t_{13}+2 \delta
   _{12} \delta _{23}^3 t_{01} t_{03} t_{12} t_{23}+\dots\right)
\end{align}
where $f_{OOO}$ is the protected OPE coefficient between three $O_2$ operators. In the second line, the superscript $\text{str}$ denotes the fact that we are only keeping the stringy corrections at order $\mathcal{O}(\lambda^{-3/2})$ and, as it is known, they simply correspond to regular terms in the Mellin variables. The dots represent other terms that we do not explicitly write here. 

Applying the factorization formulas \eqref{eq:Mellinfactorization} and \eqref{eq:residuescalar}, one finds
\begin{equation}
M_{22222}^{\text{str}}=\frac{C_{OOO}M_{222}M_{2222}^{\text{str}}}{(\delta_{45}-1)}+\dots\,,
\end{equation}
where the dots represent not only other contributions from the exchange of a 20' operator that we suppressed above but also contributions from the exchange of the R-symmetry current and the stress tensor. Here $C_{OOO}$ denotes the product of the OPE coefficient $f_{OOO}$ and constants in the factorization formula \eqref{eq:residuescalar} that we were not careful with. The R-symmetry gluing at the point $0$ can be done using the rules in~\eqref{eq:rulegluing}. We then obtain
\begin{align}
M_{22222}^{\text{str}}=&{480\zeta_3}\frac{C_{OOO}}{(\delta_{45}-1)}\left(2 \delta _{23}^4 t_{12}^2 t_{34} t_{35}
   t_{45}+\frac{2}{3} \delta _{12} \delta
   _{23}^3 t_{12} t_{23} t_{45} \left(3
   t_{15} t_{34}+3 t_{14} t_{35}-t_{13}
   t_{45}\right)\right.\nonumber\\
   &\left.-\frac{2}{3} \delta _{12}
   \delta _{23}^3 t_{12} t_{13} t_{45}
   \left(3 t_{25} t_{34}+3 t_{24}
   t_{35}-t_{23} t_{45}\right)+\dots \right)+\dots\,.
\end{align}
Once again, the two sets of dots denote other terms resultant of the 20' exchange and other spinning exchanges, respectively.
The same exercise can be done with the spinning four-point functions studied in appendix~\ref{sec:fourpoints}.  Similarly, when using the factorization formulas, we introduce the coefficients $C_{OOJ}$ and $C_{OOT}$ that account for both the OPE coefficients and the constant factors of the residue functions that appear in the factorization of the Mellin amplitude.
The values of $C_{OOO}$, $C_{OOJ}$ and $C_{OOT}$ can be found by doing this construction accounting also for the supergravity terms, and comparing those to the result of~\cite{Goncalves:2023oyx}. Using the normalizations used in that paper, we find
\begin{equation}
\label{eq:Cvalues}
C_{OOO}= 
 2 \sqrt{2}\,,\quad C_{OOT} =
 \frac{4 \sqrt{2}}{7}\,,\quad  C_{OOJ}= 
 4 \sqrt{2}\,.
 \end{equation}
 
Even though these constants are fixed by this comparison with the supergravity answer, we leave them unfixed before moving to the next steps. We will later see that the choice of the chiral algebra twist imposes a relation between $C_{OOJ} , C_{OOT}$ and $C_{OOO}$ that respects these values.

Since we have already imposed invariance of the correlator under permutation of pairs of external operators, it is enough to use factorization in a given channel (here the (45) channel) to completely determine the singular behaviour of our ansatz. It remains to fix the coefficients associated with the regular terms of the ansatz. For those, we employ the available supersymmetric constraints that arise from choosing both the Drukker-Plefka and chiral algebra twists.

\subsection{Drukker-Plefka twist}
Factorization completely fixed the singular behaviour of our ansatz. To claim success of the bootstrap, we still need to determine the regular contributions. In~\cite{Goncalves:2023oyx}, the choice of Drukker-Plefka twist, presented in section~\ref{sec:setup}, was enough to completely determine the regular piece of the supergravity limit of our correlator in terms of the polar part, fixed by factorization. Conveniently, this constraint can be imposed in Mellin space. We explain here how this can be done.

Recall that the Drukker-Plefka twist imposes
\begin{equation}
\label{eq:DPtwist}
t_{ij}=x_{ij}^2\,.
\end{equation}
In Mellin space, our scalar correlator reads
\begin{equation}
    G(x_i,t_i)_{\text{conn}}=\int [d\delta] M(\delta_{ij},t_{ij})\prod_{1\le i<j\le n} (x_{ij}^2)^{-\delta_{ij}}\, \Gamma(\delta_{ij})\,,
\end{equation}
where $M(\delta_{ij},t_{ij})$ is a linear combination of the R-symmetry structures. It is clear that for each of these terms, to impose~\eqref{eq:DPtwist} simply shifts the exponent of each $x_{ij}^2$. This shift can be absorbed in the Mellin amplitude by shifting conveniently the independent Mellin variables we are integrating over.

For concreteness, let us remove a kinematic factor and write the correlator in terms of cross ratios
\begin{equation}
\label{eq:22222mellin}
\!\!\!\!\!\!\!G(x_i,t_i)_{\text{conn}}= \frac{x_{13}^2 t_{12}^2 t_{35}^2 t_{14} t_{34}}{ x_{12}^4 x_{35}^4 x_{14}^2 x_{34}^2 t_{13}}  \int [d\delta] M(\delta_{ij},\sigma_i)\Gamma_{22222} u_1^{2-\delta _{12}} u_2^{-\delta _{23}} u_3^{2-\delta _{34}} u_4^{-\delta _{45}} u_5^{1-\delta _{15}}\,,
\end{equation}
where we choose the independent Mellin variables to be $\delta_{12}, \delta_{15},\delta_{23},\delta_{34}$ and $\delta_{45}$. The Gamma factors are given by
\begin{align}
\label{eq:gammas5pt}
\Gamma_{22222} =&\Gamma \left(\delta _{12}\right) \Gamma \left(\delta _{15}\right) \Gamma \left(\delta _{23}\right) \Gamma \left(\delta _{34}\right)  \Gamma \left(\delta
   _{45}\right)\Gamma\left(1+\delta _{15}-\delta _{23}-\delta _{34}\right)  \Gamma \left(1-\delta _{12}-\delta _{15}+\delta _{34}\right)\nonumber\\
  & \Gamma \left(1-\delta _{15}+\delta _{23}-\delta_{45}\right) \Gamma \left(1+\delta _{12}-\delta _{34}-\delta _{45}\right) \Gamma \left(1-\delta _{12}-\delta _{23}+\delta _{45}\right)\,.
\end{align}
Note that the choice of the kinematic factor is such that when we take $t_{ij}\to x_{ij}^2$, it goes to 1. Moreover, we introduced a set of cross ratios $\sigma_i$ that are obtained from the cross ratios $u_i$ of~\eqref{eq:crossratios} by replacing $x_{ij}^2$ by $t_{ij}$. 

As discussed above, in Mellin space,  the Drukker-Plefka twist choice of $t_{ij}\to x_{ij}^2$, or equivalently $\sigma_i\to u_i$, acts very simply
\begin{equation}
M(\delta_{ij},\sigma_i)=\sum_{\{n_i\}} \sigma_1^{n_1} \sigma_2^{n_2} \sigma_3^{n_3} \sigma_4^{n_4}  \sigma_5^{n_5} M_{\{n_i\}}(\delta_{ij})\longrightarrow  \sum_{\{n_i\}} u_1^{n_1} u_2^{n_2} u_3^{n_3} u_4^{n_4}  u_5^{n_5} M_{\{n_i\}}(\delta_{ij})\,,
\end{equation}
where the sum over the integer $n_i$ corresponds to the allowed R-symmetry structures of the Mellin amplitude. To recover the correct cross-ratio dependence of~\eqref{eq:22222mellin}, for each of these terms we shift the Mellin variables appropriately. This acts as difference operator for the Mellin amplitude
\begin{align}
&u_1^{n_1} u_2^{n_2} u_3^{n_3} u_4^{n_4} u_5^{n_5} {M}_{\{n_i\}}(\delta_{ij}) \to \mathbb{D}_{n_1,...,n_5} \circ {M}_{\{n_i\}}(\delta_{ij})\,,
\end{align}
where the action of the $\mathbb{D}_{n_1,...,n_5}$ in the Mellin amplitude term is given by
\begin{align}
&\mathbb{D}_{n_1,...,n_5} \circ {M}_{\{n_i\}}(\delta_{ij}) = {M}_{\{n_i\}}(\delta_{12} + n_1, \delta_{23} + n_2, ...) \times (\delta_{12})_{n_1} (\delta_{15})_{n_5} (\delta_{23})_{n_2} (\delta_{34})_{n_3} (\delta_{45})_{n_4}\nonumber\\
&(\delta_{15} - \delta_{23} - \delta_{34} + 1)_{n_5-n_2-n_3} (\delta_{23} - \delta_{15} - \delta_{45} + 1)_{n_2-n_4-n_5} (1 - \delta_{12} - \delta_{15} + \delta_{34} )_{n_3-n_1-n_5}\nonumber\\
&(1+\delta_{12}  - \delta_{34} - \delta_{45} )_{n_1-n_3-n_4} (1 - \delta_{12} - \delta_{23} + \delta_{45} )_{n_4-n_1-n_2}\,.
\end{align}
Here, the Pochhammer symbols are introduced to cancel the unwanted behaviour of the Gamma factors that is introduced by the shift of Mellin variables. 

Acting with the appropriate difference operator for each term of the Mellin amplitude amounts to impose Drukker-Plefka twist in the correlation function. As discussed in section~\ref{sec:setup}, with this choice the correlator becomes topological and it is given by a constant. In Mellin space, this translates to a vanishing Mellin amplitude according to~\cite{Rastelli:2017udc,Rastelli:2016nze}. This is consistent with the naive moving of contours when shifting Mellin variables that we applied in this section.

This implementation of the Drukker-Plefka twist fixes 19 unknown coefficients of the regular part of the Mellin amplitude. Unfortunately, this means that there are still 20 to be fixed. In the next step, we use chiral algebra to further constrain our ansatz.
 
\subsection{Chiral Algebra twist}
Contrary to the Drukker-Plefka twist that has no restriction on the position of operator insertions, chiral algebra restricts the kinematics to a 2d plane.  This limitation prevents the mapping of a generic configuration of operators to a plane via a conformal transformation for correlation functions with more than four operators. Consequently, the cross ratios develop inherent relations between themselves. In contrast, the Mellin amplitude definition assumes independence among all cross ratios. Therefore, it is unclear how such constraints can be fully applied within the Mellin space formalism.

Recently in~\cite{Goncalves:2025jcg}, the authors derived chiral algebra constraints starting from a lightcone conformal block decomposition of a six-point function of 20' operators~\cite{Bercini:2020msp,Antunes:2021kmm}. These constraints were then related to constraints for the polar part of the Mellin amplitude. Here, however, we are looking for constraints that allow us to fix the regular piece of the Mellin amplitude in terms of the already completely fixed singular part. 

To overcome these limitations, we map our Mellin amplitude to position space, where the kinematic restriction to a 2d plane offers no obstructions. We use the relation
\begin{equation}
\label{eq:DfuncsMellin}
\prod_{1\leq i<j \leq 5} (x_{ij}^2)^{-\alpha_{ij}} D_{\tilde{\Delta}_1...\tilde{\Delta}_5} \leftrightarrow {M}^{\alpha_{ij}}(\delta) = \frac{\pi^{\frac{d}{2}} \Gamma\left(\frac{\sum_i \tilde{\Delta}_i - d}{2}\right)}{\prod_i \Gamma(\tilde{\Delta}_i)} \prod_{i<j} \frac{\Gamma(\delta_{ij} - \alpha_{ij})}{\Gamma(\delta_{ij})}\,.
\end{equation}
where $\tilde{\Delta}_i$ are related to the conformal dimensions of the external operators by
\begin{equation}
\tilde{\Delta}_i + \sum_j \alpha_{ij} = \Delta_i\,.
\end{equation}
We introduced the symbol $D_{\Delta_1,\dots,\Delta_5}$ for the so-called D-functions whose definition and properties we review in Appendix~\ref{sec:Dfuncs}. We also kept the expression general in spacetime dimensions $d$, but ultimately we should set it to $d=4$.

As we discuss in the Appendix~\ref{sec:Dfuncs}, D-functions of generic integer weights can be found by taking derivatives of the seed case $D_{11112}$ and its permutations. This function can be written as
\begin{equation}
    D_{11112} = \frac{4\pi^2}{x_{14}^2 x_{35}^2 x_{25}^2} \sum_{i=1}^5 \frac{\eta_{i5} \hat{I}_4^{(i)}}{N_5}\,,
\end{equation}
where $N_5$ and $\eta_{i5}$ are determined in terms of cross ratios as can be seen in Appendix~\ref{sec:Dfuncs}. Moreover, $ \hat{I}_4^{(i)}$ are given by the one-loop scalar box diagrams with the $i$-th point omitted. 
It will be useful for us to introduce a new set of cross ratios
\begin{align}
   &u_1=\frac{(w-1) (\bar{w}-1)z \bar{z}}{(w-z) (\bar{w}-\bar{z})+\lambda}\,,\quad u_2=(z-1) (\bar{z}-1)\,,\\
   &u_3=\frac{(w-z) (\bar{w}-\bar{z})+\lambda}{(w-1) (\bar{w}-1)}\,,\quad u_4=\frac{1}{(w-1)
   (\bar{w}-1)}\,,\quad u_5=\frac{w \bar{w}}{(w-z) (\bar{w}-\bar{z})+\lambda }\,,\nonumber
\end{align}
where we use the cross ratios $u_i$ defined in~\eqref{eq:crossratios}. Taking $\lambda \to 0$ corresponds to the plane limit. In this limit, the definition of these cross ratios above becomes more transparent. It follows from a frame choice where points 1, 3 and 4 are set to 0,1 and $\infty$ and positions 2 and 5 are determined by $z,\bar{z}$ and $w, \bar{w}$, respectively.

For generic point insertions, the five $ \hat{I}_4^{(i)}$ are all independent. However, this is no longer the case when the operators are restricted to lie on a plane. Indeed, one has
\begin{equation}
\sum_{i=1}^{5} \eta_{i5}\hat{I}_4^{(i)} \bigg|_{\lambda\to0} = 0\,.
\end{equation}
At first look, this linear dependence suggests that $D_{11112}$ vanishes in the plane limit $\lambda\to0$. However, in this limit, $N_5$  also vanishes, yielding a finite and nonzero result for the seed D-function when all the points are put in a common plane. In fact, this linear dependence in the plane, which ultimately follows from the identity
\begin{equation}
\text{Li}_2\left(\frac{z w}{(1-z)(1-w)}\right) = \text{Li}_2\left(\frac{z}{1-w}\right) + \text{Li}_2\left(\frac{w}{1-z}\right) - \text{Li}_2(z) - \text{Li}_2(w) - \log(1-z)\log(1-w)\,,
\end{equation}
can be found by requiring that the evaluation of the D-functions have no singular behaviour as $\lambda\to0$.

After translating the Mellin ansatz to position space using~\eqref{eq:DfuncsMellin}, we evaluate each of the D-functions and then take the common plane limit. For large weight D-functions, the expressions in which we take $\lambda\to0$ are complicated and the evaluation of the limit is very time-consuming. For practicality, we first consider all the D-functions written in a basis of five independent box integrals, logarithms and rational terms, and then we evaluate them in a common but random frame by fixing $z, \bar{z}, w,\bar{w}$. Note that by doing this, we can still keep track of the dependence on box integrals. These terms should not suffer from ambiguities when one translates the ansatz from Mellin to position space~\cite{Rastelli:2017udc}, but should not remain after the choice of chiral algebra presented in section~\ref{sec:setup}.  Equipped with a choice of frame, we take the limit $\lambda\to 0$ and collect the terms multiplying the four remaining independent box integrals for arbitrary and independent vectors $v_i$.  The choice of chiral algebra then requires these terms to vanish. This constraint fixes the remaining ansatz up to two undetermined coefficients. Importantly, these coefficients cannot be fixed by making a different choice of frame. All frames lead to the same constraints.

A nice check of our bootstrap so far also shows up when solving the chiral algebra constraints. As detailed in the factorization section~\ref{sec:Factorizationboots}, when determining the singular behaviour of the ansatz, we left the coefficients $C_{OOJ}$ and $C_{OOT}$ unconstrained even though their form is related to the scalar case $C_{OOO}$.~\footnote{We stress that these coefficients are not just OPE coefficients.} To satisfy the chiral algebra constraints, we determine
\begin{equation}
C_{OOJ}=2 C_{OOO}\,,\quad C_{OOT}=\frac{2}{7} C_{OOO}\,,
\end{equation}
which is consistent with the determined values of these constants in~\eqref{eq:Cvalues} by comparing with the supergravity result of~\cite{Goncalves:2023oyx}.

\subsection{Protected Observables}
It remains to determine two coefficients in the regular terms of our ansatz. By using the OPE, it is possible to extract lower-point observables contained in our five-point correlator. We aim to further constrain our ansatz by focusing on protected quantities that do not get stringy corrections. For that, it will be useful to know the first operators, i.e. of lowest scaling dimension, appearing in the OPE of two 20' operators. This analysis was carried out in section 6 of~\cite{Goncalves:2019znr}.

\subsubsection{[0,4,0] sector}

We start by focusing on OPE exchanges in the [0,4,0] representation of the R-symmetry. As discussed in~\cite{Goncalves:2019znr}, in this representation, the leading OPE contribution is determined by a single exchange of a scalar double-trace operator of scaling dimension $\Delta=4$. This is a half-BPS operator of the schematic form $[O_2 O_2]$, which we will denote as $O_2^2$. Both the OPE coefficient between two 20' operators and $O_2^2$  and the four-point correlator $\langle O_2 O_2 O_2 O_2^2\rangle$ are protected and thus completely determined by their free-theory value. It follows that the Mellin amplitude of the latter vanishes. Hence, we want to explore what constraints are imposed in our ansatz by this observation.

The correlator $\langle O_2 O_2 O_2 O_2^2\rangle$ admits a writing similar to~\eqref{eq:fourpointO2}
\begin{align}
& \langle O_2 O_2 O_2 O_2^2\rangle= \frac{t_{12} t_{13} t_{14} t_{34}^2}{x_{12}^2 x_{13}^2 x_{14}^2 x_{34}^4}\mathcal{G}_{2222^2}\left(u,v;\sigma,\tau\right)\,,\\
&\mathcal{G}_{2222^2}\left(u,v;\sigma,\tau\right)= \mathcal{G}_{2222^2}^{\text{free}}\left(u,v;\sigma,\tau\right)+ R(u,v;\sigma,\tau) H(u,v)\,,
\end{align}
where $R(u,v;\sigma,\tau)$ was defined in~\eqref{eq:RSWard}. Since free-theory terms have vanishing Mellin amplitude and the protected nature of the correlator $\langle O_2 O_2 O_2 O_2^2 \rangle$ requires that the reduced correlator $H(u,v)$ vanishes, we must have that the Mellin amplitude of this correlator is indeed 0.

Starting from the five-point correlator of five 20' operators, we can extract the correlator $\langle O_2 O_2 O_2 O_2^2\rangle$ by appropriately selecting the leading OPE exchange in the [0,4,0] representation. In the supergravity limit, this exercise has already been done in~\cite{Aprile:2026uxe} and one considers
\begin{align}
&\lim_{x_5\to x_4} (x_{45}^2)^{\frac{(4-\Delta)}{2}} \lim_{t_5\to t_4}\langle O_2(x_1, t_1)O_2(x_2, t_2)O_2(x_3, t_3)O_2(x_4, t_4)O_2(x_5, t_5)\rangle\\
&=f_{O_2 O_2 O_2^2}\langle O_2(x_1, t_1)O_2(x_2, t_2)O_2(x_3, t_3)O_2^2(x_4, t_4)\rangle \nonumber
\end{align}
where $\Delta=4$ and $f_{O_2 O_2 O_2^2}$ is the OPE coefficient between the three operators. In all cases that we will be dealing with in this section, there will be no degeneracy of primary operators contributing at leading order for each R-symmetry representation.

In Mellin space, after taking the limit in the polarization vectors, one finds
\begin{align}
	\!\!\!\!\!\!\!\!\!\!\!\!\!\!\!\!\!\! \lim_{t_5\to t_4}\int [d\delta] M(\delta_{ij},t_{ij})\prod_{ i<j} \frac{\Gamma(\delta_{ij})}{x_{ij}^{2\delta_{ij}}}=t_{12} t_{13} t_{14} t_{34}^2\sum_{{ 0\le m+n\le2}} \sigma^{m}\tau^{n} \int [d\delta] M_{\sigma^m\tau^n}(\delta_{ij})\prod_{ i<j} \frac{\Gamma(\delta_{ij})}{x_{ij}^{2\delta_{ij}}}\,,
\end{align}
where the sum is over the possible remaining R-symmetry structures, i.e. $\{1,\sigma, \tau, \sigma \tau, \sigma^2,\tau^2\}$, and $M_{\sigma^m\tau^n}$ denotes the corresponding Mellin amplitudes that multiply them. Note that, up to this moment, these Mellin amplitudes are still functions of five independent Mellin variables. 

It follows from the form of $R(u,v;\sigma,\tau)$ in~\eqref{eq:RSWard}, that the reduced correlator in Mellin space can be read from $M_{\tau^2}$ alone (up to a cross ratio $u$). As it was seen in~\cite{Aprile:2026uxe}, the vanishing of the reduced correlator in the supergravity regime is kinematic in $M_{\tau^2}$ after taking the limit $t_5\to t_4$. The same is true for the stringy corrections of our ansatz. On the other hand, this is no longer the case, for instance, for the $M_1$ component and it must be the second limit $x_5 \to x_4$ that kills non-vanishing contributions. Just as in the supergravity case of~\cite{Aprile:2026uxe}, the vanishing of the Mellin amplitude is only attained after the integration over three of the Mellin variables to go from the five independent variables of a five-point Mellin amplitude to the two Mellin variables at four points. 

After taking the limit $t_5\to t_4$, one can see that the Mellin amplitude is free of poles in $\delta_{45}$. This is expected since in the [0,4,0] representation the leading OPE contribution is determined by a double-trace operator, whose contribution at large $N$ must be stored in the Gamma factors. Indeed, the limit $x_5\to x_4$ in this R-symmetry channel is regular, and the leading term is determined by the pole at $\delta_{45}=0$ in  $\Gamma(\delta_{45})$. In this limit, the first integration (out of three) is simply done by capturing the residue at this pole.

The two remaining integrals can be done after some massaging of each component of the Mellin amplitude to put its terms in the form of the first Barnes lemma.
As discussed in~\cite{Aprile:2026uxe}, the idea is to absorb the Mellin amplitude into the Gamma functions by rewriting it as a sum over products of Gamma functions with shifted arguments. In doing so, one is effectively decomposing the Mellin amplitude in sums over terms corresponding to D-functions after using the identity~\eqref{eq:DfuncsMellin}. To achieve this rewriting, it is simpler to use the unconstrained Mellin variables of the five-point Mellin amplitude and only impose the constraints~\eqref{eq:deltaconstraints} at a later stage when one starts integrating.  We then take three steps. First, if a variable appears  both in the numerator and denominator of a term in the Mellin amplitude, we split the term using
\begin{equation}
\frac{\delta}{-k + {\delta}} = 1 + \frac{k}{-k + {\delta}}\,.
\end{equation}
This step will avoid nested integrals at later stages. Next, we use the following identities to rewrite each term as a product of gamma functions
\begin{align}
&\frac{1}{-k + {\delta}} = \sum_{j=1}^{k} \frac{\Gamma[k]}{\Gamma[k - j + 1]} \frac{\Gamma[{\delta} - j]}{\Gamma[{\delta}]}\,,
&\delta^k = \sum_{j=0}^{k} (-1)^{k-j} S_{k,j} \frac{\Gamma[\delta + j]}{\Gamma[\delta]}\,,
\end{align}
where $S_{k,j} = \sum_{i=0}^{j} \frac{(-1)^{j-i} i^k}{i!(j-i)!}$ are the second Stirling numbers.

Upon using these three steps, one verifies that the two remaining integrals can be done using the first Barnes lemma,
\begin{equation}
\frac{1}{2\pi i} \int_{-i\infty}^{i\infty} \Gamma(a+s)\,\Gamma(b+s)\,\Gamma(c-s)\,\Gamma(d-s)\,ds = \frac{\Gamma(a+c)\,\Gamma(a+d)\,\Gamma(b+c)\,\Gamma(b+d)}{\Gamma(a+b+c+d)}\,.
\end{equation}

After integrating over three of the five Mellin variables in $M_{1}$, keeping only the leading contribution in the limit $x_5\to x_4$, we see that the corresponding reduced correlator vanishes independently of the two undetermined coefficients. The same analysis is true for the remaining components of the Mellin amplitude, i.e. $M_{\sigma^2}, M_{\sigma \tau}$ (trivially) and $M_{\sigma}, M_{\tau}$ (after integration) also vanish.

The same exercise could have been done in position space by translating the terms in Mellin space to a linear combination of D-functions using~\eqref{eq:DfuncsMellin}. However, this would come with extra complications and subtleties. As we have already discussed, the map between Mellin-space terms and D-functions is ambiguous and may affect lower-transcendentality terms. On the other hand, after massaging the expression of our Mellin amplitude and translating it to a symbolic expression in terms of D-functions, one observes that some of these terms lead to potential logarithmic divergences. The regularity of the OPE in this R-symmetry channel requires that such divergences cancel. These extra complications have also been encountered in~\cite{Bissi:2021hjk}, when extracting the supergravity limit of the correlator $\langle O_2 O_2 O_2 O_2^2 \rangle$ (and the one of the next subsection) from the five-point function of 20' operators of~\cite{Goncalves:2019znr}. The authors had to carefully regularize similar divergences and show that they indeed cancel.~\footnote{We also repeated the analysis of~\cite{Bissi:2021hjk} using the Mellin amplitude of~\cite{Goncalves:2023oyx} and extracted the four-point amplitudes of the correlators of this and the subsequent section. In both cases, we find that the Mellin amplitude vanishes (trivially or after integration). This method eschews the technical difficulties of position space, but is not sensitive to the precise match of these protected quantities with their free-theory value.} The computation in Mellin space offers then natural advantages as long as the OPE is regular and one can perform the first integration by just capturing the leading pole in the Gamma factors.

The analysis of this subsection provides yet another nice nontrivial check of our bootstrap, but unfortunately does not raise any new constraints on the remaining undetermined coefficients.

\subsubsection{[2,0,2] sector}
\label{sec:202sector}
Continuing following the discussion of~\cite{Goncalves:2019znr} and the low-lying spectrum in the OPE of two 20' operators, we consider now other protected quantities. 

The bottom component of the quarter-BPS multiplet in the [2,0,2] representation of the R-symmetry is realized as a scalar double-trace operator (plus a single-trace operator with a coefficient at order $\mathcal{O}(1/N)$) with $\Delta=4$. This operator, which we will denote by $Q$, can be written in a specific choice of polarizations as
\begin{equation}
Q = \text{tr}(Z^2)\text{tr}(X^2) - \text{tr}(ZX)\text{tr}(ZX) + \frac{1}{N}\text{tr}([ZX][ZX])\,,
\end{equation}
where $Z$ and $X$ are the complex scalar fields $Z=\phi^1+i \phi^2$ and $X=\phi^3+i \phi^4$. Not only is the OPE coefficient between two 20' operators and this quarter-BPS protected~\cite{Goncalves:2019znr,DHoker:2001jzy}, but also the four-point correlator $ \langle Q O_2 O_2 O_2\rangle$ is determined by its free-theory value. In~\cite{Bissi:2021hjk}, the authors computed the free-theory form of this correlator. By focusing on operator exchanges in the $[2,0,2]$ representation of the R-symmetry and then taking an OPE limit in the supergravity five-point correlation function of five 20' operators~\cite{Goncalves:2019znr}, they found perfect agreement with the free-theory result. This provided a very nontrivial check of the protectedness of this correlator, but a general argument for this is also presented in~\cite{Bissi:2021hjk}. By considering the OPE exchanges allowed by the R-symmetry selection rules, one sees that no [0,0,0] R-symmetry representation can be exchanged simultaneously in the OPE of both two 20' operators and one 20' and $Q$. As the only long multiplet that can be exchanged in the OPE of two 20' operators has a bottom component in the [0,0,0], it follows that no long operators are exchanged in $ \langle Q O_2 O_2 O_2\rangle$. This kills the possibility of having non-free-theory-like terms in the correlator, and thus it must be that our ansatz for the stringy corrections leads to a vanishing Mellin amplitude corresponding to stringy corrections of this correlator. In what follows, we use the protected nature of the correlator $ \langle Q O_2 O_2 O_2\rangle$ to further constrain our ansatz.

We start by focusing on the terms of our ansatz that contribute to the exchange of operators in the [2,0,2] by identifying which combinations of the R-symmetry structures $\mathcal{T}_{I}$ are associated with this exchange. To do so, we use an analysis entirely similar to the one done in section~\ref{sec:theansatz} regarding the eigenfunctions of the R-symmetry Casimir operator. The same idea was used in~\cite{Bissi:2021hjk} and the necessary R-symmetry structures were identified. For completeness, in Appendix~\ref{sec:Rsympoly} we list all R-symmetry structures that appear in the OPE of two 20' operators at five points. As it follows from that discussion, it is enough to consider the terms of the Mellin amplitude that multiply the R-symmetry structures $\mathbb{E}_1,\mathbb{E}_2$ and $\mathbb{E}_3$ defined in Appendix \ref{sec:Rsympoly}. 

After selecting these structures, we proceed as in the previous subsection and integrate over three of the independent Mellin variables. Just as in that case, the leading OPE contribution in the representation $[2,0,2]$ is determined by a single scalar double-trace operator with $\Delta=4$. This limit is again regular and no anomalous dimensions can be associated with the exchange of this operator. As before, the first integration is simply done by capturing the residue at $\delta_{12}=0$ in $\Gamma(\delta_{12})$. The two other integrations are then obtained by using the first Barnes lemma.~\footnote{To identify the four-point Mellin amplitude of the correlator $\langle Q O_2 O_2 O_2\rangle$, one has to divide by the factor $\prod_{ i<j}\Gamma(\delta_{ij})/{(x_{ij}^2)^{-\delta_{ij}}}$ appearing in the definition of the Mellin amplitude~\eqref{eq:Mellindef} and impose the constraints~\eqref{eq:deltaconstraints}.} The procedure is entirely similar to the one described for $O_2^2$, however, the vanishing of the Mellin amplitude requires that one of the undetermined coefficients is, in fact, determined by the other. We are now down to one undetermined coefficient in our ansatz.

\subsubsection{[0,2,0] sector}
In the low-lying spectrum allowed in the OPE of two 20' operators, we can look for other protected quantities that might help us to fully fix our ansatz. 

Besides the operators in the supermultiplet of the operator 20', whose data were already used to fix the singular part of our ansatz through use of factorization, there are other protected quantities associated with exchanges in the representations $[0,4,0], [2,0,2]$ and $[0,2,0]$, according to~\cite{Goncalves:2019znr}. The first two representations were already studied in the previous subsections and the analysis made there is consistent with the protected OPE coefficients involving $O_2^2$ and $Q$ computed in~\cite{Goncalves:2019znr}.

However, in the representation $[0,2,0]$ there is a semi-short operator, which we shall call $C$, that deserves our attention. This operator is a scalar double-trace of scaling dimension $\Delta=4$. Due to its semi-short nature, some OPE coefficients involving $C$ and other protected operators are protected as well. That is the case of the OPE coefficients $f_{O_2 O_2 C}$, $f_{O_2 O_2^2 C}$ and $f_{O_2 Q C}$. In particular, the fact that the last two OPE coefficients are protected is consistent with the protectedness of $\langle O_2^2 O_2 O_2 O_2\rangle$ and $\langle Q O_2 O_2 O_2\rangle$ we just used before. On the other hand, the OPE coefficient $f_{O_2 C C}$ at strong coupling does not match its free-theory value~\cite{Goncalves:2019znr}. This is not surprising since the correlator $\langle C O_2 O_2 O_2 \rangle$ may receive contributions from long operators as R-symmetry does not forbid them from being exchanged. We expect the stringy corrections of the OPE coefficient $f_{O_2 C C}$ to be sensitive to the last remaining coefficient of our ansatz. Unfortunately, these stringy corrections have never been computed, and therefore, we cannot follow this avenue to determine our ansatz completely. We do, however, some checks that this is indeed the case.

Contrary to the cases we studied before, $C$ does not determine the leading OPE term in its R-symmetry channel, and its contribution generally mixes with descendants of the 20'. This makes the OPE analysis of the data involving this operator harder. In particular, in this R-symmetry representation, the OPE is singular, and the simplifications we used before are no longer present.~\footnote{At leading order in the (12) OPE channel, we can still extract the leading singular behaviour of the $[0,2,0]$ R-symmetry channel following a similar strategy to the one presented before. Note, however, that these contributions will contain no information about the regular terms of our ansatz. To perform this exercise, we select contributions in the $[0,2,0]$ by focusing on the Mellin components multiplying the structures $\mathbb{E}_8$ till $\mathbb{E}_{13}$, defined in Appendix~\ref{sec:Rsympoly}. After multiplying the expression by $x_{12}^2$ to cancel the power-law singular behaviour, one can capture the residue of the amplitude at $\delta_{12}=1$, associated with the explicit poles for single-trace exchanges. Two other remaining integrations can be done using Barnes lemma. As one should, we find a perfect match with the stringy corrections of the four-point amplitude of four 20' operators that we gave as input.}

To focus on the OPE coefficient $f_{O_2 C C}$ we shall take two OPE limits, say in the (12) and (45) channels, and project out all contributions that are not given by two R-symmetry exchanges in the $[0,2,0]$, one in each channel. This can be done by, once again, changing the basis of the R-symmetry structures and decomposing the correlator in eigenfunctions of two R-symmetry Casimir equations in these channels. The change of basis is presented in Appendix~\ref{sec:Rsympoly}. In particular, we will be looking at contributions multiplying the structure $R_{12,45}^{-12,-12}$.

As we have already stated above, the exchange of two operators $C$ is not the leading contribution of this double OPE. Ideally, one would like to map our expression to position space and then perform a conformal block decomposition of the five-point correlator in this R-symmetry channel and read how the stringy corrections of the OPE coefficient $f_{O_2 C C}$ (recall that $f_{O_2 O_2 C}$ is protected) determine the last remaining coefficient of our ansatz. Unfortunately, this is not simple for two main reasons. First, our bootstrap was done in Mellin space, and a naive mapping to D-functions in position space would likely spoil a correct determination of this coefficient. On the other hand, even if one could trust completely in the map~\eqref{eq:DfuncsMellin}, the total weight of the D-functions involved would make the computation of our ansatz in position space extremely time-consuming even before starting to take the two OPE limits. 
Nonetheless, we do gather some information from position space.

Considering the contribution of each D-function in $R_{12,45}^{-12,-12}$, we analyze at what order in the OPE expansion the possible logarithmic divergences in the (12) and (45) channels can appear. We follow the asymptotic analysis of~\cite{Georgoudis:2017meq} in a given OPE. We find that, for each OPE, logarithms only start contributing to these terms at scaling dimensions $\Delta\ge6$. This motivates an approach that can be completely done in Mellin space.

In the Mellin amplitude, we would like to consider the contributions of poles corresponding to the lowest double-trace operators in this R-symmetry channel. The two scalar $C$ exchanges at $\Delta=4$ should come from the first poles of the Gamma functions $\Gamma(\delta_{12}), \Gamma(\delta_{45})$ at $\delta_{12}=\delta_{45}=0$. We stress again that these are not the leading poles and, in particular, that the OPE on both channels starts with singular single-trace contributions. Naively capturing only the residues from these Gamma functions will not be correct in general. Contrary to what happens at four points, the pinching mechanism between the poles of the Gamma functions that leads to logarithms in the correlator is not immediately transparent at five points, as one can see from~\eqref{eq:gammas5pt}. A systematic analysis of this important phenomenon at higher points is lacking in the literature, but initial steps were done in~\cite{Yuan:2018qva}.  

As we capture the residues of the poles at $\delta_{12}=\delta_{45}=0$ in $\Gamma(\delta_{12})$, $\Gamma(\delta_{45})$, we are trusting that no logarithms appear at this order. This is motivated by the explorations in position space that we just commented on above.

In a double OPE limit, one must integrate over the remaining Mellin variables. Expanding around this limit, and only after some massaging, we can integrate over the remaining Mellin variables using the first Barnes lemma once again. We find that the undetermined coefficient of the ansatz is sensitive to this quantity. We should say, however, that in doing so, we have chosen a specific order of integration and carefully manipulated the integrand to avoid some divergent behaviour of the Gamma functions that we believe to correspond to fake poles. Similar situations were found in~\cite{Yuan:2018qva}.

Since there is no available CFT data for the stringy corrections of $f_{O_2 C C}$ that can be used to fix the final coefficient of our ansatz, and the procedure for the 5-fold integration requires a more careful analysis, we refrain from presenting further details here, leaving that for future work. We stress, however, that it is very likely that this OPE coefficient is indeed sensitive to the undetermined coefficient, as follows from this analysis.
 
\subsection{Flat-space Limit}
\label{sec:Flatspacelimit}

Up to this moment, we have rigorously fixed almost entirely the ansatz for the first stringy correction of the five-point correlator $\langle O_2 O_2 O_2 O_2 O_2\rangle$ at large $N$ and large $\lambda$, except for one undetermined coefficient associated with the regular contributions of the ansatz. This coefficient multiplies dominating terms of the high-energy limit, large $\delta_{ij}$, of the Mellin amplitude scaling as $\mathcal{O}(\delta_{ij}^4)$. This limit is related to the tree-level flat-space amplitude of five gravitons of type IIB string theory in 10 dimensions. In this section, we investigate whether we can use this fact to further constrain our ansatz.

To be more explicit in what follows, we introduce a new set of Mellin variables $s_{ij}$ defined as
\begin{equation}
s_{ij}=\Delta_i+\Delta_j-2\delta_{ij}\,,
\end{equation}
which will map to flat-space Mandelstam variables $S_{ji}=-(p_i+p_j)^2$ where $p_i$ is the momentum of particle $i$. 

As shown in~\cite{Penedones:2010ue,Fitzpatrick:2011hu}, the high-energy limit of the five-point Mellin amplitude in AdS$_5\times S^5$ maps to a 10d flat-space scattering amplitude
\begin{equation}
\label{eq:fsl}
A(S_{ij},\epsilon_{\mu\nu}) = \Gamma\left(\frac{\Delta_\Sigma}{2} - 2\right) \lim_{L \to \infty} L^{15/2} V_5\int_{-i\infty}^{i\infty} \frac{d\alpha}{2\pi i} \alpha^{2-\frac{\Delta_\Sigma}{2}} e^{\alpha} M\left(s_{ij} = \frac{L^2}{2\alpha} S_{ij},t_i\right)\,,
\end{equation}
where $V_5$ is the volume of the unit $S^5$ and $\Delta_\Sigma=\sum_i \Delta_i$. On the left-hand side, the tree-level graviton flat-space amplitude one finds by taking the high-energy limit of the Mellin amplitude of five 20' operators is not in the most generic kinematics. The reason for that lies in the factorized structure of AdS$_5\times S^5$ in which the supergravitons live. We will be more precise about this below. 

To compare the Mellin and scattering amplitudes, it is also necessary to make use of the AdS/CFT dictionary relating the physical parameters of the theories on both sides. We have
\begin{equation}
	\frac{L^4}{\ell_s^4} = \lambda = g_{\text{YM}}^2 N\,, \quad g_s = \frac{g_{\text{YM}}^2}{4\pi}\,,
\end{equation}
where $g_s$ and $\ell_s=\sqrt{\alpha'}$ are the string coupling and length respectively and $L$, as previously used, denotes the radius of both AdS$_5$ and the internal $S^5$.

The closed-string tree-level amplitude for five gravitons can be computed using the KLT relations~\cite{Kawai:1985xq} and available results for open-string amplitudes~\cite{Mafra:2011nw}. We checked that the five-point graviton tree-level amplitude can be written in a $\alpha'$ expansion as~\cite{Schlotterer:2012ny,Gomez:2015uha}~\footnote{We thank Carlos Mafra for sharing with us the explicit expressions of the SYM tree amplitudes. These are made available at~\cite{Mafrapage}.}
\begin{align}
A(S_{ij},\epsilon_{\mu\nu}) &\sim k_{10}^3 \tilde{A}_{54}^T \cdot S_0 \cdot \left[ 1 + 2\zeta_3 \left( \frac{\alpha'}{2} \right)^3 M_3 + 2\zeta_5 \left( \frac{\alpha'}{2} \right)^5 M_5 + \mathcal{O}(\alpha'^6) \right] \cdot A_{45}
\end{align}
where $k_{10}\sim g_s \alpha'^2$ is the 10-dimensional gravitational coupling and $A_{45}$ and $\tilde{A}_{54}$ are two-component vectors of color-ordered tree-level SYM amplitudes
\begin{equation}
\tilde{A}_{54} \equiv \begin{pmatrix}
\tilde{A}^{\text{YM}}(1,2,3,5,4) \\
\tilde{A}^{\text{YM}}(1,3,2,5,4)
\end{pmatrix}\,, \quad A_{45} \equiv \begin{pmatrix}
A^{\text{YM}}(1,2,3,4,5) \\
A^{\text{YM}}(1,3,2,4,5)
\end{pmatrix}\,.
\end{equation}
The notation with a tilde in $\tilde{A}^{\text{YM}}$ is a remnant of the usual notation of KLT relations, and it is introduced to separate the polarization dependence of the two amplitudes. In fact, we can label the polarization vectors in each amplitude as $t_\mu$ and $\tilde{t}_\mu$.  The polarization tensors of the gravitons are then obtained from the double copy simply by considering~\cite{Alday:2021odx,Chester:2018aca}
\begin{equation}
\epsilon_{\mu\nu}=t_\mu \tilde{t}_\nu\,.
\end{equation} 
Moreover, the polarization vectors are related to the six-dimensional R-symmetry null vectors $t_I$ by adding zeros to the remaining components.
Since the polarization tensors of the graviton are symmetric in spacetime indices, we can drop the tilde notation from now on. Moreover, note that this identification of the polarization vectors with R-symmetry null vectors immediately guarantees a symmetric and traceless polarization tensor $\epsilon_{\mu\nu}$.~\footnote{Transversality $K^\mu \epsilon_{\mu\nu}=0$ is also obeyed from what we will see next.}

Above, we also use the momentum kernel
\begin{equation}
S_0 \equiv \frac{1}{4} \begin{pmatrix}
S_{12}(S_{13} + S_{23}) & S_{12}S_{13} \\
S_{12}S_{13} & S_{13}(S_{12} + S_{23})
\end{pmatrix}\,,
\end{equation}
and the matrices $M_i$ were introduced in \cite{Schlotterer:2012ny}. For our interests, it is enough to consider $M_3$
\begin{align}
&M_3 \equiv \begin{pmatrix}
m_{11} & m_{12} \\
m_{21} & m_{22}
\end{pmatrix}\,,
\end{align}
with
\begin{align}
&m_{12} = \frac{S_{13}S_{24}}{8}(S_{12} + S_{23} + S_{34} + S_{45} + S_{15})\,,\nonumber\\
&m_{11} = \frac{S_{34}}{8}\left[S_{12}(S_{12} + 2S_{23} + S_{34}) - S_{34}S_{45} - S_{45}^2\right] - \frac{S_{12}S_{15}}{8}(S_{12} + S_{15})\,,
\end{align}
and $m_{21}=m_{12}\vert_{2\leftrightarrow 3}$ and $m_{22}=m_{11}\vert_{2\leftrightarrow 3}$. 

The explict form of the SYM tree amplitudes in~\cite{Mafrapage} shows that each amplitude scales at high energy as $\mathcal{O}(S_{ij}^{-1})$. This means that the supergravity term goes in the high-energy regime  as $\mathcal{O}(S_{ij}^0)$.~\footnote{It is important for this scaling analysis that we restrict the polarization tensors built from KLT relations to the graviton case. This ensures that polarization-dependent terms of SYM amplitudes will not create any other Mandelstam variable contribution not immediately accounted by a naive inspection of these amplitudes. The same would not be true in principle for dilatons.} This is precisely the behaviour of the Mellin amplitude of our correlator in the supergravity regime at large Mellin variables $s_{ij}$~\cite{Goncalves:2019znr,Goncalves:2023oyx}, aligning with the flat-space formula where Mellin and Mandelstam variables map to each other.  We see, moreover, that the next term in the $\alpha'$ expansion multiplies a term at order $\mathcal{O}(S_{ij}^3)$.  Consequently, it seems reasonable to fix the remaining coefficient of our ansatz in order to vanish higher-order contributions that have no equivalent in flat space. As we will see, this is no longer possible after imposing that the correlator $\langle Q O_2 O_2 O_2 \rangle$ is protected as we did in section~\ref{sec:202sector}.~\footnote{So that the reader can confirm this, we also provide the result of our ansatz before this constraint in an ancillary file.}

Here comes why this is not a problem. As anticipated before, the factorized structure of AdS$_5\times S^5$ restricts the kinematics of the flat-space amplitude corresponding to the high-energy limit of the Mellin amplitude to one where the polarization vectors are orthogonal to all momenta, i.e.  $t_i\cdot p_j=0$.  With an odd number of points, it is not possible to contract the polarization vectors simply amongst themselves and, necessarily, at least one polarization vector contracts with one momentum. This ensures that in this special kinematics the SYM tree amplitudes vanish and, by the KLT relation, so does the five-point graviton amplitude.  

The vanishing of the flat-space amplitude under the flat-space limit formula~\eqref{eq:fsl} follows not from the high-energy limit of the Mellin amplitude but rather from extra powers of $L$ that suppress the CFT amplitude. For concreteness, a simple computation shows that the supergravity Mellin amplitude is suppressed by $L^{-9/2}$ when compared to the flat-space amplitude. Although it would be reasonable to anticipate that the CFT amplitude experiences no enhancement by powers of $L$ in the first stringy correction when compared to supergravity (provided we consider the subsequent order $L^6/\lambda^{3/2}$ contribution in flat space), this does not necessarily hold true. In fact, to preserve the protected nature of $\langle Q O_2 O_2 O_2 \rangle$, we observe that the first stringy is less suppressed by powers of $L$ than the supergravity result. This is not problematic because this suppression is, in any case, sufficient to yield the anticipated zero in flat space.  Actually, the left-hand side zero in \eqref{eq:fsl} could still be achieved even if we could include even higher powers of Mellin variables in the high-energy regime of the Mellin amplitude, provided the limit in the AdS radius would be enough to suppress those contributions. 
This observation, however, implies that the five-point Mellin amplitude does not yield a definite $1/L$ correction to the graviton amplitude across all stringy corrections.

The flat-space limit at five points proved very limited in use. The same is expected for every Mellin amplitude with an odd number of points. On the other hand, the flat-space formula of~\cite{Penedones:2010ue} contains much more information for an even number of points. In the future, we aim to study the high-energy limit of the six-point correlation function of 20' operators and relate it to the five-point correlators contained within. This analysis will not only shed light on some terms of the stringy corrections at six points, but it may also provide another check of our bootstrap or even an alternative approach to determine the remaining coefficient of our ansatz. 

\subsection{Result}
With the use of factorization and supersymmetry constraints, we managed to fix an ansatz for the first stringy correction of the correlation function of five 20' operators in Mellin space up to a single undetermined coefficient.
We found
\begin{equation}
\frac{M^{R^4}}{960 \sqrt{2} \zeta_3}=\sum_{a=1}^{12} M^{P}_{a}P^{a}+\sum_{a=1}^{10} M^{T}_{a}T^{a}\,,
\end{equation}
where the labels $P^{a}$ and $T^{a}$ correspond to the pentagonal $\mathcal{T}_{(ijklm)}$ and triangular $\mathcal{T}_{(ijk)(lm)}$ R-symmetry structures of~\eqref{eq:Rsymstructures}, respectively. Here we also divide by the product of $C_{OOO}$ and the constant corresponding to the first stringy correction of the four-point correlation function in~\eqref{eq:reducedMellinscalar}.

For conciseness, we write one example of the pentagon and triangle shapes. The remaining  R-symmetry structures are obtained by permutation of points. The existence of one undetermined coefficient in the regular terms leads to slightly lengthier expressions for those. We will only partially write those contributions, and we direct the reader to the ancillary file where the complete result is provided.

For the structure $P^1=t_{12}t_{34}t_{45}t_{15}$, we have
\begin{equation}
M^{P}_1= M^{P, sing}_1+M^{P, reg}_1\,,
\end{equation}
\begin{align}
& M^{P, sing}_1=-\frac{2 \left(\delta _{12}+\delta _{23}-2\right)}{\delta _{45}-1} \left(\delta _{12} \left(\delta _{23} \left(4 \delta _{23}-5 \delta _{34}-7\right)-5 \delta _{15} \left(\delta _{23}-3\right)+4
   \delta _{34}-6\right)\right.\nonumber\\
   &\left.+\delta _{12}^2 \left(4 \delta _{23}-5 \delta _{15}+1\right)+\delta _{15} \left(4 \delta _{23}-9\right)+\left(5 \left(3-\delta _{23}\right) \delta
   _{23}-9\right) \delta _{34}+\left(\delta _{23}-3\right){}^2\right)\nonumber\\
   &-\frac{2 \left(\delta _{23}+\delta _{34}-2\right)}{{\delta _{15}-1}} \left(\delta _{12} \left(4 \delta _{34}-5 \delta _{23}
   \left(\delta _{23}+\delta _{34}-3\right)-9\right)+\delta _{23} \left(\delta _{34} \left(4 \delta _{34}-5 \delta _{45}-7\right)\right.\right.\nonumber\\
   &\left.\left.+4 \delta _{45}-6\right)+\delta
   _{23}^2 \left(4 \delta _{34}+1\right)+\left(5 \left(3-\delta _{34}\right) \delta _{34}-9\right) \delta _{45}+\left(\delta _{34}-3\right){}^2\right)\nonumber\\
   &-\frac{2
   \left(\delta _{12}+\delta _{15}-2\right)}{\delta _{34}-1} \left(\delta _{12} \left(\delta _{15} \left(4 \delta _{15}-5 \delta _{23}-5 \delta _{45}-7\right)+15 \delta _{23}+4 \delta
   _{45}-6\right)\right.\nonumber\\
   &\left.+\delta _{12}^2 \left(4 \delta _{15}-5 \delta _{23}+1\right)+\delta _{15} \left(5 \left(3-\delta _{15}\right) \delta _{45}+\delta _{15}+4 \delta _{23}-6\right)-9
   \left(\delta _{23}+\delta _{45}-1\right)\right)\nonumber\\
   &-\frac{2 \left(\delta _{15}+\delta _{45}-2\right)}{\delta _{23}-1} \left(\delta _{12} \left(4 \delta _{45}-5 \delta _{15} \left(\delta
   _{15}+\delta _{45}-3\right)-9\right)+\delta _{45}^2 \left(4 \delta _{15}-5 \delta _{34}+1\right)\right.\nonumber\\
   &\left.+\delta _{45} \left(\delta _{15} \left(4 \delta _{15}-5 \delta
   _{34}-7\right)+15 \delta _{34}-6\right)+4 \delta _{15} \delta _{34}+\delta _{15}^2-6 \delta _{15}-9 \delta _{34}+9\right)\nonumber\\
   &-\frac{2 \left(\delta _{34}+\delta
   _{45}-2\right)}{\delta _{12}-1} \left(\delta _{15} \left(\delta _{34} \left(4-5 \delta _{45}\right)-5 \left(\delta _{45}-3\right) \delta _{45}-9\right)+\delta _{23} \left(5 \delta _{34}
   \left(3-\delta _{34}-\delta _{45}\right)\right.\right.\nonumber\\
   &\left.\left.+4 \delta _{45}-9\right)+\left(4 \delta _{34}+1\right) \delta _{45}^2+\left(\delta _{34} \left(4 \delta _{34}-7\right)-6\right) \delta
   _{45}+\left(\delta _{34}-3\right){}^2\right)\,,
\end{align}
\begin{align}
&M^{P, reg}_1=2 \left(\delta _{12} \left(\left(6 \delta _{45}+17\right) \delta _{15}^2+6 \left(-\left(\left(\delta
   _{34}+2\right) \delta _{45}\right)-\delta _{23} \left(\delta _{34}+\delta _{45}+2\right)-6\right) \delta
   _{15}\nonumber\right.\right.\\
&\left.\left.-7 \left(\delta _{34}-8\right) \delta _{34}+\delta _{23}^2 \left(6 \delta _{34}+17\right)-7
   \left(\delta _{45}-8\right) \delta _{45}-6 \delta _{23} \left(\delta _{34} \left(\delta
   _{45}+2\right)+6\right)+25\right)+\dots\right)\nonumber\\
 &+ \frac{1}{8} \text{reg}_1\left(4 \delta _{12}^4+4 \delta _{15}^4+4 \delta _{23}^4-32 \delta _{23}^3+87 \delta _{23}^2-163 \delta _{23}-57
   \delta _{23}^2 \delta _{34}+150 \delta _{23} \delta _{34}-163 \delta _{34}\nonumber\right.\\
 &\left.+170+4 \delta _{34}^4+6 \delta _{23} \delta _{34}^3-32 \delta _{34}^3-10 \delta _{23}^2 \delta _{34}^2-57
   \delta _{23} \delta _{34}^2+87 \delta _{34}^2+6 \delta _{23}^3 \delta _{34}\dots\right)\,,
\end{align}
where $\text{reg}_1$ is the undetermined coefficient of our ansatz.

For the structure $T^1=t_{12}^2 t_{34} t_{35} t_{45}$, we find
\begin{equation}
M^{T}_1= M^{T, sing}_1+M^{T, reg}_1\,,
\end{equation}
\begin{align}
&M^{T, sing}_1=\frac{2 \left(\delta _{23}-\delta _{15}+\delta _{34}-1\right){}^2 \left(-\delta _{12}-\delta _{15}+\delta _{23}+\delta _{34}+1\right){}^2}{\delta _{35}-1}\nonumber\\
&+\frac{2 \delta _{15}^2 \left(\delta _{12}+\delta _{15}-2\right){}^2}{\delta _{34}-1}+\frac{2 \delta _{23}^2 \left(\delta _{12}+\delta _{23}-2\right){}^2}{\delta
   _{45}-1}+\frac{2}{\delta_{12}-1} \left(\delta_{15} \left(\delta _{34} \left(20 \left(1-\delta _{45}\right) \delta _{45}-6\right)\right.\right.\nonumber\\
&\left.\left.+2 \delta _{23} \left(2 \delta _{34} \left(5 \delta _{45}-3\right)-6 \delta _{45}+3\right)+3
   \delta _{45} \left(2 \delta _{45}-1\right)\right)+2 \delta _{15}^2 \left(\delta _{45} \left(5 \delta _{45}-7\right)+3\right)\right.\nonumber\\
&\left.-\delta _{45} \left(\delta _{23} \left(20\left(\delta _{34}-1\right) \delta _{34}+6\right)+\delta _{34} \left(\delta _{34}+15\right)-11\right)+\left(\delta _{23}+1\right) \left(2 \delta _{23} \left(\delta _{34}
   \left(5 \delta _{34}-7\right)+3\right)\right.\right.\nonumber\\
   &\left.\left.-4 \delta _{34}^2+11 \delta _{34}-6\right)+\left(\delta _{34} \left(10 \delta _{34}-1\right)-4\right) \delta _{45}^2\right)\,,
\end{align}
\begin{align}
&M^{T, reg}_1=2\left( \delta _{12} \left(-6 \left(\delta _{45}+2\right) \delta _{15}^2+\left(\left(6 \delta _{34}-17\right)
   \delta _{45}+2 \delta _{23} \left(3 \delta _{34}+3 \delta _{45}+8\right)+17\right) \delta _{15}\nonumber\right.\right.\\
   &\left.\left.-6 \delta_{23}^2 \left(\delta _{34}+2\right)-\delta _{34} \left(\delta _{34}-16 \delta _{45}+9\right)-11 \delta
   _{45}+\delta _{23} \left(-2 \delta _{45}+\delta _{34} \left(6 \delta _{45}-19\right)+21\right)+6\right)+\dots\right)\nonumber\\
   &+ \frac{1}{8} \text{reg}_1\left(\delta _{15} \left(-16 \delta _{23}^3+24 \left(\delta _{45}-\delta _{34}\right) \delta _{23}^2+2 \left(8
   \delta _{34}^2+\left(84 \delta _{45}-65\right) \delta _{34}+\delta _{45} \left(8 \delta
   _{45}-65\right)\nonumber\right.\right.\right.\\
   &\left.\left.\left.+43\right) \delta _{23}+24 \delta _{34}^3+3 \left(7-4 \delta _{45}\right) \delta
   _{45}^2+20 \delta _{34}^2 \left(\delta _{45}-1\right)+\delta _{45}+\delta _{34} \left(66-108 \delta
   _{45}^2\right)-10\right)\right.\nonumber\\
   &\left.+8 \delta _{15}^4+2 \delta _{12}^3 \left(\delta _{15}+\delta _{23}+1\right)-8 \delta _{15}^3 \left(2 \delta _{23}+2 \delta _{34}-\delta _{45}+1\right)+\dots\right)\,,
\end{align}
where it was convenient to use more than five independent Mellin variables.

\section{Discussion}
\label{sec:discussion}
In this paper we explored a bootstrap approach to fix the first stringy correction to the five-point correlator of 20' operators in $\mathcal{N}=4$ SYM at large $N$ and strong coupling. 
To determine the singular behaviour of our ansatz, we made use of the factorization of Mellin amplitudes into protected three-point functions and four-point correlators whose form is known (or determined here) up to order $1/\lambda^{3/2}$. 

The regular contributions, on the other hand, are computed by requiring that the correlator satisfies certain supersymmetric constraints for two choices of twist known as Drukker-Plefka~\cite{Drukker:2009sf} and chiral algebra~\cite{Beem:2013sza} twists. These are the two known supersymmetric constraints that can be applied to higher-point correlators. It remains an interesting question for the future to fully explore the consequences of superconformal symmetry at higher points.  After imposing these constraints, we are left with two undetermined coefficients. By taking the OPE in our five-point Mellin amplitude, we study contributions from our ansatz to protected four-point correlators. These correlators do not receive stringy corrections, and this fact is used to further constrain one more coefficient. We also studied a non-protected OPE coefficient that is sensitive to the remaining freedom of our ansatz. However, its stringy corrections at order $1/\lambda^{3/2}$ have not been computed, and we cannot use this information to fully fix the ansatz at this stage.

The remaining undetermined coefficient multiplies the dominating regular terms in the high-energy limit of the Mellin amplitude. Naively, one expects this regime to be determined by the 10-dimensional flat-space type IIB graviton amplitude. However, the flat-space amplitude one obtains from the high-energy limit of the Mellin amplitude is in a specific kinematics where polarization vectors are orthogonal to all momenta~\cite{Alday:2021odx}. Ultimately, this forces the five-point amplitude to vanish. On the CFT side of the flat-space limit formula of~\cite{Penedones:2010ue}, this zero of the amplitude is understood by the suppression of the Mellin amplitude by powers of AdS radius $L$. We confirm that our answer is compatible with the zero in flat-space, but no further constraints can be extracted from there. Interestingly, however, we find that to preserve the protectedness of the aforementioned correlators, the first stringy corrections must be less suppressed by powers of $L$ than the supergravity result, while simultaneously ensuring consistency with the flat-space limit formula.

Compared to~\cite{Goncalves:2023oyx}, where the bootstrap is entirely performed in Mellin space and relies on factorization and Drukker-Plefka twist, as we consider $1/\lambda$ corrections, these steps are no longer sufficient. The necessity of imposing chiral algebra constraints requires at the moment the use of position space. This is a consequence of a restriction of the kinematics to a 2d plane when one uses the latter constraint. We write our Mellin ansatz in position space as a linear combination of D-functions. These objects can be evaluated using the methods presented in Appendix~\ref{sec:Dfuncs}. However, their complexity increases with larger external scaling dimensions. It is thus a very interesting question to find ways to impose chiral algebra constraints directly in Mellin space. This would help generalizing this bootstrap approach to arbitrary Kaluza-Klein modes and to higher orders in $1/\lambda$. Alternatively, one could also study these observables by exploring generalizations of techniques that have been successfully applied to four-point correlators, such as supersymmetric localization (e.g.~\cite{Binder:2019jwn}) or genus-0 world-sheet integrals with single-valued multiple polylogarithms (e.g.~\cite{Alday:2023mvu}).\footnote{Fixing an ansatz for such world-sheet integral would in principle also require the generalization of dispersive sum rules to higher-point functions in CFTs. This is an interesting problem on its own.} 

At five points, flat-space constraints are of very limited use due to the reasons presented above. However, one expects that for higher- and even-point functions these constraints can be much more powerful. In fact, this was already observed in the strong coupling regime of six-point functions of supergluons and supergravitons in~\cite{Alday:2023kfm,Goncalves:2025jcg}. It would be interesting to derive flat-space constraints on the stringy corrections of these correlators. Moreover, by using factorization of Mellin amplitudes at six points, one may uncover nontrivial constraints from flat space at the level of five-point correlators. We leave this for future work.

Building on the success of this bootstrap approach in determining the structure of a very nontrivial object, it is compelling to apply these and analogous techniques in new contexts. One promising direction is the computation of $1/N$ corrections to this five-point correlator at infinite coupling. Another exciting direction is the study of stringy and loop corrections to higher-point correlation functions in other theories. A good and natural example is the five-point amplitude of supergluons in AdS$_5 \times S^3$ whose strong coupling result was computed in~\cite{Alday:2022lkk}.

Finally, let us mention that our result contains a wealth of new CFT data of $\mathcal{N} = 4$ SYM including corrections up to $1/\lambda^{3/2}$. It would be interesting to use the OPE to target and extract various non-protected three- and four-point functions.

\acknowledgments

We would like to thank Carlos Mafra, Lecheng Ren, Rodolfo Russo, Congkao Wen, Xinan Zhou and especially Vasco Gonçalves for illuminating discussions and Rodolfo Russo and Vasco Gonçalves for comments on the draft. JVB is supported by the UK EPSRC grant 'CFT and Gravity: Heavy States and Black Holes' EP/W019663/1 and by FCT grant 2024.00230.CERN. No new data were generated or analysed during this study.

\appendix
\section{Four-point correlators}
\label{sec:fourpoints}
In this appendix we aim to derive the first stringy correction to the four-point correlation functions with three 20' operators and either the R-symmetry current or the stress tensor. To do so, we make use of the fact that the three operators belong to the same supermultiplet. As explained in~\cite{Belitsky:2014zha,Goncalves:2023oyx}, one can use superspace and obtain the required correlation functions by studying the correlator of four superfield operators, which are uniquely determined by correlators of the bottom components of each multiplet. 

Contrary to the main text notation, where we use the six-dimensional R-symmetry polarization vectors  $t_I$, in superspace it is natural to use instead a four component variable $y$. At last, we will only need to relate the basic two-point invariant
\begin{equation}
t_{ij}=y_{ij}^2\,.
\end{equation}
In the rest of this appendix, we use the same notation and conventions of~\cite{Goncalves:2023oyx} and we keep the review very brief. See \cite{Goncalves:2023oyx} for further details.

The superfield contains the contributions of all operators of a given multiplet and it is of the schematic form
\begin{equation}
\mathbb{O}_2=O_2+\left(\rho \sigma^\mu \bar{\rho}\right)J_\mu+\left(\rho \sigma^\mu \bar{\rho}\right)\left(\rho \sigma^\nu \bar{\rho}\right)T_{\mu\nu}+\dots\,,
\end{equation}
where we introduce the Grassmann variables $\rho$ and $\bar{\rho}$.

Note that an $n$-point correlation function of the bottom components corresponds to the $n$-point correlator of the superfield by setting the fermionic variables to 0. On the other hand, one can also extract the superdescendant contributions by acting with a differential operator in the superfield
\begin{align}
	&J=\frac{1}{2}\mathcal{D}_J \mathbb{O}_2(x,y,\rho,\bar{\rho})\Big|_{\rho,\bar{\rho}=0}\,,\\
	&T=\frac{1}{4}\mathcal{D}_T\mathbb{O}_2(x,y,\rho,\bar{\rho})\Big|_{\rho,\bar{\rho}=0}\,.
\end{align} 
Note also that here we further suppressed the spinor indices for both spacetime and R-symmetry dependences as it will become clear in the form of the of the differential operators below.

Given the charges and simetries of the two operators, one can easily write an ansatz for the differential operator that selects them. The relative coefficients are fixed by demading that non-diagonal two-point functions vanish in CFTs. One finds the operators~\cite{Belitsky:2014zha,Goncalves:2023oyx}
\begin{align}
&\mathcal{D}_{J} = \lambda^{\alpha} \bar{\lambda}^{\dot{\alpha}} v^{a} \bar{v}^{\dot{a}} 
\left( 
\frac{\partial}{\partial \bar{\rho}^{a \dot{\alpha}}} \frac{\partial}{\partial \rho^{\alpha\dot a}} 
+ \frac{1}{2} \frac{\partial}{\partial y^{a \dot{a}}} \frac{\partial}{\partial x^{\alpha \dot{\alpha}}} 
\right)\,,\\
&\mathcal{D}_{T} = \lambda^{\alpha_1} \lambda^{\alpha_2} \bar{\lambda}^{\dot{\alpha}_1} \bar{\lambda}^{\dot{\alpha}_2} 
\epsilon^{\dot{a}_1 \dot{a}_2} \epsilon^{a_1 a_2} \times \Bigg(
\frac{\partial}{\partial \bar{\rho}^{a_1 \dot{\alpha}_1}} 
\frac{\partial}{\partial \bar{\rho}^{a_2 \dot{\alpha}_2}} 
\frac{\partial}{\partial \rho^{\alpha_1 \dot{a}_1}} 
\frac{\partial}{\partial \rho^{\alpha_2 \dot{a}_2}} \nonumber
\\
&\quad -
\frac{\partial}{\partial \bar{\rho}^{a_1 \dot{\alpha}_1}} 
\frac{\partial}{\partial \rho^{\alpha_1 \dot{a}_1}} 
\frac{\partial}{\partial y^{a_2 \dot{a}_2}} 
\frac{\partial}{\partial x^{\alpha_2 \dot{\alpha}_2}} 
-\frac{1}{6}
\frac{\partial}{\partial y^{a_1 \dot{a}_1}} 
\frac{\partial}{\partial y^{a_2 \dot{a}_2}} 
\frac{\partial}{\partial x^{\alpha_1 \dot{\alpha}_1}} 
\frac{\partial}{\partial x^{\alpha_2 \dot{\alpha}_2}} 
\Bigg)\,,
\end{align}
where $\lambda$ and $v$ are introduced to saturate the spinor-index dependence of both operators regarding spacetime and R-symmetry, respectively.

These operators can be used within $n$-point correlators. 
In particular, in the four-point case we have
\begin{align}
&\langle J (1) O_{2}(2) O_{2}(3) O_{2}(4) \rangle 
= \tfrac{1}{2} \mathcal{D}_{J}(X_1)
\left\langle \mathbb{O}(X_1) \mathbb{O}_{2}(X_2) \mathbb{O}_{2}(X_3) \mathbb{O}_{2}(X_4) \right\rangle \Big|_{\rho, \bar{\rho} = 0},\\
&\langle T(1) O_{2}(2) O_{2}(3) O_{2}(4) \rangle 
= \tfrac{1}{4} \mathcal{D}_T(X_1) 
\left\langle \mathbb{O}_{2}(X_1)\mathbb{O}_{2}(X_2) \mathbb{O}_{2}(X_3) \mathbb{O}_{2}(X_4) \right\rangle \Big|_{\rho, \bar{\rho} = 0},
\end{align}
where $X$ encodes the coordinate dependence in $x, y, \rho,\bar\rho$. Importantly, as discussed in~\cite{Belitsky:2014zha,Goncalves:2023oyx}, the correlator $\left\langle \mathbb{O}_{2}(X_1)\cdots \mathbb{O}_{2}(X_4) \right\rangle$ is uniquely determined by~$\langle O_2 (1) \cdots O_{2}(4) \rangle$ after a convenient supersymmetrization of the cross-ratio dependence.

As shown in~\cite{Goncalves:2023oyx,Goncalves:2019znr}, in the end one can write these correlators in the form
\begin{equation}
\label{eq:4ptsJposition}
\langle J(1) \mathcal{O}_{2}(2) \mathcal{O}_{2}(3) \mathcal{O}_{2}(4) \rangle 
= \frac{1}{x_{12}^{4} x_{34}^{4}} 
\sum_{k=2}^4 \alpha^{(k)}(u,v; y_{ij}, Y_{1,ij}) \, 
\frac{z \cdot x_{1k}}{x_{1k}^2},
\end{equation}

\begin{equation}
\label{eq:4ptsTposition}
\langle T(1) \mathcal{O}_{2}(2) \mathcal{O}_{2}(3) \mathcal{O}_{2}(4) \rangle 
= \frac{1}{x_{12}^{4} x_{34}^{4}} 
\sum_{k,l=2}^4 \beta^{(k,l)}(u,v; y_{ij}) \, 
\frac{z \cdot x_{1k}}{x_{1k}^2} \frac{z \cdot x_{1l}}{x_{1l}^2},
\end{equation}
with
\begin{equation}
\label{eq:Ysdef}
Y_{1,ij} = y_{1i}^2 y_{1j}^2 \, V_{1,ij} \,,\quad V_{i,jk}= \bar v_i (y_{ij}^{-1}-y_{ik}^{-1})v_i\,,
\end{equation}
where ultimately both $\alpha^{(k)}$ and $\beta^{(k,l)}$ are determined by acting with the corresponding differential operators in the four-point function of 20' operators. 

In Mellin space, the spinning correlators can be written in the form of~\eqref{eq:spinningMellin} that we repeat here for reader's convenience. We have
\begin{align}
&\langle\mathcal{J}(1) {O}_{2}(2) {O}_{2}(3){O}_{2}(4)\rangle = \sum_{k=2}^{4} \frac{z \cdot x_{1k}}{x_{1k}^2} \int [d\delta] {M}^k \prod_{i=2}^{4} \frac{\Gamma(\delta_i + {\bm \delta}_i^k)}{x_{1i}^{2\delta_i}} \prod_{i<j} \frac{\Gamma(\delta_{ij})}{x_{ij}^{2\delta_{ij}}}\,,\\
&\langle T(1){O}_{2}(2){O}_{2}(3){O}_{2}(4)\rangle = \sum_{k,l=2}^{4} \frac{z \cdot x_{1k}}{x_{1k}^2} \frac{z \cdot x_{1l}}{x_{1l}^2} \int [d\delta] {M}^{kl} \prod_{i=2}^{4} \frac{\Gamma(\delta_i + {\bm \delta}_i^k + {\bm \delta}_i^l)}{x_{1i}^{2\delta_i}} \prod_{i<j} \frac{\Gamma(\delta_{ij})}{x_{ij}^{2\delta_{ij}}}\,,\nonumber
\end{align}
where ${\bm \delta}_i^j$ are Kronecker deltas. Note moreover that in this appendix, we assume the spinning operator to be at point 1 instead of 0 when comparing to~\eqref{eq:spinningMellin}.  The Mellin variables $\delta_{ij}$ are not all independent and instead obey the constraints
\begin{equation}
\delta_{ij}=\delta_{ji}\,,\quad\delta_{ii}=-\Delta_i\,,\quad  \delta_i = -\sum_{j=2}^4 \delta_{ij}, \quad \sum_{i,j=2}^4 \delta_{ij} = J - \Delta_1.
\end{equation}
For the cases at hand, $J-\Delta_1=-2$ and we can use the same set of independent Mellin variables for the scalar, spin 1 and spin 2 cases. Working with the choice made in~\eqref{eq:mellinvariables4pt} and using \eqref{eq:4ptsJposition} and \eqref{eq:4ptsTposition} above, we can read
\begin{align}
&\alpha^{(k)}(u, v; y_{ij}, Y_{1,ij}) = \int \frac{ds dt}{4} u^{\frac{s}{2}} v^{\frac{t-4}{2}} {M}^k(s, t; y_{ij}, Y_{1,ij}) \prod_{i=2}^{4} \Gamma(\delta_i + {\bm \delta}_i^k) \prod_{i<j} \Gamma(\delta_{ij})\,,\nonumber\\
&\beta^{(k,l)}(u, v; y_{ij}) = \int \frac{ds dt}{4} u^{\frac{s}{2}} v^{\frac{t-4}{2}} {M}^{kl}(s, t; y_{ij}) \prod_{i=2}^{4} \Gamma(\delta_i + {\bm \delta}_i^k + {\bm \delta}_i^l) \prod_{i<j} \Gamma(\delta_{ij})\,,
\end{align}
and conversely
\begin{align}
&{M}^k(s, t; y_{ij}, Y_{1,ij}) \prod_{i=2}^{4} \Gamma(\delta_i + {\bm\delta}_i^k) \prod_{i<j} \Gamma(\delta_{ij}) = \int_0^{\infty} du dv \, u^{-\frac{s}{2}-1} v^{1-\frac{t}{2}} \alpha^{(k)}(u, v; y_{ij}, Y_{1,ij})\,,\nonumber\\
&{M}^{kl}(s, t; y_{ij}) \prod_{i=2}^{4} \Gamma(\delta_i +{\bm\delta}_i^k +{\bm\delta}_i^l) \prod_{i<j} \Gamma(\delta_{ij}) = \int_0^{\infty} du dv \, u^{-\frac{s}{2}-1} v^{1-\frac{t}{2}}\beta^{(k,l)}(u, v; y_{ij})\,.
\end{align}

Using~\eqref{eq:reducedmellin} we can relate derivatives of the dynamical terms in $H(u,v)$ to the action of a difference operator in Mellin space
\begin{align}
&\int_0^{\infty} du \int_0^{\infty} dv \, u^{-\frac{s}{2}-1} v^{1-\frac{t}{2}} u^m v^n \frac{\partial^a}{\partial u^a} \frac{\partial^b}{\partial v^b} H(u, v) = {\mathcal{M}}(s - 2m + 2a, t - 2n + 2b)\nonumber\\
&\times (-1)^{a+b} \left(m - a - \frac{s}{2}\right)_{a} \left(n - b + \frac{4-t}{2}\right)_{b} \prod_{i<j} \Gamma(\tilde{\delta}_{ij}(s - 2m - 2a, t - 2n - 2b))\,.
\end{align}
This relation effectively allows us to compute $M^k$ and ${M}^{kl}$ in terms of the reduced scalar Mellin amplitude ${\mathcal{M}}$ whose form up to order $1/\lambda^{3/2}$ was given in~\eqref{eq:reducedMellinscalar}. Ultimately, we find the required Mellin amplitudes for the spinning correlators up the same order in $\lambda$.

For $\langle J O_2 O_2 O_2\rangle$, we have
\begin{align}
M^2=&\frac{30 \zeta _3}{\lambda ^{3/2}}\left((t-4)^2 (s+3 t-13) t_{24} t_{34} Y_{1,23} -(s+t-4)^2 (2 s+3 t-11)t_{23} t_{34} Y_{1,24}\right.\nonumber\\
  &\left. +(s-4) (2 s-9)(s+2 t-8)
   t_{23} t_{24} Y_{1,34} \right)-\frac{2 (t-4)^2 t_{24} t_{34} Y_{1,23}}{(s-2) (s+t-6)}\nonumber\\
  &+\frac{2 (s+t-4)^2 t_{23} t_{34} Y_{1,24} }{(s-2)
   (t-2)}-\frac{2 (s-4)(s+2 t-8) t_{23} t_{24} Y_{1,34} }{(t-2) (s+t-6)}\,,
\end{align}
\begin{align}
    M^3=& \frac{30 \zeta _3}{\lambda ^{3/2}}\left((s-4)^2 (2 s-t-5) t_{23} t_{24} Y_{1,34} +(t-4)^2 (s-2 t+5) t_{24} t_{34} Y_{1,23} \right.\nonumber\\
    &\left.- (s-t) (s+t-4) (2 s+2 t-7) t_{23} t_{34} Y_{1,24} \right)-\frac{2 (s-4)^2 t_{23} t_{24} Y_{1,34}}{(t-2) (s+t-6)}\nonumber\\
    &+\frac{2 (t-4)^2 t_{24} t_{34} Y_{1,23}}{(s-2)
   (s+t-6)}+\frac{2 (s-t) (s+t-4)t_{23} t_{34} Y_{1,2,4} }{(s-2) (t-2)}\,,
\end{align}
\begin{align}
   M^4=& \frac{30 \zeta _3}{\lambda ^{3/2}}\left((s+t-4)^2 (3 s+2 t-11) t_{23} t_{34} Y_{1,24}-\left((s-4)^2 (3 s+t-13) t_{23} t_{24} Y_{1,34} \right)\right.\nonumber\\
   &\left. -(t-4) (2 t-9) (2 s+t-8) t_{24} t_{34} Y_{1,23}  \right)+\frac{2 (s-4)^2 t_{23} t_{24} Y_{1,34}}{(t-2) (s+t-6)}\nonumber\\
   &+\frac{2 (t-4)(2s+t-8) t_{24} t_{34} Y_{1,23} }{(s-2) (s+t-6)}-\frac{2  (s+t-4)^2 t_{23} t_{34} Y_{1,24}}{(s-2) (t-2)}
\end{align}

\vspace{0.5cm}

For $\langle T O_2 O_2 O_2\rangle$, we have
\begin{align}
M^{22}=&-\frac{80 \zeta_3}{\lambda ^{3/2}}\left(s^2 (2 t-11)+s (t (2 t-39)+118)-23 (t-4)^2\right)t_{23} t_{24} t_{34}\nonumber\\
&+\frac{8 (s-4) (t-4)  (s+t-4) t_{23} t_{24} t_{34}}{3 (s-2) (t-2) (s+t-6)}\,,
\end{align}
\begin{align}
M^{23}=& -\frac{80 \zeta_3}{\lambda ^{3/2}}\left(s^2 (2 t-11)+s (t-8) (2 t-11)+t (13 t-35)-32\right)t_{23} t_{24} t_{34}\nonumber\\
&+\frac{8 (t-4) (s (s+t-8)+2 (t+2)) t_{23} t_{24} t_{34}}{3 (s-2) (t-2) (s+t-6)}\,,
\end{align}
\begin{align}
M^{24}=& -\frac{80 \zeta_3}{\lambda ^{3/2}}\left(s^2 (2 t+13)+s (t (2 t+21)-173)+t (13 t-173)+520\right)t_{23} t_{24} t_{34}\nonumber\\
&+\frac{8  (s+t-4) (s (t+2)+2 (t-10))t_{23} t_{24} t_{34}}{3 (s-2) (t-2) (s+t-6)}\,,
\end{align}
\begin{align}
M^{33}=&-\frac{80 \zeta_3}{\lambda ^{3/2}}\left(s^2 (2 t-11)+s (t (2 t-15)+58)+(58-11 t) t-128\right)t_{23} t_{24} t_{34}\nonumber\\
&+\frac{8 (s-4) (t-4) (s+t-4) t_{23} t_{24} t_{34}}{3 (s-2) (t-2) (s+t-6)}\,,
\end{align}
\begin{align}
M^{34}=&-\frac{80 \zeta_3}{\lambda ^{3/2}} \left(s^2 (2 t+13)+s (t (2 t-27)-35)-11 (t-8) t-32\right) t_{23} t_{24} t_{34}\nonumber\\
&+\frac{8 (s-4) (s (t+2)+(t-8) t+4)t_{23} t_{24} t_{34}}{3 (s-2) (t-2) (s+t-6)}\,,
\end{align}
\begin{align}
M^{44}=&-\frac{80 \zeta_3}{\lambda ^{3/2}}  \left((2 s-11) t^2+(s (2 s-39)+118) t-23 (s-4)^2\right)t_{23} t_{24} t_{34}\nonumber\\
&+\frac{8 (s-4) (t-4) (s+t-4)t_{23} t_{24} t_{34} }{3 (s-2) (t-2) (s+t-6)}\,.
\end{align}

The supergravity terms of the Mellin amplitudes we found agree with those of~\cite{Goncalves:2019znr,Goncalves:2023oyx}. Moreover, we also check that they obey the transversality condition~\eqref{eq:Mellintransverse}.  These Mellin amplitudes are provided in an ancillary file.

Last, note that the methods here presented can be generically applied up to any order in the large $\lambda$ expansion, provided the reduced scalar Mellin amplitude is known up to the same order.

\section{D-functions}
\label{sec:Dfuncs}
A generic $D$-function with $n$ external points is defined by the integral
\begin{equation}
\label{eq:Dfuncdef}
    D_{\Delta_1,\dots,\Delta_n} = \int \frac{dz_0\, d^d z}{z_0^{d+1}} \prod_{i=1}^{n} \left( \frac{z_0}{z_0^2 + (\vec{z} - \vec{x}_i)^2} \right)^{\Delta_i}\,,
\end{equation}
where the integration is performed over Euclidean AdS$_{d+1}$ in Poincaré coordinates.
These functions satisfy the differential identity
\begin{equation}
\label{eq:Dfuncrel}
    D_{\Delta_1,\dots,\Delta_i+1,\dots,\Delta_j+1,\dots,\Delta_n}
    = \frac{d/2 - \Sigma}{\Delta_i \Delta_j} \frac{\partial}{\partial x_{ij}^2} D_{\Delta_1,\dots,\Delta_n}\,,
\end{equation}
where $\Sigma = \frac{1}{2} \sum_{i=1}^{n} \Delta_i$. This relation, which allows one to generate higher-weight $D$-functions from seed configurations, can be derived from the Feynman parameter representation of \eqref{eq:Dfuncdef}:
\begin{equation}
    \frac{\pi^{d/2} \Gamma(\Sigma - \tfrac{d}{2}) \Gamma(\Sigma)}{2 \prod_i \Gamma(\Delta_i)} 
    \int_0^1 \prod_{j=1}^n da_j\, \delta\left(1 - \sum_j a_j \right) 
    \frac{\prod_j a_j^{\Delta_j - 1}}{\left( \sum_{i<j} a_i a_j x_{ij}^2 \right)^{\Sigma}}\,,
\end{equation}
which depends on spacetime dimensionality only through the prefactor involving $\Gamma(\Sigma - \tfrac{d}{2})$.

At four points, $D$-functions can be computed from the seed case $D_{1111}$ by repeated application of~\eqref{eq:Dfuncrel}. After the extraction of a kinematic factor, a $D$-function is determined by a $\bar{D}$-function that only depends on cross ratios~\cite{Dolan:2001tt},
\begin{equation}
\frac{\prod_{i=1}^{4} \Gamma(\Delta_i)}{\Gamma\!\left(\Sigma - \frac{1}{2}d\right)} \frac{2}{\pi^{d/2}} D_{\Delta_1 \Delta_2 \Delta_3 \Delta_4} = \frac{(x_{14}^2)^{\Sigma - \Delta_1 - \Delta_4}(x_{34}^2)^{\Sigma - \Delta_3 - \Delta_4}}{(x_{13}^2)^{\Sigma - \Delta_4}(x_{24}^2)^{\Delta_2}} \,\bar{D}_{\Delta_1 \Delta_2 \Delta_3 \Delta_4}(u, v)\,,
\end{equation}
where $u$ and $v$ were defined in~\eqref{eq:4ptcrossratios}. The seed case is determined by the one-loop scalar box integral
\begin{equation}
    \bar{D}_{1111}=I_4[1] = \Phi(z, \bar{z}) = \frac{1}{z - \bar{z}} \left[ 2 \operatorname{Li}_2(z) - 2 \operatorname{Li}_2(\bar{z}) + \log(z \bar{z}) \log \left( \frac{1 - z}{1 - \bar{z}} \right) \right]\,,
\end{equation}
where we introduce different canonical nomenclatures that we use interchangeably. This integral belongs to a broader class of integrals~\cite{Bern:1992em,Bern:1993kr},
\begin{equation}
    I_n[P(\{a_i\})] = \Gamma(n - 2) \int_0^1 \prod_{j=1}^n da_j\, \delta\left(1 - \sum_j a_j \right) 
    \frac{P(\{a_i\})}{\left( \sum_{i<j} a_i a_j x_{ij}^2 \right)^{n - 2}}\,,
\end{equation}
where $P(\{a_i\})$ is a polynomial in the Feynman parameters.

In the five-point case, all relevant $D$-functions can be obtained from the seed function $D_{11112}$ (and its permutations) that takes the form
\begin{equation}
    D_{11112} = \frac{4\pi^2}{x_{14}^2 x_{35}^2 x_{25}^2} \sum_{i=1}^5 \frac{\eta_{i5} \hat{I}_4^{(i)}}{N_5}\,,
\end{equation}
with normalization and weight factors defined via the matrix $\rho$:
\begin{equation}
    \rho = N_n \eta^{-1}, \qquad N_n = 2^{n-1} \det \rho\,,
\end{equation}
where
\begin{equation}
    \rho = \begin{pmatrix}
        0 & u_1 & 1 & 1 & u_5 \\
        u_1 & 0 & u_2 & 1 & 1 \\
        1 & u_2 & 0 & u_3 & 1 \\
        1 & 1 & u_3 & 0 & u_4 \\
        u_5 & 1 & 1 & u_4 & 0
    \end{pmatrix}\,,
\end{equation}
with the five independent conformal cross ratios defined in~\eqref{eq:crossratios}.

The quantities $\hat{I}_4^{(i)}$ are one-loop box integrals computed with the $i$-th external point omitted. For example,
\begin{equation}
    \hat{I}_4^{(5)} = \Phi(u_1 u_3, u_2)\,,
\end{equation}
with the identifications
\begin{equation}
    u_1 u_3 = \frac{x_{12}^2 x_{34}^2}{x_{13}^2 x_{24}^2} = z \bar{z}, \qquad
    u_2 = \frac{x_{14}^2 x_{23}^2}{x_{13}^2 x_{24}^2} = (1 - z)(1 - \bar{z})\,.
\end{equation}

To generate higher-weight $D$-functions from $\Phi(z, \bar{z})$, it is helpful to use the differential relations
\begin{align}
    \partial_z \Phi &= \frac{\Phi}{\bar{z} - z} + \frac{\log[(1 - z)(1 - \bar{z})]}{z(\bar{z} - z)} + \frac{\log(z \bar{z})}{(z - 1)(z - \bar{z})}\,, \\
    \partial_{\bar{z}} \Phi &= \frac{\Phi}{z - \bar{z}} + \frac{\log[(1 - z)(1 - \bar{z})]}{\bar{z}(z - \bar{z})} + \frac{\log(z \bar{z})}{(\bar{z} - 1)(\bar{z} - z)}\,.
\end{align}

\section{R-symmetry Polynomials}
\label{sec:Rsympoly}
In this section, we use the R-symmetry Casimir in~\eqref{eq:CasRsym} to write down the combinations of $R$-symmetry monomials associated with the exchange of a representation $[q,p,q]$ in the OPE of two 20' operators. 

It follows from R-symmetry selection rules that
\begin{equation}
[0,2,0]\otimes [0,2,0]=[0,0,0]\oplus[0,2,0]\oplus[1,0,1]\oplus [2,0,2]\oplus [0,4,0]\oplus [1,2,1]\,.
\end{equation}
For each of these allowed R-symmetry exchanges, we look for solutions of the eigenvalue equation in the (12) channel
\begin{equation}
\mathcal{C}_{12} R_{12}^{\lambda_c}=\lambda_c R_{12}^{\lambda_c}\,,
\end{equation}
where each $[q,p,q]$ representation has a corresponding $\lambda_c=-p^2-2 p (q+2)-2 q (q+3)$. In our cases of interest, there is no degeneracy of the Casimir eigenvalue and there is no need to consider higher-order Casimir operators.
As discussed in the main text, for each of these equations, we propose a solution of the form 
\begin{equation}
R_{12}^{\lambda_c}=\sum_{I} a_{12,I}^{\lambda_c}\mathcal{T}^I
\end{equation}
where the sum is over the 22 independent R-symmetry structures and the $a_{12,I}^{\lambda_c}$ are arbitrary coefficients. One finds the solutions
\begin{align}
R_{12}^{-20}&=a_{12,1}^{-20}\mathbb{E}_1+a_{12,2}^{-20} \mathbb{E}_2+a_{12,3}^{-20} \mathbb{E}_3\,,\\
R_{12}^{0}&=a_{12,4}^{0} \mathbb{E}_4\\
R_{12}^{-8}&=a_{12,5}^{-8} \mathbb{E}_5+ a_{12,6}^{-8}\mathbb{E}_6+ a_{12,7}^{-8}\mathbb{E}_7\,,\\
R_{12}^{-12}&=a_{12,8}^{-12} \mathbb{E}_8+ a_{12,9}^{-12}\mathbb{E}_9+ a_{12,10}^{-12}\mathbb{E}_{10}+ a_{12,11}^{-12} \mathbb{E}_{11}+ a_{12,12}^{-12}\mathbb{E}_{12}+ a_{12,13}^{-12}\mathbb{E}_{13}\,,\\
R_{12}^{-32}&=a_{12,14}^{-32} \mathbb{E}_{14}+ a_{12,15}^{-32} \mathbb{E}_{15}+ a_{12,16}^{-32}\mathbb{E}_{16}\,,\\
R_{12}^{-24}&=a_{12,17}^{-24} \mathbb{E}_{17}+ a_{12,18}^{-24}\mathbb{E}_{18}+ a_{12,19}^{-24}\mathbb{E}_{19}+ a_{12,20}^{-24} \mathbb{E}_{20}+ a_{12,21}^{-24}\mathbb{E}_{21}+ a_{12,22}^{-24}\mathbb{E}_{22}\,,
\end{align}
with arbitrary coefficients $a_{12,i}^{\lambda_c}$ and
{\allowdisplaybreaks
\begin{align}
\mathbb{E}_{1} &=10\mathcal{T}_{(123)(45)} + 20\mathcal{T}_{(145)(23)} + 20\mathcal{T}_{(245)(13)} + 2\mathcal{T}_{(345)(12)}-5\mathcal{T}_{(12345)} \nonumber\\[0.5em]
&\quad- 5\mathcal{T}_{(12354)}- 5\mathcal{T}_{(12453)} - 5\mathcal{T}_{(12543)} - 20\mathcal{T}_{(13245)} - 20\mathcal{T}_{(13254)} \,,\\[0.5em]
\mathbb{E}_{2} &= -10\mathcal{T}_{(124)(35)} - 20\mathcal{T}_{(135)(24)} - 20\mathcal{T}_{(235)(14)} - 2\mathcal{T}_{(345)(12)}+5\mathcal{T}_{(12354)} \nonumber\\[0.5em]
&\quad+ 5\mathcal{T}_{(12435)}+ 5\mathcal{T}_{(12453)} + 5\mathcal{T}_{(12534)} + 20\mathcal{T}_{(13524)} + 20\mathcal{T}_{(14235)}\,, \\[0.5em]
\mathbb{E}_{3} &=10\mathcal{T}_{(125)(34)} + 20\mathcal{T}_{(134)(25)} + 20\mathcal{T}_{(234)(15)} + 2\mathcal{T}_{(345)(12)}-5\mathcal{T}_{(12345)}\nonumber\\[0.5em]
&\quad- 5\mathcal{T}_{(12435)}- 5\mathcal{T}_{(12534)} - 5\mathcal{T}_{(12543)} - 20\mathcal{T}_{(13425)} - 20\mathcal{T}_{(14325)}\,,\\[0.5em]
\mathbb{E}_{4} &= \mathcal{T}_{(345)(12)}\,, \\[0.5em]
\mathbb{E}_{5} &= \mathcal{T}_{(12435)} - \mathcal{T}_{(12534)} \,,\\[0.5em]
\mathbb{E}_{6} &= \mathcal{T}_{(12345)} - \mathcal{T}_{(12543)} \,,\\[0.5em]
\mathbb{E}_{7} &= \mathcal{T}_{(12354)} - \mathcal{T}_{(12453)}\,, \\[0.5em]
\mathbb{E}_{8} &= \mathcal{T}_{(125)(34)}\,, \\[0.5em]
\mathbb{E}_{9} &= \mathcal{T}_{(124)(35)}\,, \\[0.5em]
\mathbb{E}_{10} &= 3\mathcal{T}_{(12354)} + 3\mathcal{T}_{(12453)} - \mathcal{T}_{(345)(12)} \,,\\[0.5em]
\mathbb{E}_{11} &= 3\mathcal{T}_{(12345)} + 3\mathcal{T}_{(12543)} - \mathcal{T}_{(345)(12)}\,, \\[0.5em]
\mathbb{E}_{12} &= \mathcal{T}_{(123)(45)}\,, \\[0.5em]
\mathbb{E}_{13} &= 3\mathcal{T}_{(12435)} + 3\mathcal{T}_{(12534)} - \mathcal{T}_{(345)(12)}\,, \\[0.5em]
\mathbb{E}_{14} &= -4\mathcal{T}_{(12345)} - 4\mathcal{T}_{(12354)} - 4\mathcal{T}_{(12453)} - 4\mathcal{T}_{(12543)} + 20\mathcal{T}_{(13245)} + 20\mathcal{T}_{(13254)}\nonumber\\[0.5em]
&\quad- 4\mathcal{T}_{(123)(45)} + 10\mathcal{T}_{(145)(23)} + 10\mathcal{T}_{(245)(13)} + \mathcal{T}_{(345)(12)}\,,\\[0.5em]
\mathbb{E}_{15} &= -4\mathcal{T}_{(12354)} - 4\mathcal{T}_{(12435)} - 4\mathcal{T}_{(12453)} - 4\mathcal{T}_{(12534)} + 20\mathcal{T}_{(13524)} + 20\mathcal{T}_{(14235)} \nonumber\\[0.5em]
&\quad- 4\mathcal{T}_{(124)(35)} + 10\mathcal{T}_{(135)(24)} + 10\mathcal{T}_{(235)(14)} + \mathcal{T}_{(345)(12)}\,,\\[0.5em]
\mathbb{E}_{16} &= -4\mathcal{T}_{(12345)} - 4\mathcal{T}_{(12435)} - 4\mathcal{T}_{(12534)} - 4\mathcal{T}_{(12543)} + 20\mathcal{T}_{(13425)} + 20\mathcal{T}_{(14325)} \nonumber\\[0.5em]
&\quad- 4\mathcal{T}_{(125)(34)} + 10\mathcal{T}_{(134)(25)} + 10\mathcal{T}_{(234)(15)} + \mathcal{T}_{(345)(12)}\,,\\[0.5em]
\mathbb{E}_{17} &= -\mathcal{T}_{(12345)} + \mathcal{T}_{(12354)} - \mathcal{T}_{(12453)} + \mathcal{T}_{(12543)} + 8\mathcal{T}_{(13245)} - 8\mathcal{T}_{(13254)}\,,\\[0.5em]
\mathbb{E}_{18} &= -\mathcal{T}_{(12345)} - \mathcal{T}_{(12354)} + \mathcal{T}_{(12453)} + \mathcal{T}_{(12543)} + 4\mathcal{T}_{(145)(23)} - 4\mathcal{T}_{(245)(13)}\,,\\[0.5em]
\mathbb{E}_{19} &= -\mathcal{T}_{(12354)} + \mathcal{T}_{(12435)} + \mathcal{T}_{(12453)} - \mathcal{T}_{(12534)} - 4\mathcal{T}_{(135)(24)} + 4\mathcal{T}_{(235)(14)}\,, \\[0.5em]
\mathbb{E}_{20} &= -\mathcal{T}_{(12354)} - \mathcal{T}_{(12435)} + \mathcal{T}_{(12453)} + \mathcal{T}_{(12534)} - 8\mathcal{T}_{(13524)} + 8\mathcal{T}_{(14235)} \,,\\[0.5em]
\mathbb{E}_{21} &= -\mathcal{T}_{(12345)} + \mathcal{T}_{(12435)} - \mathcal{T}_{(12534)} + \mathcal{T}_{(12543)} - 8\mathcal{T}_{(13425)} + 8\mathcal{T}_{(14325)}\,,\\[0.5em]
\mathbb{E}_{22} &= -\mathcal{T}_{(12345)} - \mathcal{T}_{(12435)} + \mathcal{T}_{(12534)} + \mathcal{T}_{(12543)} - 4\mathcal{T}_{(134)(25)} + 4\mathcal{T}_{(234)(15)}\,,
\end{align}}
\!\!where $\mathbb{E}_{1},\mathbb{E}_{2}$ and $\mathbb{E}_{3}$ match the structures for an exchange of a $[2,0,2]$ R-symmetry representation presented in equation (7.13) of~\cite{Bissi:2021hjk}.

At five points, we may also be interested in selecting specific combinations of R-symmetry structures corresponding to the exchange of a pair of R-symmetry representations in two different channels, say (12) and (45). The generalization of what we just did  simply amounts to looking for solutions of two Casimir equations
\begin{align}
\mathcal{C}_{12} R_{12,45}^{\lambda_c^1,\lambda_c^2}=\lambda_c^1 R_{12,45}^{\lambda_c^1,\lambda_c^2}\,,\\
\mathcal{C}_{45} R_{12,45}^{\lambda_c^1,\lambda_c^2}=\lambda_c^2 R_{12,45}^{\lambda_c^1,\lambda_c^2}\,, 
\end{align}
where $\lambda_c^1$ and $\lambda_c^2$ are the eigenvalues of the desired exchanged R-symmetry representations in channels (12) and (45), respectively. Writing $R_{12,45}^{\lambda_c^1,\lambda_c^2}$ as a linear combination of the 22 independent $\mathcal{T}_I$ with arbitrary coefficients, we find that, except for the case of two $[1,2,1]$ exchanges, there is a unique monomial combination corresponding to each pairing. We find
{\allowdisplaybreaks
\begin{align}
R_{12,45}^{0,-12} &= \mathcal{T}_{(345)(12)}\,, \\[0.5em]
R_{12,45}^{-8,-8} &= \mathcal{T}_{(12345)} - \mathcal{T}_{(12354)} + \mathcal{T}_{(12453)} - \mathcal{T}_{(12543)} \,,\\[0.5em]
R_{12,45}^{-8,-12} &= \mathcal{T}_{(12345)} + \mathcal{T}_{(12354)} - \mathcal{T}_{(12453)} - \mathcal{T}_{(12543)}\,, \\[0.5em]
R_{12,45}^{-8,-24} &= -\mathcal{T}_{(12345)} + \mathcal{T}_{(12354)} + 8\mathcal{T}_{(12435)} - \mathcal{T}_{(12453)} - 8\mathcal{T}_{(12534)} + \mathcal{T}_{(12543)}\,, \\[0.5em]
R_{12,45}^{-12,0} &= \mathcal{T}_{(123)(45)}\,,\\[0.5em]
R_{12,45}^{-12,-8} &= \mathcal{T}_{(12345)} - \mathcal{T}_{(12354)} - \mathcal{T}_{(12453)} + \mathcal{T}_{(12543)}\,, \\[0.5em]
R_{12,45}^{-12,-12} &= 3\mathcal{T}_{(12345)} + 3\mathcal{T}_{(12354)} + 3\mathcal{T}_{(12453)} + 3\mathcal{T}_{(12543)} - 2\mathcal{T}_{(123)(45)} - 2\mathcal{T}_{(345)(12)}\,, \\[0.5em]\
R_{12,45}^{-12,-20} &= -5\mathcal{T}_{(12345)} - 5\mathcal{T}_{(12354)} - 20\mathcal{T}_{(12435)} - 5\mathcal{T}_{(12453)} - 20\mathcal{T}_{(12534)} - 5\mathcal{T}_{(12543)} \nonumber\\[0.5em]
&\quad+ 2\mathcal{T}_{(123)(45)} + 20\mathcal{T}_{(124)(35)} + 20\mathcal{T}_{(125)(34)} + 10\mathcal{T}_{(345)(12)}\,, \\[0.5em]
R_{12,45}^{-12,-32} &= -4\mathcal{T}_{(12345)} - 4\mathcal{T}_{(12354)} + 20\mathcal{T}_{(12435)} - 4\mathcal{T}_{(12453)} + 20\mathcal{T}_{(12534)} - 4\mathcal{T}_{(12543)}\nonumber\\[0.5em]
&\quad + \mathcal{T}_{(123)(45)} + 10\mathcal{T}_{(124)(35)} + 10\mathcal{T}_{(125)(34)} - 4\mathcal{T}_{(345)(12)}\,, \\[0.5em]
R_{12,45}^{-12,-24} &= -\mathcal{T}_{(12345)} + \mathcal{T}_{(12354)} + \mathcal{T}_{(12453)} - \mathcal{T}_{(12543)} - 4\mathcal{T}_{(124)(35)} + 4\mathcal{T}_{(125)(34)}\,, \\[0.5em]
R_{12,45}^{-20,-12} &= -5\mathcal{T}_{(12345)} - 5\mathcal{T}_{(12354)} - 5\mathcal{T}_{(12453)} - 5\mathcal{T}_{(12543)} - 20\mathcal{T}_{(13245)} - 20\mathcal{T}_{(13254)} \nonumber\\[0.5em]
&\quad+ 10\mathcal{T}_{(123)(45)} + 20\mathcal{T}_{(145)(23)} + 20\mathcal{T}_{(245)(13)} + 2\mathcal{T}_{(345)(12)}\,,\\[0.5em]
R_{12,45}^{-20,-20} &= -3\mathcal{T}_{(12345)} - 3\mathcal{T}_{(12354)} - 4\mathcal{T}_{(12435)} - 3\mathcal{T}_{(12453)} - 4\mathcal{T}_{(12534)} - 3\mathcal{T}_{(12543)} \nonumber\\[0.5em]
&\quad - 4\mathcal{T}_{(13245)} - 4\mathcal{T}_{(13254)} - 8\mathcal{T}_{(13425)} - 8\mathcal{T}_{(13524)} - 8\mathcal{T}_{(14235)} - 8\mathcal{T}_{(14325)} \nonumber\\[0.5em]
&\quad  + 4\mathcal{T}_{(124)(35)} + 4\mathcal{T}_{(125)(34)} + 8\mathcal{T}_{(134)(25)} + 8\mathcal{T}_{(135)(24)} + 4\mathcal{T}_{(145)(23)} + 8\mathcal{T}_{(234)(15)}\nonumber\\[0.5em]
&\quad +8\mathcal{T}_{(235)(14)}+ 4\mathcal{T}_{(245)(13)} + 2\mathcal{T}_{(345)(12)} + 2\mathcal{T}_{(123)(45)}\\[0.5em]
R_{12,45}^{-20,-24} &= -\mathcal{T}_{(12345)} + \mathcal{T}_{(12354)} + \mathcal{T}_{(12453)} - \mathcal{T}_{(12543)} - 4\mathcal{T}_{(13425)} + 4\mathcal{T}_{(13524)}\nonumber\\[0.5em]
&\quad + 4\mathcal{T}_{(14235)} - 4\mathcal{T}_{(14325)} - 2\mathcal{T}_{(124)(35)} + 2\mathcal{T}_{(125)(34)} + 4\mathcal{T}_{(134)(25)} - 4\mathcal{T}_{(135)(24)}\nonumber\\[0.5em]
&\quad + 4\mathcal{T}_{(234)(15)} - 4\mathcal{T}_{(235)(14)}\,,\\[0.5em]
R_{12,45}^{-32,-12} &= -4\mathcal{T}_{(12345)} - 4\mathcal{T}_{(12354)} - 4\mathcal{T}_{(12453)} - 4\mathcal{T}_{(12543)} + 20\mathcal{T}_{(13245)} + 20\mathcal{T}_{(13254)}\nonumber\\[0.5em]
&\quad - 4\mathcal{T}_{(123)(45)} + 10\mathcal{T}_{(145)(23)} + 10\mathcal{T}_{(245)(13)} + \mathcal{T}_{(345)(12)}\,,\\[0.5em]
R_{12,45}^{-32,-32} &= -6\mathcal{T}_{(12345)} - 6\mathcal{T}_{(12354)} - 20\mathcal{T}_{(12435)} - 6\mathcal{T}_{(12453)} - 20\mathcal{T}_{(12534)} - 6\mathcal{T}_{(12543)}\nonumber\\[0.5em]
&\quad  - 20\mathcal{T}_{(13245)} - 20\mathcal{T}_{(13254)} + 50\mathcal{T}_{(13425)} + 50\mathcal{T}_{(13524)} + 50\mathcal{T}_{(14235)} + 50\mathcal{T}_{(14325)} \nonumber\\[0.5em]
&\quad + 4\mathcal{T}_{(123)(45)} - 10\mathcal{T}_{(124)(35)} - 10\mathcal{T}_{(125)(34)} + 25\mathcal{T}_{(134)(25)} + 25\mathcal{T}_{(135)(24)}\nonumber\\[0.5em]
&\quad - 10\mathcal{T}_{(145)(23)} + 25\mathcal{T}_{(234)(15)} + 25\mathcal{T}_{(235)(14)} - 10\mathcal{T}_{(245)(13)} + 4\mathcal{T}_{(345)(12)}\,,\\[0.5em]
R_{12,45}^{-32,-24} &= -2\mathcal{T}_{(12345)} + 2\mathcal{T}_{(12354)} + 2\mathcal{T}_{(12453)} - 2\mathcal{T}_{(12543)} + 10\mathcal{T}_{(13425)} - 10\mathcal{T}_{(13524)}\nonumber\\[0.5em]
&\quad - 10\mathcal{T}_{(14235)} + 10\mathcal{T}_{(14325)} + 2\mathcal{T}_{(124)(35)} - 2\mathcal{T}_{(125)(34)} + 5\mathcal{T}_{(134)(25)} - 5\mathcal{T}_{(135)(24)}\nonumber\\[0.5em]
&\quad + 5\mathcal{T}_{(234)(15)} - 5\mathcal{T}_{(235)(14)}\,, \\[0.5em]
R_{12,45}^{-24,-8} &= -\mathcal{T}_{(12345)} + \mathcal{T}_{(12354)} - \mathcal{T}_{(12453)} + \mathcal{T}_{(12543)} + 8\mathcal{T}_{(13245)} - 8\mathcal{T}_{(13254)}\,, \\[0.5em]
R_{12,45}^{-24,-12} &= -\mathcal{T}_{(12345)} - \mathcal{T}_{(12354)} + \mathcal{T}_{(12453)} + \mathcal{T}_{(12543)} + 4\mathcal{T}_{(145)(23)} - 4\mathcal{T}_{(245)(13)} \,,\\[0.5em]
R_{12,45}^{-24,-20} &= -\mathcal{T}_{(12345)} - \mathcal{T}_{(12354)} + \mathcal{T}_{(12453)} + \mathcal{T}_{(12543)} + 4\mathcal{T}_{(13425)} + 4\mathcal{T}_{(13524)} \nonumber\\[0.5em]
&\quad - 4\mathcal{T}_{(14235)} - 4\mathcal{T}_{(14325)} - 4\mathcal{T}_{(134)(25)} - 4\mathcal{T}_{(135)(24)} + 2\mathcal{T}_{(145)(23)} + 4\mathcal{T}_{(234)(15)}\nonumber\\[0.5em]
 &\quad + 4\mathcal{T}_{(235)(14)} - 2\mathcal{T}_{(245)(13)}\,, \\[0.5em]
R_{12,45}^{-24,-32} &= -2\mathcal{T}_{(12345)} - 2\mathcal{T}_{(12354)} + 2\mathcal{T}_{(12453)} + 2\mathcal{T}_{(12543)} - 10\mathcal{T}_{(13425)} - 10\mathcal{T}_{(13524)}\nonumber\\[0.5em]
&\quad + 10\mathcal{T}_{(14235)} + 10\mathcal{T}_{(14325)} - 5\mathcal{T}_{(134)(25)} - 5\mathcal{T}_{(135)(24)} - 2\mathcal{T}_{(145)(23)} + 5\mathcal{T}_{(234)(15)}\nonumber\\[0.5em]
&\quad + 5\mathcal{T}_{(235)(14)} + 2\mathcal{T}_{(245)(13)}\,,\\[0.5em]
R_{12,45}^{-24,-24,1} &= -3\mathcal{T}_{(12345)} + 3\mathcal{T}_{(12354)} - 8\mathcal{T}_{(12435)} - 3\mathcal{T}_{(12453)} + 8\mathcal{T}_{(12534)} + 3\mathcal{T}_{(12543)} - 8\mathcal{T}_{(13245)}\nonumber\\[0.5em]
&\quad + 8\mathcal{T}_{(13254)} - 16\mathcal{T}_{(134)(25)} + 16\mathcal{T}_{(135)(24)} + 16\mathcal{T}_{(234)(15)} - 16\mathcal{T}_{(235)(14)}\,, \\[0.5em]
R_{12,45}^{-24,-24,2} &= -5\mathcal{T}_{(12345)} + 5\mathcal{T}_{(12354)} + 8\mathcal{T}_{(12435)} - 5\mathcal{T}_{(12453)} - 8\mathcal{T}_{(12534)} + 5\mathcal{T}_{(12543)}+ 8\mathcal{T}_{(13245)}\nonumber\\[0.5em]
&\quad  - 8\mathcal{T}_{(13254)} - 32\mathcal{T}_{(13425)} + 32\mathcal{T}_{(13524)} - 32\mathcal{T}_{(14235)} + 32\mathcal{T}_{(14325)}\,.
\end{align}}


\bibliographystyle{JHEP}
\bibliography{biblio.bib}

@misc{Mafrapage,
  author = {Carlos Mafra},
  title = {\url{https://www.southampton.ac.uk/~crm1n16/pss.html}},
  year = {Accessed: July 2025}
}

@article{Goncalves:2019znr,
	archiveprefix = {arXiv},
	author = {Gon\c{c}alves, Vasco and Pereira, Raul and Zhou, Xinan},
	doi = {10.1007/JHEP10(2019)247},
	eprint = {1906.05305},
	journal = {JHEP},
	pages = {247},
	primaryclass = {hep-th},
	reportnumber = {PUPT-2588},
	title = {{$20'$ Five-Point Function from $AdS_5\times S^5$ Supergravity}},
	volume = {10},
	year = {2019}}

@article{Goncalves:2023oyx,
	archiveprefix = {arXiv},
	author = {Gon\c{c}alves, Vasco and Meneghelli, Carlo and Pereira, Raul and Vilas Boas, Joao and Zhou, Xinan},
	doi = {10.1007/JHEP08(2023)067},
	eprint = {2302.01896},
	journal = {JHEP},
	pages = {067},
	primaryclass = {hep-th},
	title = {{Kaluza-Klein five-point functions from AdS$_{5}$\texttimes{}S$^{5}$ supergravity}},
	volume = {08},
	year = {2023}}

@article{Beem:2013sza,
	archiveprefix = {arXiv},
	author = {Beem, Christopher and Lemos, Madalena and Liendo, Pedro and Peelaers, Wolfger and Rastelli, Leonardo and van Rees, Balt C.},
	doi = {10.1007/s00220-014-2272-x},
	eprint = {1312.5344},
	journal = {Commun. Math. Phys.},
	number = {3},
	pages = {1359--1433},
	primaryclass = {hep-th},
	reportnumber = {YITP-SB-13-45, CERN-PH-TH-2013-311, HU-EP-13-78},
	title = {{Infinite Chiral Symmetry in Four Dimensions}},
	volume = {336},
	year = {2015}}

@article{Drukker:2009sf,
	archiveprefix = {arXiv},
	author = {Drukker, Nadav and Plefka, Jan},
	doi = {10.1088/1126-6708/2009/04/052},
	eprint = {0901.3653},
	journal = {JHEP},
	pages = {052},
	primaryclass = {hep-th},
	reportnumber = {HU-EP-09-01},
	title = {{Superprotected n-point correlation functions of local operators in N=4 super Yang-Mills}},
	volume = {04},
	year = {2009}}

@article{Mack:2009mi,
	archiveprefix = {arXiv},
	author = {Mack, Gerhard},
	eprint = {0907.2407},
	month = {7},
	primaryclass = {hep-th},
	title = {{D-independent representation of Conformal Field Theories in D dimensions via transformation to auxiliary Dual Resonance Models. Scalar amplitudes}},
	year = {2009}}

@article{Penedones:2010ue,
	archiveprefix = {arXiv},
	author = {Penedones, Joao},
	doi = {10.1007/JHEP03(2011)025},
	eprint = {1011.1485},
	journal = {JHEP},
	pages = {025},
	primaryclass = {hep-th},
	title = {{Writing CFT correlation functions as AdS scattering amplitudes}},
	volume = {03},
	year = {2011}}

@article{Fitzpatrick:2011hu,
	archiveprefix = {arXiv},
	author = {Fitzpatrick, A. Liam and Kaplan, Jared},
	doi = {10.1007/JHEP10(2012)127},
	eprint = {1111.6972},
	journal = {JHEP},
	pages = {127},
	primaryclass = {hep-th},
	reportnumber = {SLAC-PUB-14841},
	title = {{Analyticity and the Holographic S-Matrix}},
	volume = {10},
	year = {2012}}

@article{Goncalves:2014rfa,
	archiveprefix = {arXiv},
	author = {Gon\c{c}alves, Vasco and Penedones, Jo\~ao and Trevisani, Emilio},
	doi = {10.1007/JHEP10(2015)040},
	eprint = {1410.4185},
	journal = {JHEP},
	pages = {040},
	primaryclass = {hep-th},
	title = {{Factorization of Mellin amplitudes}},
	volume = {10},
	year = {2015}}

@article{Costa:2011mg,
	archiveprefix = {arXiv},
	author = {Costa, Miguel S. and Penedones, Joao and Poland, David and Rychkov, Slava},
	doi = {10.1007/JHEP11(2011)071},
	eprint = {1107.3554},
	journal = {JHEP},
	pages = {071},
	primaryclass = {hep-th},
	reportnumber = {LPTENS-11-22, NSF-KITP-11-128},
	title = {{Spinning Conformal Correlators}},
	volume = {11},
	year = {2011}}

@article{Rastelli:2017udc,
	archiveprefix = {arXiv},
	author = {Rastelli, Leonardo and Zhou, Xinan},
	doi = {10.1007/JHEP04(2018)014},
	eprint = {1710.05923},
	journal = {JHEP},
	pages = {014},
	primaryclass = {hep-th},
	reportnumber = {YITP-SB-2017-44},
	title = {{How to Succeed at Holographic Correlators Without Really Trying}},
	volume = {04},
	year = {2018}}

@article{Aprile:2018efk,
	archiveprefix = {arXiv},
	author = {Aprile, Francesco and Drummond, James and Heslop, Paul and Paul, Hynek},
	doi = {10.1103/PhysRevD.98.126008},
	eprint = {1802.06889},
	journal = {Phys. Rev. D},
	number = {12},
	pages = {126008},
	primaryclass = {hep-th},
	title = {{Double-trace spectrum of $N=4$ supersymmetric Yang-Mills theory at strong coupling}},
	volume = {98},
	year = {2018}}

@article{Alday:2022uxp,
	archiveprefix = {arXiv},
	author = {Alday, Luis F. and Hansen, Tobias and Silva, Joao A.},
	doi = {10.1007/JHEP10(2022)036},
	eprint = {2204.07542},
	journal = {JHEP},
	pages = {036},
	primaryclass = {hep-th},
	title = {{AdS Virasoro-Shapiro from dispersive sum rules}},
	volume = {10},
	year = {2022}}

@article{Alday:2022xwz,
	archiveprefix = {arXiv},
	author = {Alday, Luis F. and Hansen, Tobias and Silva, Joao A.},
	doi = {10.1007/JHEP12(2022)010},
	eprint = {2209.06223},
	journal = {JHEP},
	pages = {010},
	primaryclass = {hep-th},
	title = {{AdS Virasoro-Shapiro from single-valued periods}},
	volume = {12},
	year = {2022}}

@article{Maldacena:2015waa,
	archiveprefix = {arXiv},
	author = {Maldacena, Juan and Shenker, Stephen H. and Stanford, Douglas},
	doi = {10.1007/JHEP08(2016)106},
	eprint = {1503.01409},
	journal = {JHEP},
	pages = {106},
	primaryclass = {hep-th},
	title = {{A bound on chaos}},
	volume = {08},
	year = {2016}}

@article{Binder:2019jwn,
	archiveprefix = {arXiv},
	author = {Binder, Damon J. and Chester, Shai M. and Pufu, Silviu S. and Wang, Yifan},
	doi = {10.1007/JHEP12(2019)119},
	eprint = {1902.06263},
	journal = {JHEP},
	pages = {119},
	primaryclass = {hep-th},
	reportnumber = {PUPT-2582},
	title = {{$ \mathcal{N} $ = 4 Super-Yang-Mills correlators at strong coupling from string theory and localization}},
	volume = {12},
	year = {2019}}

@article{Chester:2020dja,
	archiveprefix = {arXiv},
	author = {Chester, Shai M. and Pufu, Silviu S.},
	doi = {10.1007/JHEP01(2021)103},
	eprint = {2003.08412},
	journal = {JHEP},
	pages = {103},
	primaryclass = {hep-th},
	reportnumber = {PUPT-2616},
	title = {{Far beyond the planar limit in strongly-coupled $ \mathcal{N} $ = 4 SYM}},
	volume = {01},
	year = {2021}}

@article{Eden:2000bk,
	archiveprefix = {arXiv},
	author = {Eden, Burkhard and Petkou, Anastasios C. and Schubert, Christian and Sokatchev, Emery},
	doi = {10.1016/S0550-3213(01)00151-1},
	eprint = {hep-th/0009106},
	journal = {Nucl. Phys. B},
	pages = {191--212},
	reportnumber = {LAPTH-811-2000, KL-TH-00-06},
	title = {{Partial nonrenormalization of the stress tensor four point function in N=4 SYM and AdS / CFT}},
	volume = {607},
	year = {2001}}

@article{Nirschl:2004pa,
	archiveprefix = {arXiv},
	author = {Nirschl, M. and Osborn, H.},
	doi = {10.1016/j.nuclphysb.2005.01.013},
	eprint = {hep-th/0407060},
	journal = {Nucl. Phys. B},
	pages = {409--479},
	reportnumber = {DAMTP-04-51},
	title = {{Superconformal Ward identities and their solution}},
	volume = {711},
	year = {2005}}

@article{Dolan:2001tt,
	archiveprefix = {arXiv},
	author = {Dolan, F. A. and Osborn, H.},
	doi = {10.1016/S0550-3213(02)00096-2},
	eprint = {hep-th/0112251},
	journal = {Nucl. Phys. B},
	pages = {3--73},
	reportnumber = {DAMTP-01-82},
	title = {{Superconformal symmetry, correlation functions and the operator product expansion}},
	volume = {629},
	year = {2002}}

@article{Rastelli:2016nze,
	archiveprefix = {arXiv},
	author = {Rastelli, Leonardo and Zhou, Xinan},
	doi = {10.1103/PhysRevLett.118.091602},
	eprint = {1608.06624},
	journal = {Phys. Rev. Lett.},
	number = {9},
	pages = {091602},
	primaryclass = {hep-th},
	title = {{Mellin amplitudes for $AdS_5\times S^5$}},
	volume = {118},
	year = {2017}}

@article{Goncalves:2014ffa,
	archiveprefix = {arXiv},
	author = {Gon\c{c}alves, Vasco},
	doi = {10.1007/JHEP04(2015)150},
	eprint = {1411.1675},
	journal = {JHEP},
	pages = {150},
	primaryclass = {hep-th},
	title = {{Four point function of $\mathcal{N}=4$ stress-tensor multiplet at strong coupling}},
	volume = {04},
	year = {2015}}

@article{Alday:2023mvu,
	archiveprefix = {arXiv},
	author = {Alday, Luis F. and Hansen, Tobias},
	doi = {10.1007/JHEP10(2023)023},
	eprint = {2306.12786},
	journal = {JHEP},
	pages = {023},
	primaryclass = {hep-th},
	title = {{The AdS Virasoro-Shapiro amplitude}},
	volume = {10},
	year = {2023}}

@article{Alday:2023jdk,
	archiveprefix = {arXiv},
	author = {Alday, Luis F. and Hansen, Tobias and Silva, Joao A.},
	doi = {10.1103/PhysRevLett.131.161603},
	eprint = {2305.03593},
	journal = {Phys. Rev. Lett.},
	number = {16},
	pages = {161603},
	primaryclass = {hep-th},
	title = {{Emergent Worldsheet for the AdS Virasoro-Shapiro Amplitude}},
	volume = {131},
	year = {2023}}

@article{Mafra:2011nw,
	archiveprefix = {arXiv},
	author = {Mafra, Carlos R. and Schlotterer, Oliver and Stieberger, Stephan},
	doi = {10.1016/j.nuclphysb.2013.04.022},
	eprint = {1106.2646},
	journal = {Nucl. Phys. B},
	pages = {461--513},
	primaryclass = {hep-th},
	reportnumber = {AEI-2011-35, MPP-2011-65},
	title = {{Complete N-Point Superstring Disk Amplitude II. Amplitude and Hypergeometric Function Structure}},
	volume = {873},
	year = {2013}}

@article{Belitsky:2014zha,
	archiveprefix = {arXiv},
	author = {Belitsky, A. V. and Hohenegger, S. and Korchemsky, G. P. and Sokatchev, E.},
	doi = {10.1016/j.nuclphysb.2016.01.008},
	eprint = {1409.2502},
	journal = {Nucl. Phys. B},
	pages = {176--215},
	primaryclass = {hep-th},
	reportnumber = {CERN-PH-TH-2014-174, IPHT-T14-122, LAPTH-109-14},
	title = {{N=4 superconformal Ward identities for correlation functions}},
	volume = {904},
	year = {2016}}

@article{Goncalves:2025jcg,
	archiveprefix = {arXiv},
	author = {Goncalves, Vasco and Nocchi, Maria and Zhou, Xinan},
	eprint = {2502.10269},
	month = {2},
	primaryclass = {hep-th},
	title = {{Dissecting supergraviton six-point function with lightcone limits and chiral algebra}},
	year = {2025}}

@article{Bercini:2020msp,
	archiveprefix = {arXiv},
	author = {Bercini, Carlos and Gon\c{c}alves, Vasco and Vieira, Pedro},
	doi = {10.1103/PhysRevLett.126.121603},
	eprint = {2008.10407},
	journal = {Phys. Rev. Lett.},
	number = {12},
	pages = {121603},
	primaryclass = {hep-th},
	title = {{Light-Cone Bootstrap of Higher Point Functions and Wilson Loop Duality}},
	volume = {126},
	year = {2021}}

@article{Antunes:2021kmm,
	archiveprefix = {arXiv},
	author = {Antunes, Ant\'onio and Costa, Miguel S. and Goncalves, Vasco and Boas, Joao Vilas},
	doi = {10.1007/JHEP03(2022)139},
	eprint = {2111.05453},
	journal = {JHEP},
	pages = {139},
	primaryclass = {hep-th},
	title = {{Lightcone bootstrap at higher points}},
	volume = {03},
	year = {2022}}

@article{Bern:1992em,
	archiveprefix = {arXiv},
	author = {Bern, Zvi and Dixon, Lance J. and Kosower, David A.},
	doi = {10.1016/0370-2693(93)90400-C},
	eprint = {hep-ph/9212308},
	journal = {Phys. Lett. B},
	note = {[Erratum: Phys.Lett.B 318, 649 (1993)]},
	pages = {299--308},
	reportnumber = {SLAC-PUB-6001, CERN-TH-6756-92, UCLA-92-42},
	title = {{Dimensionally regulated one loop integrals}},
	volume = {302},
	year = {1993}}

@article{Bern:1993kr,
	archiveprefix = {arXiv},
	author = {Bern, Zvi and Dixon, Lance J. and Kosower, David A.},
	doi = {10.1016/0550-3213(94)90398-0},
	eprint = {hep-ph/9306240},
	journal = {Nucl. Phys. B},
	pages = {751--816},
	reportnumber = {SLAC-PUB-5947, SACLAY-SPH-T-92-048, UCLA-92-043},
	title = {{Dimensionally regulated pentagon integrals}},
	volume = {412},
	year = {1994}}

@article{Dolan:2003hv,
	archiveprefix = {arXiv},
	author = {Dolan, F. A. and Osborn, H.},
	doi = {10.1016/j.nuclphysb.2003.11.016},
	eprint = {hep-th/0309180},
	journal = {Nucl. Phys. B},
	pages = {491--507},
	reportnumber = {DAMTP-03-91},
	title = {{Conformal partial waves and the operator product expansion}},
	volume = {678},
	year = {2004}}

@article{Alday:2021odx,
	archiveprefix = {arXiv},
	author = {Alday, Luis F. and Behan, Connor and Ferrero, Pietro and Zhou, Xinan},
	doi = {10.1007/JHEP06(2021)020},
	eprint = {2103.15830},
	journal = {JHEP},
	pages = {020},
	primaryclass = {hep-th},
	title = {{Gluon Scattering in AdS from CFT}},
	volume = {06},
	year = {2021}}

@article{Chester:2018aca,
	archiveprefix = {arXiv},
	author = {Chester, Shai M. and Pufu, Silviu S. and Yin, Xi},
	doi = {10.1007/JHEP08(2018)115},
	eprint = {1804.00949},
	journal = {JHEP},
	pages = {115},
	primaryclass = {hep-th},
	reportnumber = {PUPT-2556},
	title = {{The M-Theory S-Matrix From ABJM: Beyond 11D Supergravity}},
	volume = {08},
	year = {2018}}

@article{Kawai:1985xq,
	author = {Kawai, H. and Lewellen, D. C. and Tye, S. H. H.},
	doi = {10.1016/0550-3213(86)90362-7},
	journal = {Nucl. Phys. B},
	pages = {1--23},
	reportnumber = {CLNS-85/667},
	title = {{A Relation Between Tree Amplitudes of Closed and Open Strings}},
	volume = {269},
	year = {1986}}

@article{Schlotterer:2012ny,
	archiveprefix = {arXiv},
	author = {Schlotterer, O. and Stieberger, S.},
	doi = {10.1088/1751-8113/46/47/475401},
	eprint = {1205.1516},
	journal = {J. Phys. A},
	pages = {475401},
	primaryclass = {hep-th},
	reportnumber = {AEI-2012-039, MPP-2012-859},
	title = {{Motivic Multiple Zeta Values and Superstring Amplitudes}},
	volume = {46},
	year = {2013}}

@article{Gomez:2015uha,
	archiveprefix = {arXiv},
	author = {Gomez, Humberto and Mafra, Carlos R. and Schlotterer, Oliver},
	doi = {10.1103/PhysRevD.93.045030},
	eprint = {1504.02759},
	journal = {Phys. Rev. D},
	number = {4},
	pages = {045030},
	primaryclass = {hep-th},
	reportnumber = {DAMTP-2015-20},
	title = {{Two-loop superstring five-point amplitude and $S$-duality}},
	volume = {93},
	year = {2016}}

@article{Alday:2020lbp,
    author = "Alday, Luis F. and Zhou, Xinan",
    title = "{All Tree-Level Correlators for M-theory on $AdS_7 \times S^4$}",
    eprint = "2006.06653",
    archivePrefix = "arXiv",
    primaryClass = "hep-th",
    doi = "10.1103/PhysRevLett.125.131604",
    journal = "Phys. Rev. Lett.",
    volume = "125",
    number = "13",
    pages = "131604",
    year = "2020"
}

@article{Alday:2020dtb,
    author = "Alday, Luis F. and Zhou, Xinan",
    title = "{All Holographic Four-Point Functions in All Maximally Supersymmetric CFTs}",
    eprint = "2006.12505",
    archivePrefix = "arXiv",
    primaryClass = "hep-th",
    doi = "10.1103/PhysRevX.11.011056",
    journal = "Phys. Rev. X",
    volume = "11",
    number = "1",
    pages = "011056",
    year = "2021"
}

@article{Rastelli:2019gtj,
    author = "Rastelli, Leonardo and Roumpedakis, Konstantinos and Zhou, Xinan",
    title = "{$\mathbf{AdS_3\times S^3}$ Tree-Level Correlators: Hidden Six-Dimensional Conformal Symmetry}",
    eprint = "1905.11983",
    archivePrefix = "arXiv",
    primaryClass = "hep-th",
    doi = "10.1007/JHEP10(2019)140",
    journal = "JHEP",
    volume = "10",
    pages = "140",
    year = "2019"
}

@article{Giusto:2020neo,
    author = "Giusto, Stefano and Russo, Rodolfo and Tyukov, Alexander and Wen, Congkao",
    title = "{The CFT$_6$ origin of all tree-level 4-point correlators in AdS$_3 \times S^3$}",
    eprint = "2005.08560",
    archivePrefix = "arXiv",
    primaryClass = "hep-th",
    reportNumber = "QMUL-PH-20-10",
    doi = "10.1140/epjc/s10052-020-8300-4",
    journal = "Eur. Phys. J. C",
    volume = "80",
    number = "8",
    pages = "736",
    year = "2020"
}

@article{Alday:2024ksp,
    author = "Alday, Luis F. and Hansen, Tobias",
    title = "{Single-valuedness of the AdS Veneziano amplitude}",
    eprint = "2404.16084",
    archivePrefix = "arXiv",
    primaryClass = "hep-th",
    doi = "10.1007/JHEP08(2024)108",
    journal = "JHEP",
    volume = "08",
    pages = "108",
    year = "2024"
}

@article{Alday:2024yax,
    author = "Alday, Luis F. and Chester, Shai M. and Hansen, Tobias and Zhong, De-liang",
    title = "{The AdS Veneziano amplitude at small curvature}",
    eprint = "2403.13877",
    archivePrefix = "arXiv",
    primaryClass = "hep-th",
    doi = "10.1007/JHEP05(2024)322",
    journal = "JHEP",
    volume = "05",
    pages = "322",
    year = "2024"
}

@article{Alday:2024rjs,
    author = "Alday, Luis F. and Giribet, Gaston and Hansen, Tobias",
    title = "{On the AdS$_{3}$ Virasoro-Shapiro amplitude}",
    eprint = "2412.05246",
    archivePrefix = "arXiv",
    primaryClass = "hep-th",
    doi = "10.1007/JHEP03(2025)002",
    journal = "JHEP",
    volume = "03",
    pages = "002",
    year = "2025"
}

@article{Fardelli:2023fyq,
    author = "Fardelli, Giulia and Hansen, Tobias and Silva, Joao A.",
    title = "{AdS Virasoro-Shapiro amplitude with KK modes}",
    eprint = "2308.03683",
    archivePrefix = "arXiv",
    primaryClass = "hep-th",
    doi = "10.1007/JHEP11(2023)064",
    journal = "JHEP",
    volume = "11",
    pages = "064",
    year = "2023"
}

@article{Alday:2017xua,
    author = "Alday, Luis F. and Bissi, Agnese",
    title = "{Loop Corrections to Supergravity on $AdS_5 \times S^5$}",
    eprint = "1706.02388",
    archivePrefix = "arXiv",
    primaryClass = "hep-th",
    doi = "10.1103/PhysRevLett.119.171601",
    journal = "Phys. Rev. Lett.",
    volume = "119",
    number = "17",
    pages = "171601",
    year = "2017"
}

@article{Aharony:2016dwx,
    author = "Aharony, Ofer and Alday, Luis F. and Bissi, Agnese and Perlmutter, Eric",
    title = "{Loops in AdS from Conformal Field Theory}",
    eprint = "1612.03891",
    archivePrefix = "arXiv",
    primaryClass = "hep-th",
    doi = "10.1007/JHEP07(2017)036",
    journal = "JHEP",
    volume = "07",
    pages = "036",
    year = "2017"
}

@article{Aprile:2017bgs,
    author = "Aprile, F. and Drummond, J. M. and Heslop, P. and Paul, H.",
    title = "{Quantum Gravity from Conformal Field Theory}",
    eprint = "1706.02822",
    archivePrefix = "arXiv",
    primaryClass = "hep-th",
    doi = "10.1007/JHEP01(2018)035",
    journal = "JHEP",
    volume = "01",
    pages = "035",
    year = "2018"
}

@article{Drummond:2019hel,
    author = "Drummond, J. M. and Paul, H.",
    title = "{One-loop string corrections to AdS amplitudes from CFT}",
    eprint = "1912.07632",
    archivePrefix = "arXiv",
    primaryClass = "hep-th",
    doi = "10.1007/JHEP03(2021)038",
    journal = "JHEP",
    volume = "03",
    pages = "038",
    year = "2021"
}

@article{Drummond:2020uni,
    author = "Drummond, J. M. and Glew, R. and Paul, H.",
    title = "{One-loop string corrections for AdS Kaluza-Klein amplitudes}",
    eprint = "2008.01109",
    archivePrefix = "arXiv",
    primaryClass = "hep-th",
    doi = "10.1007/JHEP12(2021)072",
    journal = "JHEP",
    volume = "12",
    pages = "072",
    year = "2021"
}

@article{Aprile:2019rep,
    author = "Aprile, Francesco and Drummond, James and Heslop, Paul and Paul, Hynek",
    title = "{One-loop amplitudes in AdS$_{5}${\texttimes}S$^{5}$ supergravity from $ \mathcal{N} $ = 4 SYM at strong coupling}",
    eprint = "1912.01047",
    archivePrefix = "arXiv",
    primaryClass = "hep-th",
    doi = "10.1007/JHEP03(2020)190",
    journal = "JHEP",
    volume = "03",
    pages = "190",
    year = "2020"
}

@article{Alday:2018pdi,
    author = "Alday, Luis F. and Bissi, Agnese and Perlmutter, Eric",
    title = "{Genus-One String Amplitudes from Conformal Field Theory}",
    eprint = "1809.10670",
    archivePrefix = "arXiv",
    primaryClass = "hep-th",
    doi = "10.1007/JHEP06(2019)010",
    journal = "JHEP",
    volume = "06",
    pages = "010",
    year = "2019"
}

@article{Alday:2019nin,
    author = "Alday, Luis F. and Zhou, Xinan",
    title = "{Simplicity of AdS Supergravity at One Loop}",
    eprint = "1912.02663",
    archivePrefix = "arXiv",
    primaryClass = "hep-th",
    reportNumber = "PUPT-2604",
    doi = "10.1007/JHEP09(2020)008",
    journal = "JHEP",
    volume = "09",
    pages = "008",
    year = "2020"
}

@article{Alday:2018kkw,
    author = "Alday, Luis F.",
    title = "{On genus-one string amplitudes on $AdS_5 \times S^5$}",
    eprint = "1812.11783",
    archivePrefix = "arXiv",
    primaryClass = "hep-th",
    doi = "10.1007/JHEP04(2021)005",
    journal = "JHEP",
    volume = "04",
    pages = "005",
    year = "2021"
}

@article{Drummond:2020dwr,
    author = "Drummond, J. M. and Paul, H. and Santagata, M.",
    title = "{Bootstrapping string theory on AdS5{\texttimes}S5}",
    eprint = "2004.07282",
    archivePrefix = "arXiv",
    primaryClass = "hep-th",
    doi = "10.1103/PhysRevD.108.026020",
    journal = "Phys. Rev. D",
    volume = "108",
    number = "2",
    pages = "026020",
    year = "2023"
}

@article{Alday:2021ymb,
    author = "Alday, Luis F. and Chester, Shai M. and Raj, Himanshu",
    title = "{ABJM at strong coupling from M-theory, localization, and Lorentzian inversion}",
    eprint = "2107.10274",
    archivePrefix = "arXiv",
    primaryClass = "hep-th",
    doi = "10.1007/JHEP02(2022)005",
    journal = "JHEP",
    volume = "02",
    pages = "005",
    year = "2022"
}

@article{Alday:2021vfb,
    author = "Alday, Luis F. and Chester, Shai M. and Hansen, Tobias",
    title = "{Modular invariant holographic correlators for $ \mathcal{N} $ = 4 SYM with general gauge group}",
    eprint = "2110.13106",
    archivePrefix = "arXiv",
    primaryClass = "hep-th",
    doi = "10.1007/JHEP12(2021)159",
    journal = "JHEP",
    volume = "12",
    pages = "159",
    year = "2021"
}

@article{Chester:2025ssu,
    author = "Chester, Shai M. and Ferrero, Pietro and Pavarini, Daniele R.",
    title = "{Modular invariant gluon-graviton scattering in AdS at one loop}",
    eprint = "2504.10319",
    archivePrefix = "arXiv",
    primaryClass = "hep-th",
    month = "4",
    year = "2025"
}

@article{Alday:2020tgi,
    author = "Alday, Luis F. and Chester, Shai M. and Raj, Himanshu",
    title = "{6d (2,0) and M-theory at 1-loop}",
    eprint = "2005.07175",
    archivePrefix = "arXiv",
    primaryClass = "hep-th",
    doi = "10.1007/JHEP01(2021)133",
    journal = "JHEP",
    volume = "01",
    pages = "133",
    year = "2021"
}

@article{Chester:2019pvm,
    author = "Chester, Shai M.",
    title = "{Genus-2 holographic correlator on AdS$_{5}${\texttimes} S$^{5}$ from localization}",
    eprint = "1908.05247",
    archivePrefix = "arXiv",
    primaryClass = "hep-th",
    doi = "10.1007/JHEP04(2020)193",
    journal = "JHEP",
    volume = "04",
    pages = "193",
    year = "2020"
}

@article{Alday:2022rly,
    author = "Alday, Luis F. and Chester, Shai M. and Raj, Himanshu",
    title = "{M-theory on AdS$_{4}${\texttimes} S$^{7}$ at 1-loop and beyond}",
    eprint = "2207.11138",
    archivePrefix = "arXiv",
    primaryClass = "hep-th",
    doi = "10.1007/JHEP11(2022)091",
    journal = "JHEP",
    volume = "11",
    pages = "091",
    year = "2022"
}

@article{Bissi:2020wtv,
    author = "Bissi, Agnese and Fardelli, Giulia and Georgoudis, Alessandro",
    title = "{Towards all loop supergravity amplitudes on AdS5{\texttimes}S5}",
    eprint = "2002.04604",
    archivePrefix = "arXiv",
    primaryClass = "hep-th",
    doi = "10.1103/PhysRevD.104.L041901",
    journal = "Phys. Rev. D",
    volume = "104",
    number = "4",
    pages = "L041901",
    year = "2021"
}

@article{Bissi:2020woe,
    author = "Bissi, Agnese and Fardelli, Giulia and Georgoudis, Alessandro",
    title = "{All loop structures in supergravity amplitudes on AdS5 {\texttimes} S5 from CFT}",
    eprint = "2010.12557",
    archivePrefix = "arXiv",
    primaryClass = "hep-th",
    doi = "10.1088/1751-8121/ac0ebf",
    journal = "J. Phys. A",
    volume = "54",
    number = "32",
    pages = "324002",
    year = "2021"
}

@article{Alday:2021ajh,
    author = "Alday, Luis F. and Bissi, Agnese and Zhou, Xinan",
    title = "{One-loop gluon amplitudes in AdS}",
    eprint = "2110.09861",
    archivePrefix = "arXiv",
    primaryClass = "hep-th",
    reportNumber = "UUITP-51/21",
    doi = "10.1007/JHEP02(2022)105",
    journal = "JHEP",
    volume = "02",
    pages = "105",
    year = "2022"
}

@article{Chester:2019jas,
    author = "Chester, Shai M. and Green, Michael B. and Pufu, Silviu S. and Wang, Yifan and Wen, Congkao",
    title = "{Modular invariance in superstring theory from $ \mathcal{N} $ = 4 super-Yang-Mills}",
    eprint = "1912.13365",
    archivePrefix = "arXiv",
    primaryClass = "hep-th",
    reportNumber = "PUPT-2607, QMUL-PH-19-37",
    doi = "10.1007/JHEP11(2020)016",
    journal = "JHEP",
    volume = "11",
    pages = "016",
    year = "2020"
}

@article{Komatsu:2020sag,
    author = "Komatsu, Shota and Paulos, Miguel F. and Van Rees, Balt C. and Zhao, Xiang",
    title = "{Landau diagrams in AdS and S-matrices from conformal correlators}",
    eprint = "2007.13745",
    archivePrefix = "arXiv",
    primaryClass = "hep-th",
    reportNumber = "CPHT-RR119.122020",
    doi = "10.1007/JHEP11(2020)046",
    journal = "JHEP",
    volume = "11",
    pages = "046",
    year = "2020"
}

@article{Alday:2022lkk,
    author = "Alday, Luis F. and Gon{\c{c}}alves, Vasco and Zhou, Xinan",
    title = "{Supersymmetric Five-Point Gluon Amplitudes in AdS Space}",
    eprint = "2201.04422",
    archivePrefix = "arXiv",
    primaryClass = "hep-th",
    doi = "10.1103/PhysRevLett.128.161601",
    journal = "Phys. Rev. Lett.",
    volume = "128",
    number = "16",
    pages = "161601",
    year = "2022"
}

@article{Alday:2023kfm,
    author = "Alday, Luis F. and Gon{\c{c}}alves, Vasco and Nocchi, Maria and Zhou, Xinan",
    title = "{Six-point AdS gluon amplitudes from flat space and factorization}",
    eprint = "2307.06884",
    archivePrefix = "arXiv",
    primaryClass = "hep-th",
    doi = "10.1103/PhysRevResearch.6.L012041",
    journal = "Phys. Rev. Res.",
    volume = "6",
    number = "1",
    pages = "L012041",
    year = "2024"
}

@article{Huang:2024dxr,
    author = "Huang, Zhongjie and Wang, Bo and Yuan, Ellis Ye and Zhang, Jiarong",
    title = "{All Five-Point Kaluza-Klein Correlators and Hidden 8D Symmetry in AdS5{\texttimes}S3}",
    eprint = "2408.12260",
    archivePrefix = "arXiv",
    primaryClass = "hep-th",
    doi = "10.1103/PhysRevLett.134.161601",
    journal = "Phys. Rev. Lett.",
    volume = "134",
    number = "16",
    pages = "161601",
    year = "2025"
}

@article{Huang:2021xws,
    author = "Huang, Zhongjie and Yuan, Ellis Ye",
    title = "{Graviton scattering in AdS$_{5}${\texttimes} S$^{5}$ at two loops}",
    eprint = "2112.15174",
    archivePrefix = "arXiv",
    primaryClass = "hep-th",
    doi = "10.1007/JHEP04(2023)064",
    journal = "JHEP",
    volume = "04",
    pages = "064",
    year = "2023"
}

@article{Cao:2024bky,
    author = "Cao, Qu and He, Song and Li, Xiang and Tang, Yichao",
    title = "{Supergluon scattering in AdS: constructibility, spinning amplitudes, and new structures}",
    eprint = "2406.08538",
    archivePrefix = "arXiv",
    primaryClass = "hep-th",
    doi = "10.1007/JHEP10(2024)040",
    journal = "JHEP",
    volume = "10",
    pages = "040",
    year = "2024"
}

@article{Cao:2023cwa,
    author = "Cao, Qu and He, Song and Tang, Yichao",
    title = "{Constructibility of AdS Supergluon Amplitudes}",
    eprint = "2312.15484",
    archivePrefix = "arXiv",
    primaryClass = "hep-th",
    doi = "10.1103/PhysRevLett.133.021605",
    journal = "Phys. Rev. Lett.",
    volume = "133",
    number = "2",
    pages = "021605",
    year = "2024"
}

@article{Meltzer:2019nbs,
    author = "Meltzer, David and Perlmutter, Eric and Sivaramakrishnan, Allic",
    title = "{Unitarity Methods in AdS/CFT}",
    eprint = "1912.09521",
    archivePrefix = "arXiv",
    primaryClass = "hep-th",
    reportNumber = "CALT-TH-2019-052",
    doi = "10.1007/JHEP03(2020)061",
    journal = "JHEP",
    volume = "03",
    pages = "061",
    year = "2020"
}

@article{Meltzer:2020qbr,
    author = "Meltzer, David and Sivaramakrishnan, Allic",
    title = "{CFT unitarity and the AdS Cutkosky rules}",
    eprint = "2008.11730",
    archivePrefix = "arXiv",
    primaryClass = "hep-th",
    reportNumber = "CALT-TH-2020-032",
    doi = "10.1007/JHEP11(2020)073",
    journal = "JHEP",
    volume = "11",
    pages = "073",
    year = "2020"
}

@article{Zhou:2021gnu,
    author = "Zhou, Xinan",
    title = "{Double Copy Relation in AdS Space}",
    eprint = "2106.07651",
    archivePrefix = "arXiv",
    primaryClass = "hep-th",
    doi = "10.1103/PhysRevLett.127.141601",
    journal = "Phys. Rev. Lett.",
    volume = "127",
    number = "14",
    pages = "141601",
    year = "2021"
}

@article{Farrow:2018yni,
    author = "Farrow, Joseph A. and Lipstein, Arthur E. and McFadden, Paul",
    title = "{Double copy structure of CFT correlators}",
    eprint = "1812.11129",
    archivePrefix = "arXiv",
    primaryClass = "hep-th",
    doi = "10.1007/JHEP02(2019)130",
    journal = "JHEP",
    volume = "02",
    pages = "130",
    year = "2019"
}

@article{Armstrong:2020woi,
    author = "Armstrong, Connor and Lipstein, Arthur E. and Mei, Jiajie",
    title = "{Color/kinematics duality in AdS$_{4}$}",
    eprint = "2012.02059",
    archivePrefix = "arXiv",
    primaryClass = "hep-th",
    doi = "10.1007/JHEP02(2021)194",
    journal = "JHEP",
    volume = "02",
    pages = "194",
    year = "2021"
}

@article{Albayrak:2020fyp,
    author = "Albayrak, Soner and Kharel, Savan and Meltzer, David",
    title = "{On duality of color and kinematics in (A)dS momentum space}",
    eprint = "2012.10460",
    archivePrefix = "arXiv",
    primaryClass = "hep-th",
    doi = "10.1007/JHEP03(2021)249",
    journal = "JHEP",
    volume = "03",
    pages = "249",
    year = "2021"
}

@article{Jain:2021qcl,
    author = "Jain, Sachin and John, Renjan Rajan and Mehta, Abhishek and Nizami, Amin A. and Suresh, Adithya",
    title = "{Double copy structure of parity-violating CFT correlators}",
    eprint = "2104.12803",
    archivePrefix = "arXiv",
    primaryClass = "hep-th",
    doi = "10.1007/JHEP07(2021)033",
    journal = "JHEP",
    volume = "07",
    pages = "033",
    year = "2021"
}

@article{Diwakar:2021juk,
    author = "Diwakar, Pranav and Herderschee, Aidan and Roiban, Radu and Teng, Fei",
    title = "{BCJ amplitude relations for Anti-de Sitter boundary correlators in embedding space}",
    eprint = "2106.10822",
    archivePrefix = "arXiv",
    primaryClass = "hep-th",
    reportNumber = "LCTP-21-15",
    doi = "10.1007/JHEP10(2021)141",
    journal = "JHEP",
    volume = "10",
    pages = "141",
    year = "2021"
}

@article{Cheung:2022pdk,
    author = "Cheung, Clifford and Parra-Martinez, Julio and Sivaramakrishnan, Allic",
    title = "{On-shell correlators and color-kinematics duality in curved symmetric spacetimes}",
    eprint = "2201.05147",
    archivePrefix = "arXiv",
    primaryClass = "hep-th",
    reportNumber = "CALT-TH-2022-002",
    doi = "10.1007/JHEP05(2022)027",
    journal = "JHEP",
    volume = "05",
    pages = "027",
    year = "2022"
}

@article{Herderschee:2022ntr,
    author = "Herderschee, Aidan and Roiban, Radu and Teng, Fei",
    title = "{On the differential representation and color-kinematics duality of AdS boundary correlators}",
    eprint = "2201.05067",
    archivePrefix = "arXiv",
    primaryClass = "hep-th",
    reportNumber = "LCTP-22-01",
    doi = "10.1007/JHEP05(2022)026",
    journal = "JHEP",
    volume = "05",
    pages = "026",
    year = "2022"
}

@article{Drummond:2022dxd,
    author = "Drummond, J. M. and Glew, R. and Santagata, M.",
    title = "{Bern-Carrasco-Johansson relations in AdS5{\texttimes}S3 and the double-trace spectrum of super gluons}",
    eprint = "2202.09837",
    archivePrefix = "arXiv",
    primaryClass = "hep-th",
    doi = "10.1103/PhysRevD.107.L081901",
    journal = "Phys. Rev. D",
    volume = "107",
    number = "8",
    pages = "L081901",
    year = "2023"
}

@article{Bissi:2022wuh,
    author = "Bissi, Agnese and Fardelli, Giulia and Manenti, Andrea and Zhou, Xinan",
    title = "{Spinning correlators in $ \mathcal{N} $ = 2 SCFTs: Superspace and AdS amplitudes}",
    eprint = "2209.01204",
    archivePrefix = "arXiv",
    primaryClass = "hep-th",
    reportNumber = "UUITP-36/22",
    doi = "10.1007/JHEP01(2023)021",
    journal = "JHEP",
    volume = "01",
    pages = "021",
    year = "2023"
}

@article{Armstrong:2022mfr,
    author = "Armstrong, Connor and Gomez, Humberto and Lipinski Jusinskas, Renann and Lipstein, Arthur and Mei, Jiajie",
    title = "{New recursion relations for tree-level correlators in anti{\textendash}de Sitter spacetime}",
    eprint = "2209.02709",
    archivePrefix = "arXiv",
    primaryClass = "hep-th",
    doi = "10.1103/PhysRevD.106.L121701",
    journal = "Phys. Rev. D",
    volume = "106",
    number = "12",
    pages = "L121701",
    year = "2022"
}

@article{Lee:2022fgr,
    author = "Lee, Hayden and Wang, Xinkang",
    title = "{Cosmological double-copy relations}",
    eprint = "2212.11282",
    archivePrefix = "arXiv",
    primaryClass = "hep-th",
    doi = "10.1103/PhysRevD.108.L061702",
    journal = "Phys. Rev. D",
    volume = "108",
    number = "6",
    pages = "L061702",
    year = "2023"
}

@article{Li:2022tby,
    author = "Li, Yue-Zhou",
    title = "{Flat-space structure of gluons and gravitons in AdS spacetime}",
    eprint = "2212.13195",
    archivePrefix = "arXiv",
    primaryClass = "hep-th",
    doi = "10.1103/PhysRevD.107.125018",
    journal = "Phys. Rev. D",
    volume = "107",
    number = "12",
    pages = "125018",
    year = "2023"
}

@article{Caron-Huot:2018kta,
    author = "Caron-Huot, Simon and Trinh, Anh-Khoi",
    title = "{All tree-level correlators in AdS$_{5}${\texttimes}S$_{5}$ supergravity: hidden ten-dimensional conformal symmetry}",
    eprint = "1809.09173",
    archivePrefix = "arXiv",
    primaryClass = "hep-th",
    doi = "10.1007/JHEP01(2019)196",
    journal = "JHEP",
    volume = "01",
    pages = "196",
    year = "2019"
}

@article{Chen:2025yxg,
    author = "Chen, Junding and Jiang, Yunfeng and Zhou, Xinan",
    title = "{Giant Graviton Correlators as Defect Systems}",
    eprint = "2503.22987",
    archivePrefix = "arXiv",
    primaryClass = "hep-th",
    reportNumber = "USTC-ICTS/PCFT-25-14",
    month = "3",
    year = "2025"
}

@article{Liu:2025uqu,
    author = "Liu, James and Minasian, Ruben and Savelli, Raffaele and Schachner, Andreas",
    title = "{Type IIB at eight derivatives: Five-Point Axio-Dilaton Couplings}",
    eprint = "2507.07934",
    archivePrefix = "arXiv",
    primaryClass = "hep-th",
    reportNumber = "LITP-25-10, LMU-ASC 19/25",
    month = "7",
    year = "2025"
}

@article{Aprile:2024lwy,
    author = "Aprile, Francesco and Giusto, Stefano and Russo, Rodolfo",
    title = "{Holographic correlators with BPS bound states in $\mathcal{N} = 4$ SYM}",
    eprint = "2409.12911",
    archivePrefix = "arXiv",
    primaryClass = "hep-th",
    doi = "10.1103/PhysRevLett.134.091602",
    journal = "Phys. Rev. Lett.",
    volume = "134",
    number = "9",
    pages = "091602",
    year = "2025"
}

@article{Aprile:2025hlt,
    author = "Aprile, Francesco and Giusto, Stefano and Russo, Rodolfo",
    title = "{Four-point correlators with BPS bound states in AdS$_3$ and AdS$_5$}",
    eprint = "2503.02855",
    archivePrefix = "arXiv",
    primaryClass = "hep-th",
    month = "3",
    year = "2025"
}

@article{Wang:2025pjo,
    author = "Wang, Bo and Wu, Di and Yuan, Ellis Ye",
    title = "{Kaluza-Klein AdS Virasoro-Shapiro Amplitude near Flat Space}",
    eprint = "2503.01964",
    archivePrefix = "arXiv",
    primaryClass = "hep-th",
    doi = "10.1103/v72s-rv7y",
    journal = "Phys. Rev. Lett.",
    volume = "135",
    number = "4",
    pages = "041603",
    year = "2025"
}

@article{Wang:2025owf,
    author = "Wang, Bo",
    title = "{Bootstrap AdS Veneziano Amplitude with Arbitrary Kaluza-Klein Modes}",
    eprint = "2508.14968",
    archivePrefix = "arXiv",
    primaryClass = "hep-th",
    month = "8",
    year = "2025"
}

@article{Green:2020eyj,
    author = "Green, Michael B. and Wen, Congkao",
    title = "{Maximal U(1)$_{Y}$-violating n-point correlators in $ \mathcal{N} $ = 4 super-Yang-Mills theory}",
    eprint = "2009.01211",
    archivePrefix = "arXiv",
    primaryClass = "hep-th",
    reportNumber = "QMUL-PH-20-23",
    doi = "10.1007/JHEP02(2021)042",
    journal = "JHEP",
    volume = "02",
    pages = "042",
    year = "2021"
}

@article{Fernandes:2025eqe,
    author = "Fernandes, Bruno and Goncalves, Vasco and Huang, Zhongjie and Tang, Yichao and Vilas Boas, Joao and Yuan, Ellis Ye",
    title = "{AdS$\times$S Mellin Bootstrap, Hidden 10d Symmetry and Five-point Kaluza-Klein Functions in $\mathcal{N}=4$ SYM}",
    eprint = "2507.14124",
    archivePrefix = "arXiv",
    primaryClass = "hep-th",
    month = "7",
    year = "2025"
}

@article{Aprile:2026uxe,
    author = "Aprile, Francesco and Giusto, Stefano and Russo, Rodolfo and Vilas Boas, Jo{\~a}o",
    title = "{Multi-particle correlators with higher KK modes I: a bootstrap approach}",
    eprint = "2601.16085",
    archivePrefix = "arXiv",
    primaryClass = "hep-th",
    month = "1",
    year = "2026"
}

@article{Bissi:2021hjk,
    author = "Bissi, Agnese and Fardelli, Giulia and Manenti, Andrea",
    title = "{Rebooting quarter-BPS operators in $ \mathcal{N} $ = 4 super Yang-Mills}",
    eprint = "2111.06857",
    archivePrefix = "arXiv",
    primaryClass = "hep-th",
    reportNumber = "UUITP-55/21",
    doi = "10.1007/JHEP04(2022)016",
    journal = "JHEP",
    volume = "04",
    pages = "016",
    year = "2022"
}

@article{DHoker:2001jzy,
    author = "D'Hoker, Eric and Ryzhov, Anton V.",
    title = "{Three point functions of quarter BPS operators in N=4 SYM}",
    eprint = "hep-th/0109065",
    archivePrefix = "arXiv",
    reportNumber = "UCLA-01-TEP-22",
    doi = "10.1088/1126-6708/2002/02/047",
    journal = "JHEP",
    volume = "02",
    pages = "047",
    year = "2002"
}

@article{Georgoudis:2017meq,
    author = "Georgoudis, Alessandro and Goncalves, Vasco and Pereira, Raul",
    title = "{Konishi OPE coefficient at the five loop order}",
    eprint = "1710.06419",
    archivePrefix = "arXiv",
    primaryClass = "hep-th",
    doi = "10.1007/JHEP11(2018)184",
    journal = "JHEP",
    volume = "11",
    pages = "184",
    year = "2018"
}

@article{Yuan:2018qva,
    author = "Yuan, Ellis Ye",
    title = "{Simplicity in AdS Perturbative Dynamics}",
    eprint = "1801.07283",
    archivePrefix = "arXiv",
    primaryClass = "hep-th",
    month = "1",
    year = "2018"
}

\end{document}